\documentclass[11pt,oneolumn,floatfix,altaffilletter,superscriptaddress,tightenlines,showpacs,showkeys,preprintnumbers,nofootinbib]{revtex4-2}
\pdfoutput=1
\usepackage[colorlinks=true,citecolor=blue,linkcolor=blue,breaklinks=true,linktoc=none]{hyperref}
\usepackage{amsmath,amssymb}
\usepackage{epsfig} 
\usepackage{graphicx}
\usepackage{url}
\usepackage{color}
\usepackage{multirow}
\usepackage{placeins}
\usepackage{float}
\usepackage[dvipsnames]{xcolor}
 \usepackage{booktabs} 
\usepackage{braket}
\usepackage{float}
\usepackage{slashed,tabularray}
\hypersetup{
  colorlinks=true,
  linkcolor=red,
  filecolor=magenta,   
  urlcolor=blue,
  citecolor=blue,
}

\hypersetup{colorlinks,linkcolor={blue},citecolor={teal},urlcolor={violet}}

\allowdisplaybreaks

\bibpunct{[}{]}{,}{n}{}{,}

\def\beq{\begin{equation}}
\def\eeq{\end{equation}}
\def\bea{\begin{eqnarray}}
\def\eea{\end{eqnarray}}
\makeatletter
\@addtoreset{equation}{section}

\newcommand{\INFN}{INFN - Sezione di Napoli, Complesso Universitario Monte S. Angelo, I-80126 Napoli, Italy}

\newcommand{\SSM}{Scuola Superiore Meridionale, Università degli studi di Napoli ``Federico II'', Largo San Marcellino 10, 80138 Napoli, Italy}
\newcommand{\NAN}{Department of Physics and Institute of Theoretical Physics,
Nanjing Normal University, Nanjing, 210023, China}

\begin{document}

\title{LISA Reconstruction Landscape for Metastable Cosmic Strings} 
\author{Satyabrata Datta}
    \email{amisatyabrata703@gmail.com}
    \affiliation{\NAN}
 \author{Rome Samanta}
  \email{samanta@na.infn.it}
  \affiliation{\SSM}
  \affiliation{\INFN}

\begin{abstract}
We investigate the prospects for reconstructing stochastic gravitational-wave
backgrounds from metastable cosmic strings with the Laser Interferometer Space
Antenna (LISA). We focus on the standard vacuum-tunneling benchmark, in which
the string network decays through zero-temperature nucleation of
monopole--antimonopole pairs on the string worldsheet. The network initially
follows ordinary cosmic-string scaling, but loop production is suppressed after
a characteristic decay time associated with efficient loop breaking and network
collapse. This finite lifetime imprints an infrared tail and a broad
transition toward a stable-string-like high-frequency plateau. Using synthetic LISA data including instrumental noise and unresolved
astrophysical foregrounds, we perform a Bayesian reconstruction analysis in the
$(G\mu,\kappa_{\rm CS})$ parameter space. We map detectability, marginalized
uncertainties, posterior correlations, covariance anisotropy, and posterior
localization in order to distinguish stochastic-background detection from
genuine parameter reconstruction. We find that reconstruction is controlled
mainly by the lifetime-dependent spectral structure accessible in the LISA band.
When LISA samples the broad transition between the infrared tail and the
plateau-dominated regime, the data contain both amplitude and shape information,
allowing the string tension and metastability scale to be reconstructed. In
this region the posterior can remain strongly correlated, reflecting an
amplitude--lifetime trade-off, but still occupy a small area in parameter space.
By contrast, when the spectrum is effectively featureless in the LISA band, the
data constrain only limited combinations of the parameters or become prior
dominated. We also find that high-SNR plateau-like spectra can retain partial
sensitivity to $\kappa_{\rm CS}$ through residual finite-lifetime dependence of
the full signal, even when the main transition is not visually prominent.
Finally, we quantify the dependence of these conclusions on Galactic foreground
modeling by comparing a conservative flexible template with a reduced tanh
template. Our results identify where LISA can move beyond detecting a
stochastic background and begin to probe the lifetime of the underlying string
network.
\end{abstract}

\maketitle
\tableofcontents
\section{Introduction}
\label{sec:introduction}

The direct detection of gravitational waves (GWs) by the LIGO--Virgo
collaboration opened a new observational window on the Universe and established
gravitational-wave astronomy as a precision probe of compact objects,
astrophysics, and fundamental physics \cite{LIGOScientific:2016aoc}. Beyond
transient signals from compact-binary mergers, a major target of current and
future GW searches is the stochastic gravitational-wave background (SGWB). Such
a background can arise from the incoherent superposition of many unresolved
astrophysical sources, but it can also be generated by processes in the early
Universe \cite{ligoo5}. The recent evidence from pulsar timing arrays (PTAs) for
a nanohertz common-spectrum signal has further sharpened the importance of
broad-band SGWB searches across many decades in frequency
\cite{ng1,ng2,ng3,ng4,ng5}.

In the millihertz band, the central future observatory is the Laser
Interferometer Space Antenna (LISA) \cite{lisa,Armano:2018kix}. LISA will
consist of three spacecraft forming a triangular constellation with arm lengths
of order \(2.5\times10^6\,{\rm km}\), following a heliocentric orbit behind the
Earth. This configuration gives LISA access to a frequency range that is
inaccessible to ground-based interferometers and complementary to PTAs. While
LISA will observe a rich astrophysical source population, it will also provide
one of the most sensitive probes of cosmological stochastic backgrounds in the
mHz band.

Cosmological SGWBs are especially interesting because their spectra can retain
information about high-energy physics that is otherwise inaccessible. Examples
include first-order phase transitions, topological defects, nonstandard
cosmological histories, and other phenomena beyond the Standard Model. For such
sources, detectability is only the first question. If a cosmological background
is observed, one would like to know whether the underlying physical parameters
can actually be reconstructed. This is a more demanding problem than a standard
sensitivity forecast, because parameter recovery depends not only on the signal
amplitude, but also on the spectral morphology, the location of characteristic
features relative to the detector band, and degeneracies with instrumental noise
and astrophysical foregrounds.

For LISA, foregrounds are an essential part of this inference problem. The
Galactic population of compact double-white-dwarf binaries is expected to
produce a strong foreground over a significant part of the mHz band. Although
many binaries will be individually resolvable and can in principle be
subtracted, the unresolved population remains as a confusion foreground
\cite{Adams:2013qma,Boileau:2020rpg,Boileau:2021sni,Korol:2021pun,Korol:2020lpq,Liu:2023qap}.
Unresolved extragalactic compact-binary populations provide additional
stochastic components
\cite{Phinney:2001di,Regimbau:2011rp,Babak:2023lro,Lehoucq:2023zlt}. A
realistic LISA SGWB analysis must therefore treat the data as a superposition of
instrumental noise, astrophysical foregrounds, and a possible cosmological
signal. The question is not simply whether the cosmological component is loud
enough, but whether its spectral information can be separated from these
competing contributions.

From the cosmology side, in this work, we focus on metastable cosmic strings
\cite{Preskill:1992ck,Vilenkin:1982hm,Monin:2008mp,Buchmuller:2021mbb,Buchmuller:2023aus},
which have attracted renewed interest following the summer 2023 PTA
announcements of evidence for a nanohertz common-spectrum gravitational-wave
signal, see, e.g.,
Refs.~\cite{Antusch:2023zjk,Lazarides:2023rqf,Afzal:2023cyp,Ahmed:2023pjl,Afzal:2023kqs,Lazarides:2024niy,Ahmed:2024iyd,Antusch:2024nqg,Datta:2025owx,Maji:2024cwv,Pallis:2024joc,Chitose:2023dam,Chitose:2024pmz}.
In the standard zero-temperature vacuum-tunneling scenario \cite{Preskill:1992ck,Vilenkin:1982hm,Monin:2008mp,Buchmuller:2021mbb,Buchmuller:2023aus}, the string network
initially follows the usual scaling evolution, but later decays through
monopole--antimonopole pair nucleation on the string worldsheet. This decay
suppresses the abundance of long-lived loops and imprints an infrared turnover
and a broad lifetime-dependent transition in the SGWB spectrum
\cite{Buchmuller:2021mbb,Buchmuller:2023aus}. The overall amplitude of the
background is controlled primarily by the string tension \(G\mu\), while the
decay time of the network is governed by the metastability parameter
\(\kappa_{\rm CS}\). Metastable strings therefore provide a useful target for
reconstruction studies: the signal contains amplitude information through
\(G\mu\), and, when lifetime-dependent spectral structure is accessible to
LISA, information about the decay history through \(\kappa_{\rm CS}\).

\begin{figure}
    \centering
    \includegraphics[width=1\linewidth]{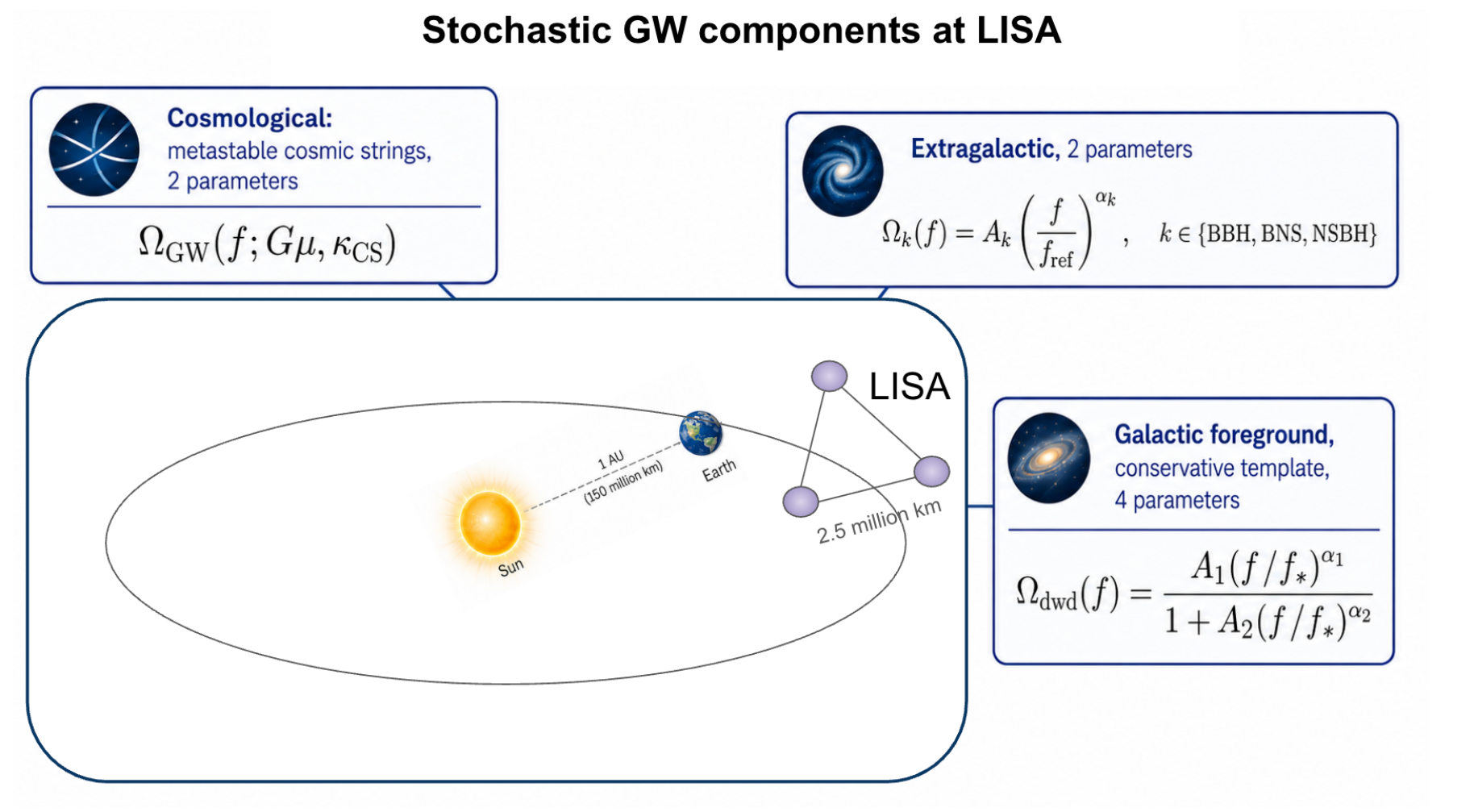}
    \caption{
    Schematic overview of the stochastic gravitational-wave components included
    in the LISA analysis. The cosmological contribution is modeled as a
    metastable cosmic-string background, parametrized by the string tension
    \(G\mu\) and the metastability parameter \(\kappa_{\rm CS}\). The data model
    also includes unresolved extragalactic compact-binary backgrounds, described
    by the power-law spectra of Eq.~\eqref{eq:astro_power_law}, and a
    conservative Galactic double-white-dwarf foreground template, defined in
    Eq.~\eqref{eq:dwd_model}. The Galactic template allows both amplitude and
    frequency-shape freedom, which helps avoid overestimating the reconstruction
    precision and accuracy of the cosmological signal. In
    Sec.~\ref{sec:foreground_template_comparison}, we also study an optimistic
    foreground scenario based on the alternative tanh template of
    Eq.~\eqref{gal_tanh}, in which only the overall Galactic normalization is
    varied. The central panel illustrates the LISA constellation in heliocentric
    orbit, with three spacecraft connected by laser links and trailing Earth
    around the Sun.
    }
    \label{fig:lisa_sgwb_components}
\end{figure}

The reconstruction problem is controlled by which part of the metastable-string
spectrum is visible in the LISA band. If LISA observes only the high-frequency
plateau-like part of the signal, the data mainly constrain the overall
normalization, and the metastability parameter can become weakly identifiable.
If LISA observes only the infrared tail, the spectrum can again lack enough
independent shape information to separate the effects of \(G\mu\) and
\(\kappa_{\rm CS}\). The most direct reconstruction occurs when the broad
lifetime-dependent transition between the infrared tail and the plateau-dominated
regime is sampled by the LISA band. In that case, the data can use both the
amplitude and the spectral morphology to distinguish the string tension from the
network lifetime.

The posterior geometry in this transition region need not be isotropic. In fact,
one of the characteristic features of the metastable-string reconstruction
problem is that well-localized posteriors can remain strongly correlated. The
data may determine a narrow amplitude--lifetime combination very accurately,
while the remaining uncertainty is oriented along a trade-off direction between
\(G\mu\) and \(\kappa_{\rm CS}\).

There is also a subtler high-SNR regime near the transition toward the
stable-network limit. The metastable-string spectrum does not contain a sharp
break from an infrared \(f^2\) tail to a plateau, but instead exhibits a broad
lifetime-dependent transition \cite{Buchmuller:2021mbb,Buchmuller:2023aus}. As
a result, the total spectrum can appear approximately plateau-like across the
LISA band while still retaining small coherent deviations from the strict
stable-string limit. If the signal is sufficiently loud, these residual
finite-lifetime distortions can contribute to the reconstruction of
\(\kappa_{\rm CS}\). Thus, sensitivity to the metastability parameter is not
restricted to cases in which the most visually apparent transition lies directly
inside the LISA band.

\begin{figure}
    \centering
    \includegraphics[width=.9\linewidth]{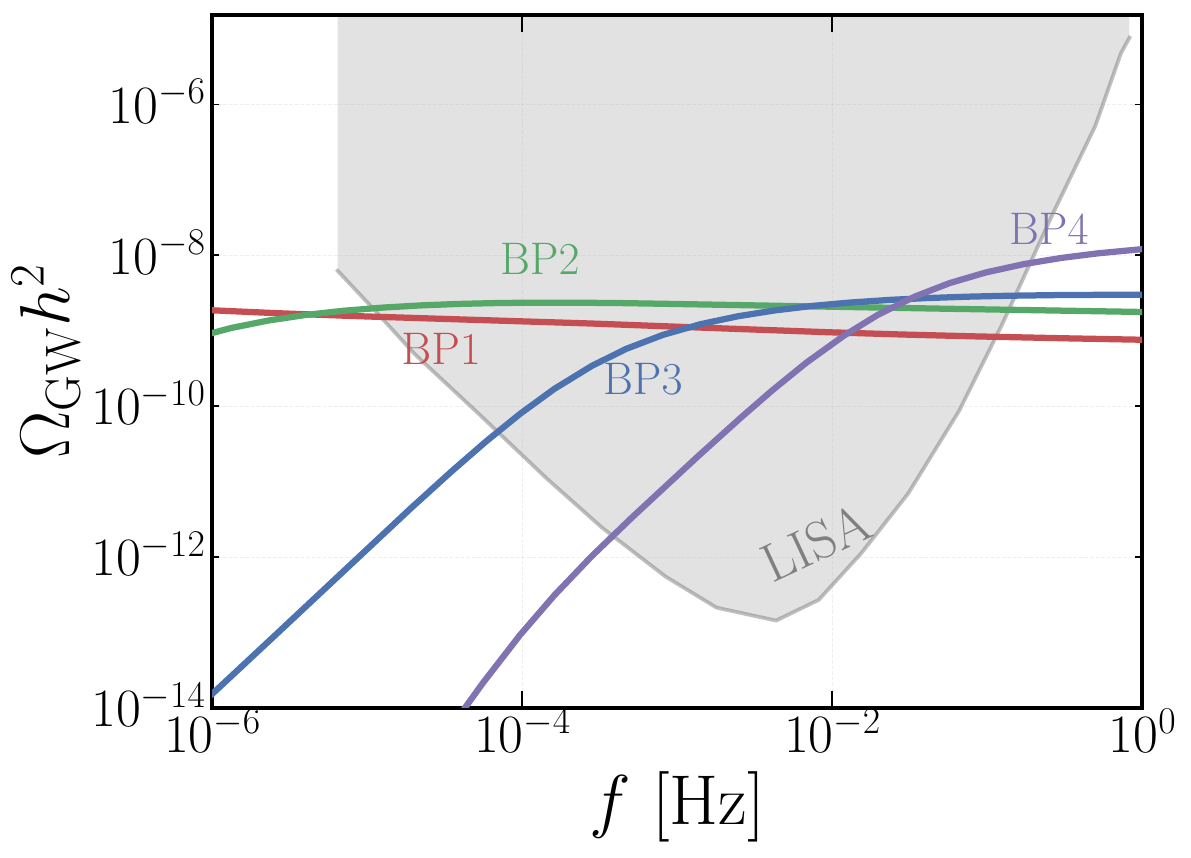}
    \caption{
    Representative metastable cosmic-string spectra for the benchmark points
    used in the reconstruction analysis. The examples illustrate different
    locations of the lifetime-dependent spectral transition relative to the LISA
    sensitivity band, ranging from visibly curved spectra to cases that appear
    closer to a plateau. The corresponding posterior distributions are shown in
    Figs.~\ref{fig:corner_bp1}--\ref{fig:corner_bp4}.
    }
    \label{fig:bp_spectra}
\end{figure}

The main goal of this paper is to construct a reconstruction landscape for
metastable cosmic strings under realistic LISA observing conditions. We perform
a Bayesian analysis in the parameter space \((G\mu,\kappa_{\rm CS})\), including
instrumental noise, unresolved Galactic foregrounds, and unresolved
extragalactic compact-binary backgrounds. We quantify not only detectability,
but also marginalized reconstruction uncertainties, posterior correlations,
covariance anisotropy, posterior localization, and reconstruction accuracy.
This allows us to identify where LISA can genuinely reconstruct both parameters,
where it measures a compact but correlated amplitude--lifetime combination, and
where the posterior is prior dominated or effectively one-dimensional.

The present study is restricted to the standard zero-temperature
vacuum-tunneling benchmark \cite{Monin:2008mp}. In this scenario, the network
decay is controlled by a single characteristic timescale, and the SGWB spectrum
is parametrized by \(G\mu\) and \(\kappa_{\rm CS}\). More general
metastable-string scenarios may involve finite-temperature effects, distinct
loop-breaking and network-collapse timescales, modified loop-production
histories, or additional contributions from finite string segments
\cite{Buchmuller:2021mbb,Buchmuller:2023aus,Tranchedone:2026lav,Asl:2026zpj,deGiorgi:2026fyx,Chitose:2025qyt}.
Such effects can shift the turnover, modify the infrared tail, or introduce new
degeneracies between the microscopic parameters. A full model-discrimination
study among these generalized templates is beyond the scope of this work. Our
aim here is instead to establish the baseline LISA reconstruction landscape for
the conventional metastable-string benchmark, against which more general models
can later be compared. More broadly, this work contributes to the growing effort to move from SGWB
detectability forecasts toward inference-level reconstruction of cosmological
source parameters in the LISA band; see, e.g.,
Refs.~\cite{Caprini:2019pxz,Flauger:2020qyi,Caprini:2024hue,Blanco-Pillado:2024aca,LISACosmologyWorkingGroup:2025vdz,Gowling:2021gcy,Giese:2021dnw,Gowling:2022pzb,Samanta:2025jec,Chen:2023zkb,Guan:2025idx,Kume:2024xvh,Dimitriou:2025bvq,Dimitriou:2026agw}
for related discussions.

Figure~\ref{fig:lisa_sgwb_components} provides a schematic summary of the LISA
inference setup used throughout this work, including the metastable
cosmic-string signal, unresolved astrophysical foregrounds, and instrumental
noise. To connect this setup with the spectral reconstruction problem,
Fig.~\ref{fig:bp_spectra} shows four representative metastable-string spectra
used as benchmark points in the Bayesian analysis. These examples span
different locations of the lifetime-dependent spectral transition relative to
the LISA sensitivity band and are used below to interpret the corresponding
posterior distributions in Figs.~\ref{fig:corner_bp1}--\ref{fig:corner_bp4}.

The paper is organized as follows. In Sec.~\ref{sec:gw_metastable_strings}, we
review the gravitational-wave spectrum from metastable cosmic strings in the
standard vacuum-tunneling benchmark. In
Sec.~\ref{sec:lisa_inference_framework}, we describe the LISA data model,
including instrumental noise and astrophysical foregrounds. In
Sec.~\ref{sec:posterior_covariance_metrics}, we introduce the posterior
diagnostics used to distinguish detection, posterior anisotropy, parameter
degeneracy, posterior localization, and genuine reconstruction. In
Sec.~\ref{sec:results_discussion}, we present the main reconstruction maps and
benchmark posteriors. In Sec.~\ref{sec:foreground_template_comparison}, we study
how the results change when the Galactic foreground is modeled with a reduced
tanh template. We summarize our conclusions in Sec.~\ref{sec:conclusions}.

\section{Gravitational waves from metastable cosmic strings}
\label{sec:gw_metastable_strings}

In this work we adopt the standard metastable cosmic-string benchmark used in earlier gravitational-wave studies \cite{Buchmuller:2021mbb,Buchmuller:2023aus}. In this scenario, the network evolves as an ordinary stable scaling string network until a characteristic decay time $t_s$. After this time, the long-string network no longer acts as an efficient source of new loops, and the loop population is progressively depleted by metastable decay. One can identify this characteristic time with both the onset of efficient loop breaking and the collapse of the loop-producing long-string network \cite{Asl:2026zpj},
\begin{equation}
    t_s \equiv t_{\rm LB} \equiv t_{\rm NC}.
\end{equation}
Here $t_{\rm LB}$ denotes the time at which closed loops break efficiently through monopole--antimonopole nucleation, while $t_{\rm NC}$ denotes the time at which the long-string network, or equivalently the loop-producing long-segment network, ceases to remain in the scaling regime. Within this one-timescale approximation, loop production proceeds as in a stable cosmic-string network for $t<t_s$, whereas for $t>t_s$ no significant new loop production is included. The gravitational-wave signal then receives only the residual contribution from loops produced before $t_s$ that continue to radiate while being depleted by metastable decay.

The decay is assumed to be controlled by zero-temperature quantum tunneling on the string worldsheet \cite{Preskill:1992ck,Monin:2008mp}. A string breaks through the nucleation of a monopole--antimonopole pair, with decay rate per unit length schematically given by
\begin{equation}
    \Gamma_{\rm q}
    \sim
    \frac{\mu}{2\pi}\,
    e^{-S_4},
    \label{eq:vacuum_decay_rate}
\end{equation}
where $S_4$ is the four-dimensional Euclidean bounce action. In the thin-defect approximation, the action is controlled by the ratio of the monopole mass to the string tension,
\begin{equation}
    \kappa_{\rm CS}
    \equiv
    \frac{m_M^2}{\mu},
    \qquad
    S_4 \sim \pi \kappa_{\rm CS}.
    \label{eq:kappa_cs_def}
\end{equation}
Thus larger $\kappa_{\rm CS}$ exponentially suppresses the nucleation rate and corresponds to a longer-lived metastable-string network. This vacuum-tunneling picture provides the benchmark gravitational-wave template used in the present analysis.

Recent developments indicate, however, that this benchmark may not be the most general possible description of metastable-string evolution \cite{Tranchedone:2026lav,Asl:2026zpj,deGiorgi:2026fyx}. More generally, the decay rate can be written schematically as \cite{Monin:2008uj,Monin:2009ch,Monin:2009gi}
\begin{equation}
    \Gamma(T)
    \sim
    A(T)\,
    \exp[-S_B(T)],
\end{equation}
where both the prefactor $A(T)$ and the bounce action $S_B(T)$ may depend on temperature. Finite-temperature effects can enhance monopole-pair nucleation relative to the vacuum estimate and may lead to an earlier disruption of the network unless $\kappa_{\rm CS}$ is sufficiently large \cite{Tranchedone:2026lav}. In such cases, the time at which the loop-producing network collapses need not coincide with the time at which closed loops begin to break efficiently. The one-timescale identification $t_s\equiv t_{\rm LB}\equiv t_{\rm NC}$ can therefore fail. More general templates allowing different hierarchies among $t_{\rm NC}$, $t_{\rm LB}$, and $t_0$ can lead to modified spectra, especially in the infrared tail, whose shape is sensitive to the history of loop production and network collapse \cite{Asl:2026zpj}. We defer a systematic reconstruction study of these generalized scenarios to an upcoming publication.

Restricting now to the benchmark case $t_s \equiv t_{\rm LB} \equiv t_{\rm NC}$, we briefly summarize the gravitational-wave signal from loops. Loops are produced when long strings self-intersect and chop off closed segments \cite{Vilenkin:1981bx,Vachaspati:1984gt}. The long-string network is characterized by the correlation length
\begin{equation}
    L
    =
    \sqrt{\frac{\mu}{\rho_\infty}},
\end{equation}
where $\rho_\infty$ is the long-string energy density. The string tension is given by \cite{Hill:1987qx}
\begin{equation}
    \mu
    =
    \pi v_\Phi^2\,h(\lambda,g^\prime),
    \label{eq:string_tension}
\end{equation}
with $h(\lambda,g^\prime)\simeq 1$ unless the scalar and gauge couplings are strongly hierarchical \cite{Emond:2021vts}.

A loop formed at time $t_i$ with initial length $l_i=\alpha t_i$ shrinks through gravitational-wave emission according to
\begin{equation}
    l(t)
    =
    l_i
    -
    \Gamma_{\rm GW}G\mu\,(t-t_i),
    \label{eq:loop_evolution}
\end{equation}
where $\Gamma_{\rm GW}\simeq 50$ \cite{Vilenkin:1981bx,Vachaspati:1984gt}, $\alpha\simeq 0.1$ \cite{Blanco-Pillado:2013qja,Blanco-Pillado:2017oxo}, and $G$ is Newton's constant. The loop emits into a set of normal modes labeled by $j=1,2,\dots,j_{\rm max}$. The emitted and observed frequencies are related by
\begin{equation}
    f_{\rm em}^{(j)}
    =
    \frac{2j}{l_j},
    \qquad
    f
    =
    \frac{a(t)}{a(t_0)}\,
    f_{\rm em}^{(j)},
\end{equation}
where $a(t)$ is the scale factor and $t_0$ denotes the present time. Equivalently, the loop length contributing to the observed frequency $f$ in the $j$th harmonic is
\begin{equation}
    l_j(t)
    =
    \frac{2j}{f}\,
    \frac{a(t)}{a(t_0)}.
\end{equation}

Summing over harmonics, the present-day gravitational-wave energy density from stable loops can be written as \cite{Blanco-Pillado:2013qja,Blanco-Pillado:2017oxo}
\begin{equation}
\Omega_{\rm GW}(f)
=
\sum_{j=1}^{j_{\rm max}}
\frac{2j\,G\mu^2\,\Gamma_j}{f\,\rho_c}
\int_{t_F}^{t_0} dt\,
\left[\frac{a(t)}{a(t_0)}\right]^5
\frac{\mathcal{F}_\alpha A_\beta(t_i)}
     {t_i^4\,\alpha(\alpha+\Gamma_{\rm GW}G\mu)}
\left[\frac{a(t_i)}{a(t)}\right]^3
\Theta(t_i-t_{\rm scl})
\Theta\!\left(t_i-\frac{l_c}{\alpha}\right).
\label{eq:omega_gw_stable}
\end{equation}
Here $\rho_c$ is the critical density, $\mathcal{F}_\alpha\simeq 0.1$ is an efficiency factor, $t_{\rm scl}$ is the time at which the network enters the scaling regime, and $l_c$ denotes the critical length above which gravitational-wave emission dominates over particle emission. Using Eq.\eqref{eq:loop_evolution}, the corresponding loop number density can be obtained as
\begin{equation}
    n_\omega(t,l_j)
    =
    \frac{A_\beta}{\alpha}\,
    \frac{(\alpha+\Gamma_{\rm GW}G\mu)^{3(1-\beta)}}
    {\left[l_j(t)+\Gamma_{\rm GW}G\mu t\right]^{4-3\beta}\,
    t^{3\beta}},
    \label{eq:loop_density_stable}
\end{equation}
where $\beta=2/[3(1+\omega)]$, with $\omega$ the equation-of-state parameter. In the scaling regime, $A_\beta=5.4$ in radiation domination and $A_\beta=0.39$ in matter domination \cite{Martins:1996jp,Martins:2000cs,Sousa:2013aaa,Auclair:2019wcv}. The power emitted into the $j$th harmonic is parametrized as
\begin{equation}
    \Gamma_j
    =
    \frac{\Gamma_{\rm GW}\,j^{-\delta}}
    {\zeta(\delta)},
\end{equation}
with $\delta=4/3$ for cusp-dominated emission and $\delta=5/3$ for kink-dominated emission \cite{Damour:2001bk}.

The Heaviside factors in Eq.~\eqref{eq:omega_gw_stable} impose the onset of scaling and the high-frequency cutoff associated with particle emission from small loops. For loop lengths $l>l_c$, gravitational-wave emission dominates over particle production, as suggested by recent numerical simulations \cite{Matsunami:2019fss,Auclair:2019jip}. The critical length may be estimated as
\begin{equation}
    l_c
    \simeq
    \delta_w
    (\Gamma_{\rm GW}G\mu)^{-\gamma},
\end{equation}
where $\delta_w=(\sqrt{\lambda}v_\Phi)^{-1}$ is the string width and $\gamma=2$ ($\gamma=1$) for loops containing cusps (kinks). In the parameter space considered here, these cutoffs occur at frequencies well above the LISA band and do not affect the reconstruction analysis. They can become important, however, in scenarios where the high-frequency falloff is modified, for example by an early matter-dominated era or by dilution of the string network during inflation \cite{Cui:2018rwi,Guedes:2018afo,Gouttenoire:2019kij,Datta:2025owx,Antusch:2024ypp,Borah:2022byb,Chianese:2024gee}.

For metastable strings, the stable-loop number density in Eq.~\eqref{eq:loop_density_stable} is multiplied by an additional survival factor \cite{Buchmuller:2021mbb,Buchmuller:2023aus},
\begin{equation}
    D(l,t)
    =
    \exp\!\left[
    -\Gamma_d
    \left(
    l(t-t_i)
    +
    \frac{1}{2}\Gamma_{\rm GW}G\mu\,(t-t_i)^2
    \right)
    \right]
    \Theta(t_s-t_i).
    \label{eq:metastable_decay_factor}
\end{equation}
The exponential factor accounts for the probability that a loop survives
monopole--antimonopole pair nucleation, while the Heaviside function implements
the suppression of efficient loop production after the network decay time
\(t_s\). This depletion of late loop production introduces a characteristic
infrared scale \cite{Buchmuller:2021mbb,Buchmuller:2023aus},
\begin{equation}
    f_{\rm low}\simeq
    10^{-4}\,{\rm Hz}
    \left(\frac{10^{-15}}{G\mu}\right)^{1/2}
    \exp\left[
    -\frac{\pi}{4}\left(\kappa_{\rm CS}-64\right)
    \right],
    \label{eq:flow_metastable}
\end{equation}
below which the metastable-string background enters the asymptotic infrared
regime. In contrast to stable cosmic strings, this low-frequency tail behaves as \cite{Buchmuller:2021mbb}
\begin{equation}
    \Omega_{\rm GW}(f)
    \propto
    f^2 .
    \label{eq:metastable_ir_tail}
\end{equation}
At higher frequencies the spectrum gradually rises away from this infrared tail
and approaches a stable-string-like plateau, as shown in
Refs.~\cite{Buchmuller:2021mbb,Buchmuller:2023aus}.

For reconstruction, the important point is that \(f_{\rm low}\) sets the lower
edge of a broad lifetime-dependent transition, rather than a sharp spectral
break. We will also use \(f_{\rm cross}\) (defined in Eq.\eqref{eq:fcross_numeric_covariance}) to denote the approximate frequency at
which the total spectrum becomes plateau dominated. The most effective
two-parameter reconstruction occurs when LISA has access to the broad region
between the infrared \(f^2\) tail and the plateau-dominated regime, where both
the amplitude and the lifetime-dependent spectral morphology are visible. If
LISA observes only the asymptotic infrared tail or only the plateau-like part of
the spectrum, the data can instead constrain mainly an effective
amplitude--lifetime combination of \(G\mu\) and \(\kappa_{\rm CS}\).

As an aside, let us mention that the steep \(f^2\) infrared scaling is phenomenologically important because it
provides a spectral feature that can help distinguish metastable strings from
stable cosmic-string backgrounds and from other stochastic sources. This scaling
is also interesting in view of PTA analyses, where an approximate \(f^2\)
spectrum is among the few possibilities when the data are described by a pure
power-law signal; see, for example,
Refs.~\cite{Babichev:2023pbf,Balaji:2023ehk}.

In the present work we restrict attention to gravitational-wave emission from closed loops. Monopole nucleation can also produce finite string segments, which may radiate gravitational waves if the monopoles do not carry unconfined flux \cite{Buchmuller:2021mbb,Leblond:2009fq}. Recent studies, however, have questioned the efficiency of this channel. Ref.~\cite{Chitose:2025qyt} argued that collapsing segments can dissipate most of their energy into particle radiation through repeated interactions with thermal fluctuations along the string, rather than entering a long-lived oscillatory regime. This is consistent with the numerical simulations of Ref.~\cite{Dvali:2022vwh}, which find efficient particle emission and no long-lived oscillatory phase for the segments studied there. We therefore neglect gravitational-wave emission from string segments in the present analysis. A more complete understanding of segment dynamics remains important for determining whether any observable segment-induced gravitational-wave signal can survive in more general metastable-string scenarios.

\section{LISA data model, foregrounds, and Bayesian inference setup}
\label{sec:lisa_inference_framework}

This section summarizes the LISA detector response, instrumental-noise model, astrophysical foreground model, and Bayesian likelihood used in the metastable-string inference. The analysis is formulated in the time-delay interferometry (TDI) basis \cite{Tinto:2004wu,McNamara:2008zz,Armano:2018kix,Caprini:2019pxz,Flauger:2020qyi,Caprini:2024hue}, where delayed linear combinations of the inter-spacecraft phase measurements suppress laser-frequency noise and yield approximately noise-orthogonal data channels. Throughout this work, the stochastic gravitational-wave background (SGWB) is modeled as an incoherent, stationary, Gaussian signal whose spectral shape is described by the dimensionless energy-density spectrum $\Omega_{\rm GW}(f)$. The measured LISA spectra are modeled as the sum of instrumental noise, unresolved astrophysical foregrounds, and a possible cosmological contribution from metastable cosmic strings.

\subsection{LISA configuration, TDI observables, and instrumental noise}
\label{subsec:lisa_tdi_noise}

LISA is a triangular space-based interferometer consisting of three spacecraft in heliocentric orbit. We adopt the nominal arm length
\begin{equation}
    L = 2.5\times 10^9~{\rm m},
\end{equation}
with corresponding transfer frequency
\begin{equation}
    f_\ast = \frac{c}{2\pi L}.
\end{equation}
The raw inter-spacecraft Doppler measurements are dominated by laser-frequency noise, which is many orders of magnitude larger than the expected gravitational-wave signal. Time-delay interferometry suppresses this contribution by forming delayed linear combinations of the one-way phase measurements. In the equal-arm approximation used here, the resulting TDI observables can be transformed into the approximately orthogonal channels
\begin{equation}
    \tilde d_a(f), \qquad a\in\{A,E,T\}.
\end{equation}
The $A$ and $E$ channels provide the dominant sensitivity to an isotropic SGWB. The $T$ channel has a strongly suppressed gravitational-wave response at frequencies below the transfer frequency and therefore acts as an approximate null channel in the low-frequency regime. More generally, it provides an internal consistency channel for constraining the instrumental-noise model.

For each TDI channel, the one-sided power spectral density is written as
\begin{equation}
    P_a(f;\boldsymbol{\theta})
    =
    S_a(f;\boldsymbol{\theta})
    +
    N_a(f;\boldsymbol{\theta}),
    \qquad
    a\in\{A,E,T\},
\end{equation}
where $S_a$ denotes the gravitational-wave contribution and $N_a$ denotes the instrumental-noise contribution. The signal spectrum in channel $a$ is related to the underlying energy-density spectrum by
\begin{equation}
    S_a(f;\boldsymbol{\theta})
    =
    \frac{3H_0^2}{4\pi^2}
    \frac{\Omega_{\rm GW}(f;\boldsymbol{\theta})}{f^3}
    \mathcal R_a(f),
    \label{eq:lisa_signal_psd}
\end{equation}
where $\mathcal R_a(f)$ is the sky- and polarization-averaged LISA response function.

The response functions used in this analysis are \cite{Cornish:2001bb,Smith:2019wny}
\begin{equation}
    \mathcal R_A(f)=\mathcal R_E(f)
    =
    \frac{9}{5}\,|W(f)|^2
    \left[
    1+\left(\frac{f}{4f_\ast/3}\right)^2
    \right]^{-1},
    \label{eq:lisa_response_AE}
\end{equation}
and
\begin{equation}
    \mathcal R_T(f)
    =
    \frac{1}{1008}\,|W(f)|^2
    \left(\frac{f}{f_\ast}\right)^6
    \left[
    1+\frac{5}{16128}
    \left(\frac{f}{f_\ast}\right)^8
    \right]^{-1},
    \label{eq:lisa_response_T}
\end{equation}
with
\begin{equation}
    W(f)=1-e^{-2if/f_\ast}.
\end{equation}
The factor $W(f)$ encodes the finite-arm transfer associated with the delayed TDI combinations. The steep suppression of $\mathcal R_T(f)$ at low frequencies motivates the null-channel approximation used below, in which the $T$ channel is treated as noise dominated over the frequency range relevant for the SGWB search.

The instrumental noise is modeled in terms of residual test-mass acceleration noise and optical metrology system noise. The acceleration noise amplitude is
\begin{equation}
    \sqrt{S_{\rm acc}(f)}
    =
    N_{\rm acc}
    \left[
    1+\left(\frac{0.4~{\rm mHz}}{f}\right)^2
    \right]^{1/2}
    \left[
    1+\left(\frac{f}{8~{\rm mHz}}\right)^4
    \right]^{1/2}
    \left(
    \frac{{\rm m}}{{\rm s}^2\sqrt{\rm Hz}}
    \right),
    \label{eq:acc_noise}
\end{equation}
with fiducial value
\begin{equation}
    N_{\rm acc}=3\times 10^{-15}.
\end{equation}
The optical metrology noise is modeled as
\begin{equation}
    \sqrt{S_{\rm OMS}(f)}
    =
    \delta x
    \left[
    1+\left(\frac{2~{\rm mHz}}{f}\right)^4
    \right]^{1/2}
    \left(
    \frac{{\rm m}}{\sqrt{\rm Hz}}
    \right),
    \label{eq:oms_noise}
\end{equation}
with the LISA fiducial displacement noise
\begin{equation}
    \delta x=15\times10^{-12}.
\end{equation}

In the orthogonal TDI basis, the instrumental-noise spectra are expressed as
\begin{equation}
    N_A(f)=N_E(f)=N_1(f)-N_2(f),
    \qquad
    N_T(f)=N_1(f)+2N_2(f),
    \label{eq:noise_AET}
\end{equation}
where
\begin{equation}
    N_1(f)
    =
    \frac{|W(f)|^2}{L^2}
    \left\{
    4S_{\rm OMS}(f)
    +
    8\left[
    1+\cos^2\left(\frac{f}{f_\ast}\right)
    \right]
    \frac{S_{\rm acc}(f)}{(2\pi f)^4}
    \right\},
    \label{eq:N1}
\end{equation}
and
\begin{equation}
    N_2(f)
    =
    -\frac{|W(f)|^2}{L^2}
    \left[
    2S_{\rm OMS}(f)
    +
    8\frac{S_{\rm acc}(f)}{(2\pi f)^4}
    \right]
    \cos\left(\frac{f}{f_\ast}\right).
    \label{eq:N2}
\end{equation}

The leading instrumental-noise amplitudes are treated as nuisance parameters rather than fixed constants. This is important because a weak cosmological background can be partially degenerate with smooth variations in the noise level, particularly near the boundaries of the LISA band. Marginalizing over the noise parameters avoids artificially optimistic constraints, while the $T$ channel provides an internal consistency check on the inferred instrumental spectrum.

For diagnostic purposes, we define the effective sensitivity of each TDI channel as
\begin{equation}
    \mathcal S_a(f)
    =
    \left[
    \frac{N_a(f)}{\mathcal R_a(f)}
    \right]^{1/2},
    \qquad
    a\in\{A,E,T\}.
\end{equation}
The corresponding channel-wise signal-to-noise ratio is
\begin{equation}
    {\rm SNR}_a
    =
    \left[
    2T_{\rm obs}
    \int_{f_{\rm min}}^{f_{\rm max}}
    \left(
    \frac{S_a(f)}{N_a(f)}
    \right)^2df
    \right]^{1/2},
    \label{eq:snr_channel}
\end{equation}
where $T_{\rm obs}$ is the total observation time and the integration limits specify the analyzed frequency band.

\subsection{Astrophysical foregrounds and metastable-string signal model}
\label{subsec:lisa_foregrounds}

A cosmological SGWB in the LISA band must be inferred in the presence of
astrophysical foregrounds. The dominant contribution at millihertz frequencies
is expected to arise from the Galactic population of double-white-dwarf binaries \cite{Korol:2021pun,Korol:2020lpq,Liu:2023qap}.
While the loudest systems may be individually resolved and subtracted, the
unresolved population produces a confusion foreground. This component is itself
an astrophysical gravitational-wave signal, but for cosmological searches it
acts as a structured foreground whose amplitude and spectral shape must be
marginalized over. In addition, unresolved extragalactic compact-binary
populations contribute a smoother stochastic background across the LISA band \cite{Phinney:2001di,Regimbau:2011rp,Babak:2023lro,Lehoucq:2023zlt}.

The total gravitational-wave energy density is modeled as
\begin{equation}
    \Omega_{\rm GW}(f;\boldsymbol{\theta})
    =
    \Omega_{\rm DWD}(f;\boldsymbol{\theta})
    +
    \Omega_{\rm ast}(f;\boldsymbol{\theta})
    +
    \Omega_{\rm CS}(f;\boldsymbol{\theta}),
    \label{eq:total_omega}
\end{equation}
where $\Omega_{\rm DWD}$ denotes the unresolved Galactic double-white-dwarf
foreground, $\Omega_{\rm ast}$ denotes the unresolved extragalactic
astrophysical background, and $\Omega_{\rm CS}$ is the metastable cosmic-string
contribution.

For our baseline analysis we adopt a conservative phenomenological model for
the Galactic foreground \cite{Korol:2021pun,Korol:2020lpq,Liu:2023qap,Chen:2023zkb,Datta:2026ffs,Datta:2026fav},
\begin{equation}
    \Omega_{\rm DWD}(f)
    =
    \frac{
    A_1\left(f/f_{\rm ref}\right)^{\alpha_1}
    }{
    1+
    A_2\left(f/f_{\rm ref}\right)^{\alpha_2}
    },
    \label{eq:dwd_model}
\end{equation}
where $A_1$, $A_2$, $\alpha_1$, and $\alpha_2$ control the normalization,
low-frequency slope, turnover behavior, and high-frequency suppression of the
unresolved Galactic foreground. The reference frequency $f_{\rm ref}$ is fixed
and used only to define dimensionless frequency ratios. This template is
intentionally flexible: by allowing both amplitude and shape freedom, it can
absorb broad spectral features in the mHz band. This makes the reconstruction
forecast conservative, since part of the spectral curvature associated with a
cosmological signal can be partially degenerate with the Galactic foreground.

The smooth extragalactic contribution is modeled as a power law,
\begin{equation}
    \Omega_{\rm ast}(f)
    =
    \Omega_{\rm ast}
    \left(
    \frac{f}{f_{\rm ref}}
    \right)^{\varepsilon},
    \label{eq:astro_power_law}
\end{equation}
where $\Omega_{\rm ast}$ is the amplitude and $\varepsilon$ is the spectral
index. Over the restricted frequency interval relevant for LISA, this form
provides an effective description of the unresolved extragalactic compact-binary
background.

The cosmological component $\Omega_{\rm CS}(f)$ is specified by the
metastable-string model. In the standard vacuum-tunneling benchmark considered
here, the signal is parameterized by the string tension $G\mu$ and the
metastability parameter $\kappa_{\rm CS}$, which controls the characteristic
decay scale of the string network. The full parameter vector used in the
baseline inference is therefore
\begin{equation}
    \boldsymbol{\theta}
    =
    \left\{
    N_{\rm acc},
    \delta x,
    A_1,
    \alpha_1,
    A_2,
    \alpha_2,
    \Omega_{\rm ast},
    \varepsilon,
    G\mu,
    \kappa_{\rm CS}
    \right\}.
    \label{eq:theta_full_lisa}
\end{equation}
Here $N_{\rm acc}$ and $\delta x$ denote the instrumental-noise parameters,
$(A_1,\alpha_1,A_2,\alpha_2)$ describe the conservative Galactic foreground
template, $(\Omega_{\rm ast},\varepsilon)$ describe the extragalactic
astrophysical background, and $(G\mu,\kappa_{\rm CS})$ are the metastable-string
parameters of interest.

Since the Galactic double-white-dwarf foreground is one of the main limitations
for LISA SGWB reconstruction, the choice of foreground template can have a
significant impact on the inferred cosmological parameters. The flexible model
in Eq.~\eqref{eq:dwd_model} is therefore used as our default foreground
description throughout the main reconstruction analysis. In
Sec.~\ref{sec:foreground_template_comparison}, we separately study a reduced
tanh foreground template, Eq.~\eqref{gal_tanh}, in which the Galactic spectral
shape is fixed by the observation time and only the overall normalization is
allowed to vary \cite{Caprini:2024hue,Blanco-Pillado:2024aca,Karnesis:2021tsh,Samanta:2025jec}. This comparison brackets the foreground-model dependence of the
reconstruction: the broken-power-law template gives a conservative estimate of
the LISA reconstruction capability, while the tanh template illustrates a more
optimistic scenario in which the Galactic foreground shape is assumed to be
better characterized. In Table \ref{tab:lisa_parameters}, we show the prior ranges used in the analysis.

\begin{table}[htbp]
\centering
\begin{tblr}
	{
    hlines,
    vlines,
    row{1} = {bg=gray7, fg=white, font=\bfseries},
column{1} = {bg=gray9},
cell{1}{1} = {bg=gray7, fg=white},
}

\textbf{Parameter} & \textbf{Fiducial value} & \textbf{Uniform prior} \\
$\log_{10}(N_{\rm acc})$      & $-14.523$ & $(-16.0,-13.7)$ \\
$\log_{10}(\delta x)$         & $-11.097$ & $(-13.0,-10.7)$ \\
$\log_{10}(A_1)$              & $-15.4$   & $(-18.0,-5.0)$ \\
$\alpha_1$                    & $-5.7$    & $(-15.0,-3.0)$ \\
$\log_{10}(A_2)$              & $-6.32$   & $(-10.0,5.0)$ \\
$\alpha_2$                    & $-6.2$    & $(-10.0,-1.0)$ \\
$\log_{10}(\Omega_{\rm ast})$ & $-11.0$   & $(-15.0,-8.0)$ \\
$\varepsilon$                 & $0.67$    & $(0.0,1.0)$ \\
$\log_{10}(G\mu)$             & grid      & $(-20,-4)$ \\
$\log_{10}(\kappa_{\rm CS})$  & grid      & $(30,90)$ \\
\end{tblr}
\caption{Model parameters, fiducial values, and prior ranges used in the LISA-only inference. The first two parameters describe the leading instrumental-noise amplitudes, the next six describe astrophysical foregrounds, and the final two determine the metastable cosmic-string signal.}
\label{tab:lisa_parameters}
\end{table}

\subsection{Frequency-domain likelihood and synthetic data generation}
\label{subsec:lisa_likelihood}

The likelihood is constructed from the complex Fourier coefficients of the TDI data \cite{Boileau:2020rpg,Romano:2016dpx,Smith:2019wny,Guan:2025idx,Datta:2026fav,Datta:2026ffs}. The observation is divided into $N_{\rm seg}$ approximately stationary segments of duration $T$, indexed by $r=1,\ldots,N_{\rm seg}$. The sampling interval is $\Delta t$, the sampling frequency is $f_s=1/\Delta t$, and each segment contains
\begin{equation}
    N=\frac{T}{\Delta t}.
\end{equation}
The discrete Fourier frequencies are
\begin{equation}
    f_k=k\Delta f,
    \qquad
    \Delta f=\frac{1}{T},
    \qquad
    k=0,1,\ldots,\frac{N}{2}.
\end{equation}
For each segment and positive-frequency bin, we define the data vector
\begin{equation}
    \mathbf d_{rk}
    =
    \begin{pmatrix}
        \tilde d_A^r(f_k)\\
        \tilde d_E^r(f_k)\\
        \tilde d_T^r(f_k)
    \end{pmatrix}.
\end{equation}

We assume that $\mathbf d_{rk}$ is drawn from a zero-mean complex Gaussian distribution with covariance matrix
\begin{equation}
    C_k(\boldsymbol{\theta})
    =
    \frac{Tf_s^2}{2}
    \begin{pmatrix}
        P_A(f_k;\boldsymbol{\theta}) & 0 & 0\\
        0 & P_E(f_k;\boldsymbol{\theta}) & 0\\
        0 & 0 & P_T(f_k;\boldsymbol{\theta})
    \end{pmatrix}.
    \label{eq:lisa_covariance}
\end{equation}
The diagonal form follows from the use of the approximately noise-orthogonal $A,E,T$ basis and the assumption that residual inter-channel correlations are negligible. In the null-channel approximation,
\begin{equation}
    P_A(f;\boldsymbol{\theta})
    =
    N_A(f;\boldsymbol{\theta})+S_A(f;\boldsymbol{\theta}),
\end{equation}
\begin{equation}
    P_E(f;\boldsymbol{\theta})
    =
    N_E(f;\boldsymbol{\theta})+S_E(f;\boldsymbol{\theta}),
\end{equation}
and
\begin{equation}
    P_T(f;\boldsymbol{\theta})
    =
    N_T(f;\boldsymbol{\theta}).
\end{equation}
This approximation is valid when the response-suppressed stochastic contribution in the $T$ channel satisfies $S_T(f)\ll N_T(f)$ over the analyzed band. If this condition is not imposed, one should instead use the more general expression $P_T=N_T+S_T$.

For a three-dimensional complex Gaussian vector, the single-bin probability density is
\begin{equation}
    p(\mathbf d_{rk}|\boldsymbol{\theta})
    =
    \frac{1}{\pi^3\det C_k(\boldsymbol{\theta})}
    \exp\left[
    -\mathbf d_{rk}^{\dagger}
    C_k^{-1}(\boldsymbol{\theta})
    \mathbf d_{rk}
    \right].
\end{equation}
Assuming statistical independence among segments and Fourier bins, the full likelihood is
\begin{equation}
    \mathcal L(\boldsymbol{\theta})
    =
    \prod_{r=1}^{N_{\rm seg}}
    \prod_{k=1}^{N/2}
    p(\mathbf d_{rk}|\boldsymbol{\theta}).
\end{equation}
Dropping additive constants independent of $\boldsymbol{\theta}$, the log-likelihood becomes \cite{Boileau:2020rpg,Romano:2016dpx,Smith:2019wny,Guan:2025idx,Datta:2026fav,Datta:2026ffs}
\begin{equation}
\begin{aligned}
    \ln\mathcal L(\boldsymbol{\theta})
    =
    -\sum_{r=1}^{N_{\rm seg}}
    \sum_{k=1}^{N/2}
    \Bigg\{
        &\ln\left[
        P_A(f_k;\boldsymbol{\theta})
        P_E(f_k;\boldsymbol{\theta})
        P_T(f_k;\boldsymbol{\theta})
        \right]
        \\
        &+
        \frac{2}{Tf_s^2}
        \left[
        \frac{|\tilde d_A^r(f_k)|^2}{P_A(f_k;\boldsymbol{\theta})}
        +
        \frac{|\tilde d_E^r(f_k)|^2}{P_E(f_k;\boldsymbol{\theta})}
        +
        \frac{|\tilde d_T^r(f_k)|^2}{P_T(f_k;\boldsymbol{\theta})}
        \right]
    \Bigg\}.
\end{aligned}
\label{eq:lisa_loglike}
\end{equation}
Equivalently, using the null-channel approximation explicitly,
\begin{equation}
\begin{aligned}
    \ln\mathcal L(\boldsymbol{\theta})
    =
    -\sum_{r=1}^{N_{\rm seg}}
    \sum_{k=1}^{N/2}
    \Bigg\{
        &\ln\left[
        \big(N_A+S_A\big)
        \big(N_E+S_E\big)
        N_T
        \right]_{f=f_k}
        \\
        &+
        \frac{2}{Tf_s^2}
        \left[
        \frac{|\tilde d_A^r(f_k)|^2}
        {N_A(f_k;\boldsymbol{\theta})+S_A(f_k;\boldsymbol{\theta})}
        +
        \frac{|\tilde d_E^r(f_k)|^2}
        {N_E(f_k;\boldsymbol{\theta})+S_E(f_k;\boldsymbol{\theta})}
        +
        \frac{|\tilde d_T^r(f_k)|^2}
        {N_T(f_k;\boldsymbol{\theta})}
        \right]
    \Bigg\}.
\end{aligned}
\label{eq:lisa_loglike_null}
\end{equation}

Synthetic LISA data are generated by selecting an injected parameter vector $\boldsymbol{\theta}_{\rm inj}$ and drawing Fourier coefficients from the corresponding covariance matrix. For each segment and frequency bin,
\begin{equation}
    \mathbf d_{rk}^{\rm mock}
    \sim
    \mathcal N_{\mathbb C}
    \left(
    0,
    C_k(\boldsymbol{\theta}_{\rm inj})
    \right).
\end{equation}
Since the covariance is diagonal in the adopted approximation, the draw can be performed independently in each channel:
\begin{equation}
    \tilde d_a^r(f_k)
    =
    \left[
    \frac{Tf_s^2}{4}
    P_a(f_k;\boldsymbol{\theta}_{\rm inj})
    \right]^{1/2}
    \left(
    x_{a,rk}
    +
    i y_{a,rk}
    \right),
    \label{eq:mock_draw}
\end{equation}
where
\begin{equation}
    x_{a,rk},y_{a,rk}
    \sim
    \mathcal N(0,1)
\end{equation}
are independent standard normal variates. This normalization ensures that
\begin{equation}
    \left\langle
    |\tilde d_a^r(f_k)|^2
    \right\rangle
    =
    \frac{Tf_s^2}{2}
    P_a(f_k;\boldsymbol{\theta}_{\rm inj}).
\end{equation}
Repeating this procedure over all segments, frequency bins, and TDI channels yields the mock dataset
\begin{equation}
    \left\{
    \tilde d_A^r(f_k),
    \tilde d_E^r(f_k),
    \tilde d_T^r(f_k)
    \right\}.
\end{equation}

The posterior distribution is
\begin{equation}
    p(\boldsymbol{\theta}|d)
    =
    \frac{
    \mathcal L(d|\boldsymbol{\theta})\pi(\boldsymbol{\theta})
    }{
    Z
    },
    \label{eq:posterior}
\end{equation}
where $\pi(\boldsymbol{\theta})$ is the prior and
\begin{equation}
    Z=
    \int d\boldsymbol{\theta}\,
    \mathcal L(d|\boldsymbol{\theta})\pi(\boldsymbol{\theta})
\end{equation}
is the Bayesian evidence. Posterior samples are used to obtain marginalized constraints on the instrumental, astrophysical, and metastable-string parameters. In the numerical implementation, the posterior is explored with nested sampling, using \texttt{dynesty} \cite{speagle2020dynesty} interfaced through \texttt{Bilby} \cite{Ashton:2018jfp}.
\section{Posterior covariance and reconstruction diagnostics}
\label{sec:posterior_covariance_metrics}
\subsection{Covariance diagnostics in the signal-parameter sector}
To quantify the reconstruction performance across the metastable-string
parameter space, we characterize the geometry of the posterior distribution
using its sample covariance matrix. Let
\begin{equation}
\boldsymbol{\theta}
=
(\theta_1,\theta_2,\dots,\theta_N)^T
\end{equation}
denote the vector of sampled parameters, and define
\begin{equation}
C
=
{\rm Cov}(\boldsymbol{\theta}) .
\end{equation}
Although the posterior need not be exactly Gaussian, the covariance matrix
provides a useful second-moment summary of its local geometry. Since \(C\) is
real and symmetric, it admits the eigendecomposition
\begin{equation}
C
=
E\Lambda E^T,
\qquad
\Lambda
=
{\rm diag}(\lambda_1,\lambda_2,\dots,\lambda_N),
\end{equation}
where the columns of \(E\) are orthonormal principal directions and the
eigenvalues \(\lambda_i\) are the posterior variances along those directions.

In the present analysis, the main focus is the two-dimensional signal sector
spanned by
\begin{equation}
\theta_1
=
\log_{10}(G\mu),
\qquad
\theta_2
=
\log_{10}\kappa_{\rm CS},
\end{equation}
where \(G\mu\) is the string tension and \(\kappa_{\rm CS}\) controls the
metastability scale. The logarithmic parametrization provides a dimensionless
coordinate system for the covariance analysis and is particularly useful for
diagnosing multiplicative reconstruction uncertainties. All covariance
diagnostics below are therefore defined in this sampled log-parameter basis.

In this two-parameter subspace, we denote the larger and smaller eigenvalues of
the covariance matrix by
\begin{equation}
\lambda_\parallel
\equiv
\lambda_+,
\qquad
\lambda_\perp
\equiv
\lambda_-,
\end{equation}
corresponding respectively to the long and short principal axes of the posterior.
The ratio
\begin{equation}
\kappa_{\rm deg}
\equiv
\frac{\lambda_\parallel}{\lambda_\perp}
\end{equation}
is used as our measure of posterior anisotropy. Large values,
\(\kappa_{\rm deg}\gg1\), indicate that the posterior is strongly elongated and
that the data constrain one parameter combination much more accurately than the
orthogonal one. By contrast, \(\kappa_{\rm deg}\sim1\) corresponds to a nearly
isotropic posterior in the sampled coordinates. In likelihood-dominated regions,
this may indicate balanced two-parameter reconstruction, if the
posterior is also sufficiently compact. In prior-dominated regions, the value of
\(\kappa_{\rm deg}\) instead reflects the geometry of the prior rather than
information supplied by the data.

The smaller eigenvalue \(\lambda_\perp\) is particularly useful because it
measures the variance along the most tightly constrained principal direction. In
strongly anisotropic regions, the broader direction associated with
\(\lambda_\parallel\) can be affected by prior volume or by a weakly constrained
degeneracy direction, whereas the narrower direction is more directly controlled
by the likelihood. For this reason, \(\lambda_\perp\) provides a diagnostic of
local constraining power along the best-measured parameter combination~\cite{Datta:2026fav,Datta:2026ffs}. This
combination need not coincide with either \(G\mu\) or \(\kappa_{\rm CS}\)
separately.
\subsection{Correlation, anisotropy, and posterior localization}
It is useful to distinguish posterior anisotropy from posterior localization.
The eigenvalue ratio \(\kappa_{\rm deg}\) measures the shape of the posterior
ellipse, but not its overall size. In the Gaussian approximation, the area of a
constant-probability ellipse scales as
\begin{equation}
A_{\rm post}
\propto
\sqrt{\det C}
=
\sqrt{\lambda_\parallel\lambda_\perp}.
\end{equation}
Thus a strongly correlated and anisotropic posterior can still occupy a smaller
area than an uncorrelated posterior with \(\kappa_{\rm deg}\sim1\), provided the
best-constrained direction is very narrow and the broad direction remains
bounded. Conversely, an approximately isotropic posterior does not by itself
imply good reconstruction if both eigenvalues are large. Therefore
\(\kappa_{\rm deg}\) measures posterior shape, while
\(\sqrt{\lambda_\parallel\lambda_\perp}\), \(\lambda_\perp\), and the
marginalized widths measure complementary aspects of the total localization.

In addition to the principal-component widths, we compute the marginalized
standard deviations in the original parameter basis,
\begin{equation}
\sigma_1
=
\sqrt{C_{11}},
\qquad
\sigma_2
=
\sqrt{C_{22}},
\end{equation}
as well as the Pearson correlation coefficient
\begin{equation}
\rho
=
\frac{C_{12}}{\sqrt{C_{11}C_{22}}}.
\end{equation}
It is important to distinguish the information contained in \(\rho\) from that
contained in \(\kappa_{\rm deg}\). The coefficient \(\rho\) measures the linear
correlation between the two chosen coordinates. Equivalently, it diagnoses
whether the posterior ellipse is tilted relative to the coordinate axes. It is
therefore a basis-dependent measure of parameter mixing, not a general measure
of posterior elongation. On the other hand, \(\kappa_{\rm deg}\) measures the
hierarchy between the principal variances and is insensitive to the orientation
of the ellipse.

These two diagnostics coincide only in special cases. As discussed in
Refs.~\cite{Datta:2026fav,Datta:2026ffs}, in the domain-wall case the
ultraviolet or infrared tail in the detector band can depend approximately on a
single combination of the microscopic parameters. In logarithmic variables, this
produces an oblique posterior ridge, so a large eigenvalue hierarchy is
accompanied by \(|\rho|\simeq1\). In that situation, maps of
\(\kappa_{\rm deg}\) and \(1-\rho^2\) can appear qualitatively similar. This
correspondence, however, is not generic.

A useful limiting example illustrates this point. If the two marginal variances
are comparable,
\begin{equation}
C
\simeq
\sigma^2
\begin{pmatrix}
1 & \rho \\
\rho & 1
\end{pmatrix},
\end{equation}
then
\begin{equation}
\kappa_{\rm deg}
=
\frac{1+|\rho|}{1-|\rho|}.
\end{equation}
Thus a posterior can have a sizeable correlation while remaining only moderately
elongated. For example, \(\kappa_{\rm deg}\simeq10\) corresponds to
\(|\rho|\simeq0.82\) in this simplified case, with an axis ratio in standard
deviations of only \(\sqrt{\kappa_{\rm deg}}\simeq3.2\). Such a posterior is
tilted, but it need not have a large area if both principal widths are small.

The interpretation of \(\kappa_{\rm deg}\) also requires care in low-SNR
regions. When the likelihood is uninformative, the posterior approaches the
prior. For a uniform prior in a parameter \(\theta_i\) with width
\(\Delta\theta_i\), the prior variance is
\begin{equation}
{\rm Var}(\theta_i)
=
\frac{\Delta\theta_i^2}{12}.
\end{equation}
Therefore, for a prior-dominated rectangular posterior in the
\((\log_{10}G\mu,\log_{10}\kappa_{\rm CS})\) plane, one expects
\begin{equation}
\kappa_{\rm deg}^{\rm prior}
\sim
\left(
\frac{\Delta\theta_{\rm wide}}
     {\Delta\theta_{\rm narrow}}
\right)^2 .
\end{equation}
If the prior range in \(\log_{10}G\mu\) is much wider than the prior range in
\(\log_{10}\kappa_{\rm CS}\), this prior geometry alone can produce
\(\kappa_{\rm deg}\gg1\), even in the absence of any physical degeneracy.
Consequently, saturated low-SNR regions in the \(\kappa_{\rm deg}\) map should
be interpreted as prior dominated rather than as likelihood-driven posterior
anisotropy.
\subsection{Physical interpretation and reconstruction regimes}
For metastable cosmic strings, the posterior geometry can realize several
distinct physical regimes. When LISA observes mainly a plateau-like part of the
spectrum, the signal may be sensitive primarily to the overall amplitude, while
the metastability parameter remains weakly constrained. The posterior can then
be elongated along a direction nearly aligned with one of the coordinate axes.
In such a case,
\begin{equation}
\kappa_{\rm deg}
\gg
1,
\qquad
|\rho|
\not\simeq
1 .
\end{equation}
This corresponds to poor identifiability of one parameter, typically
\(\kappa_{\rm CS}\), but not necessarily to a strong correlated trade-off
between \(G\mu\) and \(\kappa_{\rm CS}\).

A different situation occurs when the LISA-band spectrum is sensitive to the
broad metastable transition, but the amplitude and decay scale are still partly
degenerate. Since \(G\mu\) controls the overall normalization, while
\(\kappa_{\rm CS}\) controls the decay time through the nucleation rate,
schematically
\begin{equation}
\Gamma_d
\propto
e^{-\pi\kappa_{\rm CS}},
\end{equation}
the two parameters can compensate each other. Increasing \(G\mu\) raises the
stochastic-background amplitude, whereas decreasing \(\kappa_{\rm CS}\) shortens
the network lifetime and suppresses the signal. The posterior then develops an
anti-correlated ridge,
\begin{equation}
G\mu \uparrow
\quad
\Longleftrightarrow
\quad
\kappa_{\rm CS}\downarrow ,
\end{equation}
or equivalently
\begin{equation}
\rho
<
0,
\qquad
|\rho|
\simeq
1 .
\end{equation}
In this case, a large value of \(\kappa_{\rm deg}\) indicates a genuine
amplitude--lifetime trade-off: the data constrain one local combination of
parameters much better than the orthogonal direction.

It is important, however, that this correlated regime is not necessarily a poor
reconstruction regime. If the covariance area
\begin{equation}
A_{\rm cov}
=
\sqrt{\lambda_\parallel\lambda_\perp}
\end{equation}
is small, then the posterior is compact even though it is elongated. Such a
case corresponds to compact but correlated reconstruction: LISA localizes the
allowed region in the \((G\mu,\kappa_{\rm CS})\) plane, but the remaining
uncertainty is oriented along an amplitude--lifetime trade-off direction. This
behavior is typical when the broad metastable transition is sampled by the LISA
band, so that the signal contains substantial shape information, but the two
parameters are not completely orthogonal in their spectral effects.

The most balanced reconstruction occurs when the data contain enough independent
spectral information to reduce both the total posterior area and the principal
axis hierarchy. In this case one expects
\begin{equation}
\kappa_{\rm deg}
\sim
1,
\qquad
|\rho|
\ll
1,
\end{equation}
together with small marginalized widths and small covariance area. This is the
clearest case of independent two-parameter reconstruction. More generally,
however, genuine reconstruction should not be identified with
\(\kappa_{\rm deg}\sim1\) alone. A posterior with large \(\kappa_{\rm deg}\) can
still be informative if \(A_{\rm cov}\) is small, while a posterior with
\(\kappa_{\rm deg}\sim1\) can be uninformative if both eigenvalues are large or
if one direction remains prior dominated.

There is an additional high-SNR regime that is specific to metastable
strings. Sensitivity to \(\kappa_{\rm CS}\) is not restricted to cases in which
the most visually apparent transition lies directly inside the LISA band. A
useful way to understand this regime is to introduce an approximate crossover
frequency \(f_{\rm cross}\), which marks the onset of the plateau-dominated
part of the total spectrum. Equivalently, in a schematic decomposition of the
metastable-string background (cf. Eq.\eqref{eq:omega_gw_stable} and Eq.\eqref{eq:metastable_decay_factor}),
\begin{equation}
    \Omega_{\rm GW}(f)
    \simeq
    \Omega_{\rm stable-like}^{\,t<t_s}(f;G\mu)
    +
    \Omega_{\rm decay}^{\,t>t_s}(f;G\mu,\kappa_{\rm CS}) ,\label{split}
\end{equation}
\(f_{\rm cross}\) corresponds roughly to the frequency at which the
stable-like contribution begins to dominate over the decay-sensitive part. This
decomposition is meant only as a guide to the spectral morphology; the actual
inference is controlled by the \(\kappa_{\rm CS}\)-dependence of the full
spectrum.

An estimate of \(f_{\rm cross}\) can be obtained by comparing the loop length
that contributes to an observed frequency \(f\) at the network decay epoch,
\begin{equation}
    \ell(f,t_s)
    \simeq
    \frac{2}{f}\frac{a_s}{a_0},
\end{equation}
with the gravitational-radiation length scale
\begin{equation}
    \ell_{\rm GW}(t_s)
    \sim
    \Gamma_{\rm GW}G\mu\,t_s .
\end{equation}
Setting \(\ell(f_{\rm cross},t_s)\sim \ell_{\rm GW}(t_s)\) gives
\begin{equation}
    f_{\rm cross}
    \sim
    \frac{2}{\Gamma_{\rm GW}G\mu\,t_s(1+z_s)} .
\end{equation}
During radiation domination, \(H_s\simeq1/(2t_s)\), and therefore
\begin{equation}
    f_{\rm cross}
    \sim
    \frac{4H_s}{\Gamma_{\rm GW}G\mu(1+z_s)}
    =
    \frac{4H_0\sqrt{\Omega_r}}{\Gamma_{\rm GW}}
    \frac{1+z_s}{G\mu}.
    \label{eq:fcross_zs_covariance}
\end{equation}
For \(\Gamma_{\rm GW}\simeq50\), this becomes
\begin{equation}
    f_{\rm cross}
    \simeq
    1.7\times10^{-21}\,{\rm Hz}
    \left(\frac{50}{\Gamma_{\rm GW}}\right)
    \frac{1+z_s}{G\mu}.
    \label{eq:fcross_numeric_covariance}
\end{equation}
Using the standard vacuum-tunneling relation
\begin{equation}
    1+z_s
    \propto
    T_s
    \propto
    (G\mu)^{1/4}
    \exp\left(-\frac{\pi\kappa_{\rm CS}}{4}\right),
\end{equation}
one obtains
\begin{equation}
    f_{\rm cross}
    \simeq
    f_0
    \left(\frac{10^{-15}}{G\mu}\right)^{3/4}
    \exp\left[
    -\frac{\pi}{4}
    \left(\kappa_{\rm CS}-64\right)
    \right],
    \label{eq:fcross_gmu_kappa_covariance}
\end{equation}
where \(f_0\) is an order-Hz normalization. Its precise value depends on the
operational definition of the plateau onset, numerical conventions, and the
sum over emitted harmonics. The robust part of Eq.~\eqref{eq:fcross_gmu_kappa_covariance}
is therefore the scaling with \(G\mu\) and \(\kappa_{\rm CS}\).

The condition \(f_{\rm cross}\sim f_{\rm LISA}^{\rm low}\) marks the approximate
onset of the most informative reconstruction window. At this point the
crossover into the plateau-dominated regime begins to enter the LISA band, so
the decay-sensitive part of the spectrum starts to become directly accessible.
As the parameters vary, \(f_{\rm cross}\) moves through the LISA band, and the
detector samples an increasing fraction of the broad lifetime-dependent
transition. The most favorable reconstruction is thus expected while this transition is
contained within the LISA band, roughly between the boundaries
\[
    f_{\rm cross}\sim f_{\rm LISA}^{\rm low}
    \qquad \text{and} \qquad
    f_{\rm low}\sim f_{\rm LISA}^{\rm high}.
\]
In this window, LISA observes the spectral curvature connecting the infrared
\(f^2\) regime to the plateau-dominated regime, so the data contain both
amplitude information and lifetime-dependent shape information.

The regime \(f_{\rm cross}\lesssim f_{\rm LISA}^{\rm low}\) is qualitatively
different. There the onset of the plateau-dominated regime has already moved
below the LISA band, so the observed spectrum can appear approximately
plateau-like and the transition is no longer directly visible. This does not
imply that all lifetime information is lost: in high-SNR regions the full
metastable-string spectrum can still retain small coherent finite-lifetime
distortions relative to a strict stable-string plateau. For example, for
\(G\mu=10^{-8}\) and \(\kappa_{\rm CS}=70\), a simple SNR estimate for the
decay-sensitive contribution alone gives
\({\rm SNR}\simeq6\times10^3\), even though this contribution is visually
subdominant in the total spectrum. Thus, reduced uncertainty in
\(\kappa_{\rm CS}\), or small to moderate \(\kappa_{\rm deg}\), in this
plateau-like regime can be attributed to residual sensitivity to the
finite-lifetime dependence of the full signal, rather than as direct
observation of the main spectral transition inside the LISA band.

Thus, a large value of \(\kappa_{\rm deg}\) should not be interpreted in a
unique way without also considering \(\rho\), the signal-to-noise ratio, the
posterior area, the marginalized widths, and the spectral morphology. The
combination
\begin{equation}
\kappa_{\rm deg}
\gg
1,
\qquad
|\rho|
\simeq
1
\end{equation}
signals a correlated trade-off degeneracy, while
\begin{equation}
\kappa_{\rm deg}
\gg
1,
\qquad
|\rho|
\ll
1
\end{equation}
signals an axis-aligned loss of identifiability. In addition, a posterior may
have sizeable \(|\rho|\) while remaining only mildly elongated, or even
occupying a small total area, if both parameters are still reasonably well
localized. Finally, in low-SNR regions, \(\kappa_{\rm deg}\) is controlled
mainly by the prior geometry and should not be interpreted as a physical
reconstruction diagnostic.

Since the signal parameters are sampled logarithmically, the principal widths
\(\sqrt{\lambda_\parallel}\) and \(\sqrt{\lambda_\perp}\) are measured in dex.
In regions where the posterior is approximately isotropic, the marginalized
uncertainties \(\sigma_1,\sigma_2\) become comparable to the principal widths.
In anisotropic regions, however, the principal widths, marginalized widths, and
posterior area encode different information: \(\lambda_\perp\) measures the
best-constrained linear combination, \(\lambda_\parallel\) measures the poorly
constrained principal direction, \(\sqrt{\lambda_\parallel\lambda_\perp}\)
measures the overall covariance area, and \(\sigma_1\) and \(\sigma_2\) measure
the projected uncertainties on \(G\mu\) and \(\kappa_{\rm CS}\) separately.
Comparing heat maps of these quantities therefore allows us to distinguish
genuine two-parameter reconstruction, compact but correlated reconstruction,
coordinate-aligned loss of identifiability, strongly correlated parameter
degeneracy, and prior-dominated regions.

To define a practical reconstruction threshold, we use the width of the
constrained principal direction. A positive displacement \(\delta\) in a
logarithmic parameter corresponds to a multiplicative fractional uncertainty
\begin{equation}
\Delta_+
=
10^\delta
-
1 .
\end{equation}
Requiring the positive-side uncertainty associated with the constrained
principal mode to be below \(25\%\) gives
\begin{equation}
10^{\sqrt{\lambda_\perp}}
-
1
<
0.25 .
\end{equation}
Equivalently,
\begin{equation}
\sqrt{\lambda_\perp}
<
\log_{10}(1.25)
\simeq
0.097~{\rm dex},
\end{equation}
or
\begin{equation}
\log_{10}\!\left(\sqrt{\lambda_\perp}\right)
\lesssim
-1.01 .
\end{equation}
The quantities \(\sigma_i\) and \(\sqrt{\lambda_i}\) are
posterior standard deviations, i.e. second-moment measures of the spread of the
posterior samples. In the Gaussian limit, these widths coincide with the usual
one-sigma uncertainties. For non-Gaussian, skewed, prior-truncated, or
multimodal posteriors, however, they should be understood as RMS-width
diagnostics rather than exact credible intervals. Exact credible levels can
instead be obtained directly from posterior quantiles or enclosed posterior
mass.

In summary, the metastable-string scenario provides a particularly instructive
case in which reconstruction requires a careful distinction between posterior
shape, posterior size, and reconstruction accuracy. A large
\(\kappa_{\rm deg}\) indicates that the posterior is anisotropic, but it does
not by itself determine either the physical origin of the anisotropy or the
degree of localization. When accompanied by \(|\rho|\simeq1\), a large
\(\kappa_{\rm deg}\) reflects a correlated amplitude--lifetime trade-off between
\(G\mu\) and \(\kappa_{\rm CS}\). In the metastable-string case, this regime can
still correspond to compact reconstruction if the covariance area is small: the
posterior is elongated, but the allowed region in the
\((G\mu,\kappa_{\rm CS})\) plane is well localized. When large
\(\kappa_{\rm deg}\) is instead accompanied by small or moderate \(|\rho|\), it
more often reflects weak identifiability of one coordinate direction, typically
\(\kappa_{\rm CS}\). In saturated low-SNR regions, large values of
\(\kappa_{\rm deg}\) can also arise from prior geometry rather than from
likelihood-driven parameter information. On the other hand, small or moderate values of \(\kappa_{\rm deg}\) should not
automatically be interpreted as optimal two-parameter reconstruction. Such
posteriors are more balanced in shape, but they may still have a larger
covariance area or be partially influenced by the prior. They indicate strong
reconstruction only when accompanied by small marginalized widths, small
\(\lambda_\perp\), small covariance area, and accurate recovery of the injected
parameters. The joint interpretation of \(\rho\), \(\kappa_{\rm deg}\),
\(\lambda_\perp\), the covariance area, marginalized widths, and reconstruction
bias is therefore essential for interpreting the LISA reconstruction maps. 
\section{Results and discussion}
\label{sec:results_discussion}

We now present the LISA reconstruction landscape for metastable cosmic-string
stochastic backgrounds across the $(G\mu,\kappa_{\rm CS})$ plane. The posterior
diagnostics defined in Sec.~\ref{sec:posterior_covariance_metrics} are used to
identify the physical origin of the reconstruction performance. Readers interested in the overall summary of the results, can directly go to Sec.\ref{sec:sumpr}.

\subsection{Reconstruction landscape}

\begin{figure}
    \centering
    \includegraphics[width=.52\linewidth]{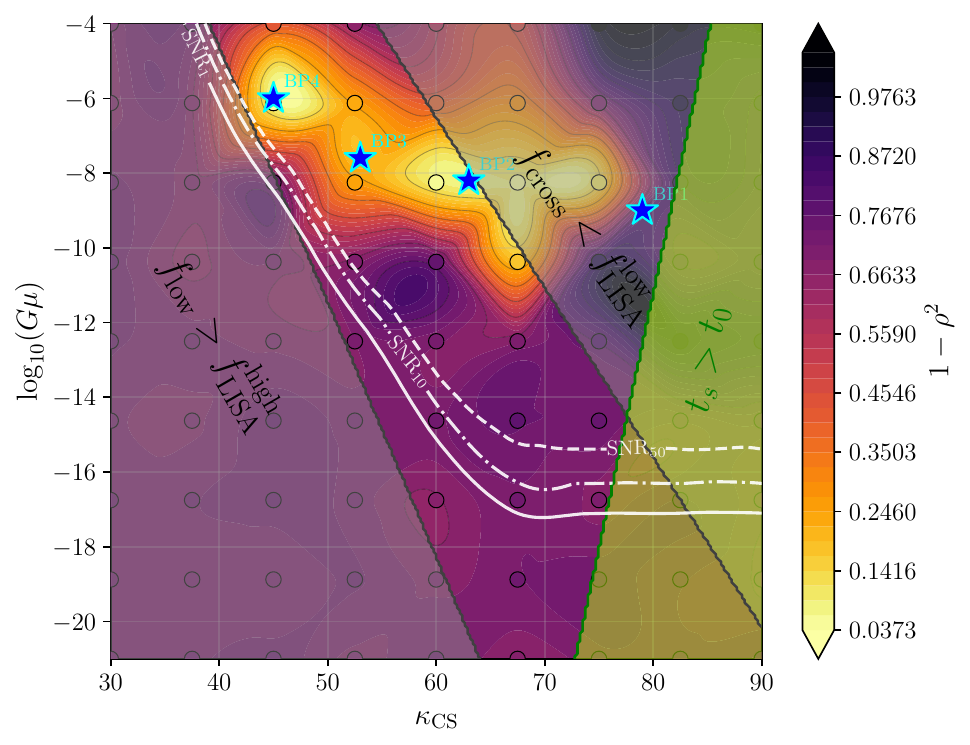}\includegraphics[width=.52\linewidth]{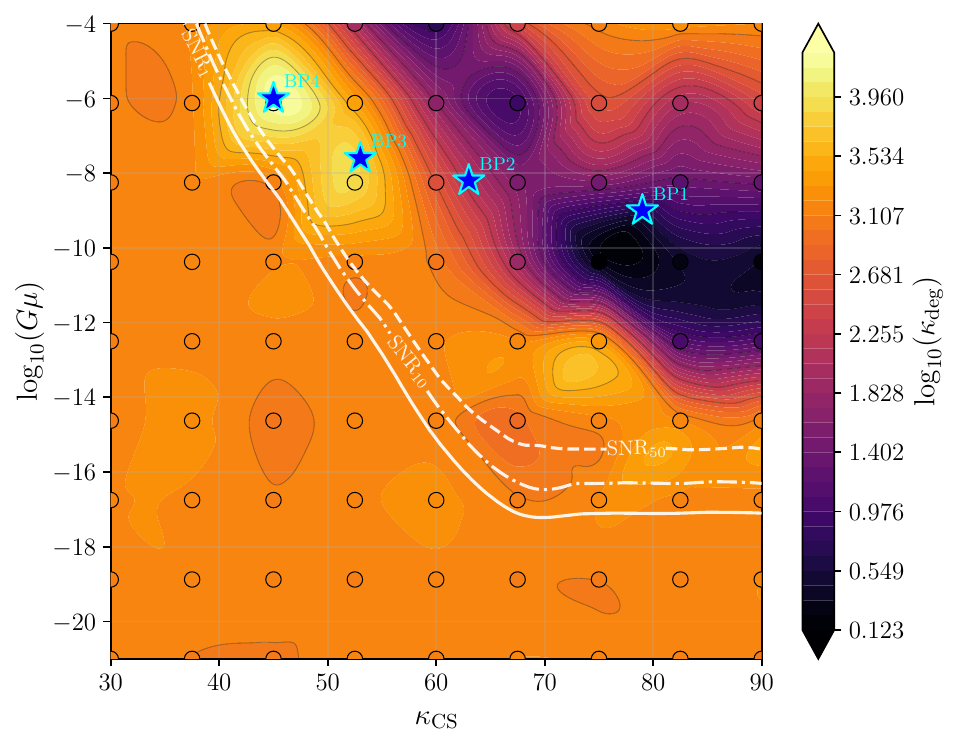}\\ \includegraphics[width=.52\linewidth]{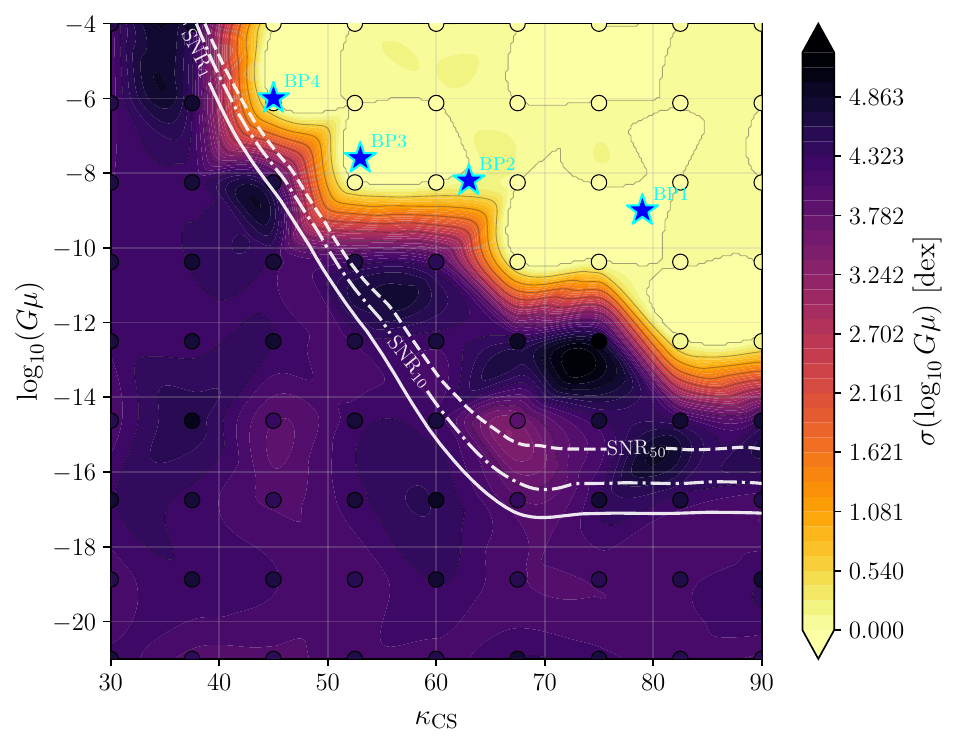}\includegraphics[width=.52\linewidth]{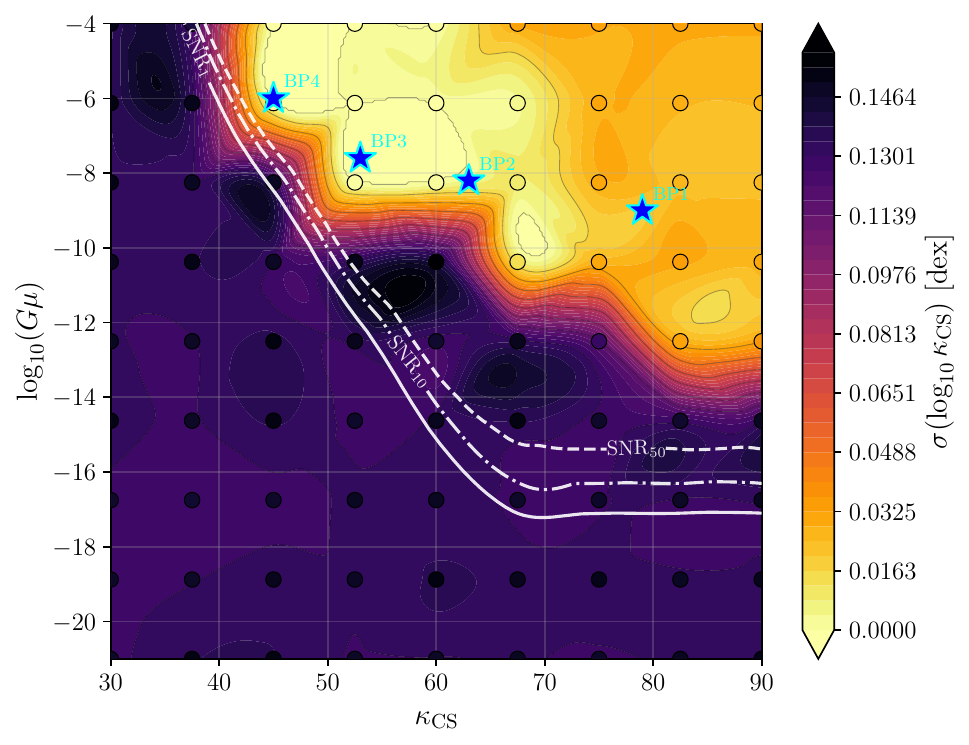}
 \caption{
Posterior-covariance diagnostics on the $(G\mu,\kappa_{\rm CS})$ plane. 
The top-left panel shows $1-\rho^2$, where small values indicate a strong
linear correlation between $G\mu$ and $\kappa_{\rm CS}$. This panel also shows
the approximate spectral-regime boundaries used to guide the physical
interpretation. The region labeled $f_{\rm low}>f_{\rm LISA}^{\rm high}$
corresponds to cases in which the LISA band lies mainly on the infrared side of
the metastable-string spectrum. The region labeled
$f_{\rm cross}<f_{\rm LISA}^{\rm low}$ corresponds to cases in which the onset
of the plateau-dominated regime occurs below the LISA band. The boundaries are
computed using Eq.~\eqref{eq:flow_metastable} for $f_{\rm low}$ and
Eq.~\eqref{eq:fcross_numeric_covariance} for $f_{\rm cross}$. The green region
marks the stable-network boundary $t_s>t_0$, where the network has not decayed
before the present epoch and lifetime reconstruction should be interpreted with
care. These spectral-regime and stability boundaries are shown only in the
$1-\rho^2$ panel to avoid repetition. The correlated band running from large $G\mu$ and small $\kappa_{\rm CS}$ to
small $G\mu$ and large $\kappa_{\rm CS}$ reflects the trade-off between the
overall signal amplitude and the metastable decay scale. Along this band, the
correlation is strongest when LISA samples only part of the broad
lifetime-dependent transition, while it is reduced when the spectral morphology
is better resolved. The top-right panel shows the posterior-anisotropy
diagnostic $\kappa_{\rm deg}=\lambda_\parallel/\lambda_\perp$, which measures
the hierarchy between the principal posterior widths and highlights regions
where one parameter combination is much more weakly constrained than the
orthogonal one. The bottom panels show the marginalized reconstruction
uncertainties for $G\mu$ and $\kappa_{\rm CS}$. As expected, the uncertainty on
$G\mu$ improves for larger signal amplitudes, whereas the reconstruction of
$\kappa_{\rm CS}$ degrades when the spectrum becomes effectively
one-dimensional in the LISA band. The four starred benchmark points illustrate
representative posterior geometries, with the corresponding corner plots shown
in Figs.~\ref{fig:corner_bp1}--\ref{fig:corner_bp4}. Dark saturated regions,
especially those with ${\rm SNR}<1$, are prior dominated. In these regions,
large values of $\kappa_{\rm deg}$ should not be interpreted as
likelihood-driven parameter degeneracy or genuine two-parameter reconstruction,
but rather as a consequence of prior geometry.
}
\label{fig:two_panel}
\end{figure}

Figure~\ref{fig:two_panel} summarizes the main posterior-covariance diagnostics
on the $(G\mu,\kappa_{\rm CS})$ plane. The $1-\rho^2$ map shows a diagonal band
extending from larger $G\mu$ and smaller $\kappa_{\rm CS}$ toward smaller
$G\mu$ and larger $\kappa_{\rm CS}$. Along this band, the posterior is strongly
correlated, indicating an amplitude--lifetime trade-off. Physically, increasing
$G\mu$ raises the stochastic-background amplitude, while decreasing
$\kappa_{\rm CS}$ shortens the network lifetime and suppresses the signal. When
LISA observes only part of the broad lifetime-dependent spectral transition,
these effects can partially compensate each other, producing an elongated
posterior direction in the $(G\mu,\kappa_{\rm CS})$ plane.

The spectral-regime boundaries overlaid on the $1-\rho^2$ panel provide a useful
guide to this structure. The boundary $f_{\rm cross}=f_{\rm LISA}^{\rm low}$
marks the approximate entry of the plateau-onset scale into the LISA band. As
$f_{\rm cross}$ moves through the band, the decay-sensitive part of the spectrum
becomes increasingly accessible, and LISA begins to sample the broad transition
toward the plateau-dominated regime. The other boundary,
$f_{\rm low}=f_{\rm LISA}^{\rm high}$, marks the point at which the infrared
side of the transition reaches the high-frequency edge of the LISA band.
Between these two boundaries, LISA typically observes the spectral curvature
connecting the infrared $f^2$ tail to the stable-string-like plateau. This
transition window provides the most direct spectral information on
$\kappa_{\rm CS}$ in addition to the amplitude information carried by $G\mu$.

The $\kappa_{\rm deg}$ map shows that this informative region is not necessarily
isotropic. In fact, for metastable strings, some of the smallest covariance
areas occur in regions where both $\kappa_{\rm deg}$ and $|\rho|$ are large.
These regions should be interpreted as compact but correlated reconstruction:
LISA localizes the allowed region in the $(G\mu,\kappa_{\rm CS})$ plane, but the
remaining uncertainty is oriented along an amplitude--lifetime trade-off
direction. Thus, a large $\kappa_{\rm deg}$ does not by itself imply poor
reconstruction. Its interpretation depends on the covariance area, the
marginalized widths, and whether the elongation is likelihood driven or prior
dominated.

Conversely, regions with smaller or moderate $\kappa_{\rm deg}$ and weak
correlation are more balanced in posterior shape, but they are not necessarily
better localized unless the covariance area is also small. This distinction is
important for the metastable-string reconstruction landscape. The strongest
constraints need not correspond to the most isotropic posteriors; rather, they
often correspond to regions where the data measure a narrow
amplitude--lifetime combination with high precision.

At large $G\mu$ and large $\kappa_{\rm CS}$, typically toward the top-right
corner of the parameter plane, the signal can have high SNR while the dominant
LISA-band spectrum appears approximately plateau-like. In this regime, LISA
measures the overall amplitude accurately and therefore constrains $G\mu$, while
the direct identifiability of $\kappa_{\rm CS}$ can be reduced. However, the
condition $f_{\rm cross}<f_{\rm LISA}^{\rm low}$ should not be interpreted as a
sharp loss of lifetime information. Below the stable-network boundary, the full
metastable-string spectrum can retain small coherent finite-lifetime
distortions even when the visually dominant spectrum is plateau-like. In
sufficiently high-SNR regions, these residual shape differences can still
contribute to the reconstruction of $\kappa_{\rm CS}$, as reflected in the
posterior-covariance diagnostics.

The saturated low-SNR region should be interpreted separately. There the signal
is not sufficiently distinguished from instrumental noise and astrophysical
foregrounds, and the posterior is controlled mainly by the prior. Apparent
structures in $\rho$, $\kappa_{\rm deg}$, the covariance area, or the
marginalized widths in this region should therefore not be interpreted as
physical reconstruction. In particular, large values of $\kappa_{\rm deg}$ in
low-SNR regions can arise from prior geometry rather than from a
likelihood-driven parameter degeneracy.
\begin{figure}
    \includegraphics[width=.52\linewidth]{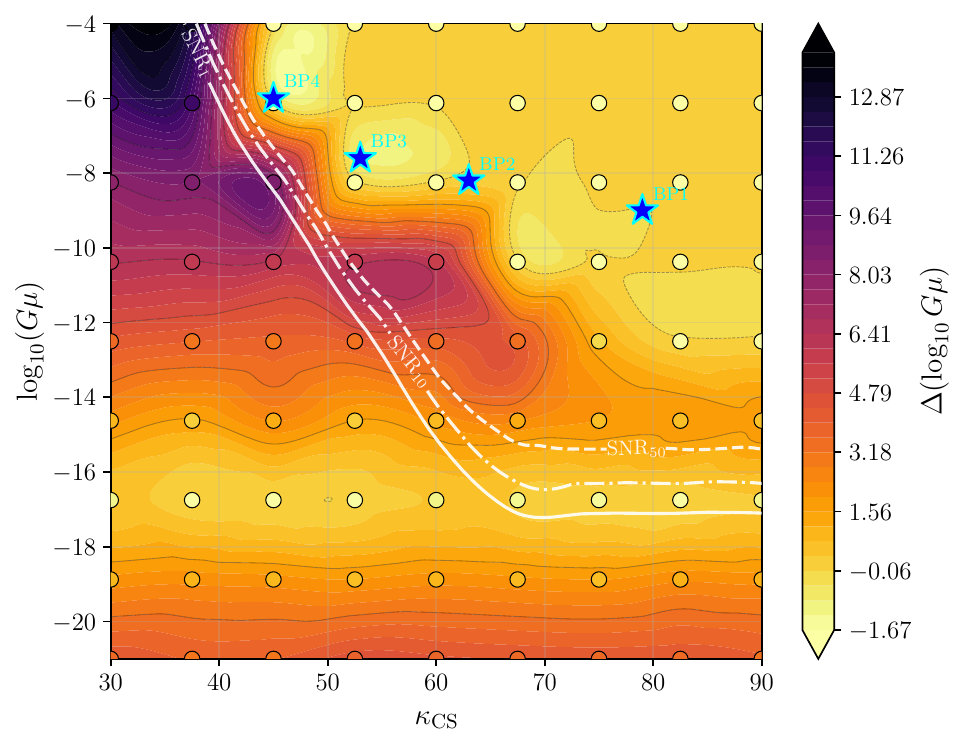}\includegraphics[width=.52\linewidth]{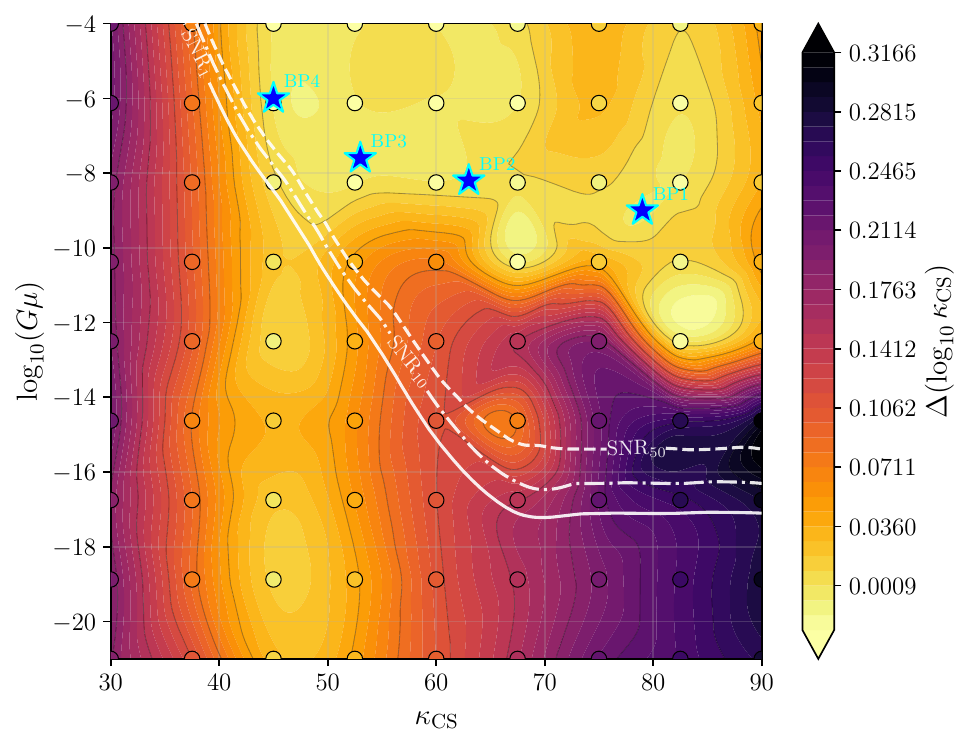}
   \caption{Heat maps of $\Delta(\log_{10}G\mu)$ and $\Delta(\log_{10}\kappa_{\rm CS})$, where $\Delta(\log_{10}X)\equiv |\mathrm{median}(\log_{10}X)-\log_{10}X_{\rm inj}|$ quantifies the offset of the reconstructed median from the injected value. Values $\lesssim 0.1$ correspond approximately to biases below $25\%$. These maps complement the posterior-width maps by tracking reconstruction accuracy rather than precision alone.}
\label{fig:rec_acc}
\end{figure}
\subsection{Marginalized uncertainties and reconstruction accuracy}

The bottom panels of Fig.~\ref{fig:two_panel} show the marginalized
uncertainties in $G\mu$ and $\kappa_{\rm CS}$. The uncertainty in $G\mu$
generally decreases as the signal amplitude increases, reflecting the fact that
$G\mu$ primarily controls the overall normalization of the cosmic-string
background. The behavior of $\kappa_{\rm CS}$ is more selective. Its uncertainty
improves only when the LISA-band spectrum contains measurable information about
the finite lifetime of the string network.

The most direct sensitivity to $\kappa_{\rm CS}$ occurs when the signal is both
loud and spectrally informative. This includes the transition window in which
the broad metastable feature is sampled by the LISA band, beginning roughly when
$f_{\rm cross}$ enters the band near $f_{\rm LISA}^{\rm low}$ and extending
until the infrared side of the transition approaches the high-frequency edge of
the band, $f_{\rm low}\sim f_{\rm LISA}^{\rm high}$. In this regime, varying
$\kappa_{\rm CS}$ shifts the decay epoch and changes the spectral morphology in
a way that cannot be absorbed by a simple rescaling of the signal amplitude
through $G\mu$. The data therefore contain both normalization information and
lifetime-dependent shape information.

A subtler behavior appears as one moves from this transition window toward the
stable-network limit. In part of this regime, the total SNR has already begun to
saturate and the LISA-band spectrum can appear approximately plateau-like, even
though the network lifetime remains finite and the decay history has not
completely disappeared from the signal. This illustrates an important
distinction: SNR saturation is not equivalent to loss of parameter sensitivity.
Schematically, the SNR is controlled by the weighted norm of the signal,
\begin{equation}
    {\rm SNR}^2
    \sim
    \int df\,
    \frac{\Omega_{\rm GW}^2(f)}
         {\sigma_\Omega^2(f)} ,
\end{equation}
whereas the information on parameters $\theta_i$ is controlled by local spectral
derivatives,
\begin{equation}
    F_{ij}
    \sim
    \int df\,
    \frac{1}{\sigma_\Omega^2(f)}
    \frac{\partial \Omega_{\rm GW}(f)}{\partial \theta_i}
    \frac{\partial \Omega_{\rm GW}(f)}{\partial \theta_j} .
    \label{eq:fisher_derivative}
\end{equation}
Thus the total signal power may change only weakly with increasing
$\kappa_{\rm CS}$, while the likelihood can still retain nonzero curvature in
the $\kappa_{\rm CS}$ direction if the full spectrum has a coherent residual
dependence on the network lifetime.

This point can be understood using the crossover scale introduced in
Sec.~\ref{sec:posterior_covariance_metrics}. The condition
$f_{\rm cross}\lesssim f_{\rm LISA}^{\rm low}$ indicates that the onset of the
plateau-dominated regime occurs below the LISA band. It does not, however, imply
that the metastable spectrum is exactly identical to a stable-string plateau.
Below the stable-network boundary, the full spectrum can still retain small
finite-lifetime distortions. In addition, as discussed in the previous section,
the decay-sensitive contribution can itself have a large weighted significance
even when it is visually subdominant in the total spectrum. For example, for
$G\mu=10^{-8}$ and $\kappa_{\rm CS}=70$, a simple SNR estimate for the
decay-sensitive loop contribution alone gives ${\rm SNR}\simeq6\times10^3$.
This illustrates why high-SNR plateau-like spectra can still carry information
about $\kappa_{\rm CS}$ through coherent residual differences in the full
metastable-string signal.

In the deeper plateau-like regime, $f_{\rm cross}\ll f_{\rm LISA}^{\rm low}$,
the leading stable-like contribution is close to its asymptotic plateau form
across the LISA band, and its amplitude is therefore only weakly dependent on
$\kappa_{\rm CS}$. The remaining $\kappa_{\rm CS}$ information is then expected
to be driven primarily by the derivative of the decay-sensitive finite-lifetime
contribution, schematically
\begin{equation}
    \frac{\partial \Omega_{\rm GW}}{\partial \kappa_{\rm CS}}
    \simeq
    \frac{\partial \Omega_{\rm decay}^{\,t>t_s}}
         {\partial \kappa_{\rm CS}}
    \qquad
    \left(f_{\rm cross}\ll f_{\rm LISA}^{\rm low}\right),
\end{equation}
up to residual $\kappa_{\rm CS}$ dependence associated with the approach to the
plateau. This is why a component that is visually subdominant in
$\Omega_{\rm GW}$ can nevertheless control the $\kappa_{\rm CS}$ sensitivity:
the likelihood responds to parameter derivatives, not only to the dominant
contribution to the total amplitude.
Therefore, a reduced marginalized uncertainty in $\kappa_{\rm CS}$, or small to
moderate $\kappa_{\rm deg}$, in a high-SNR plateau-like region should not be
interpreted as evidence that the main transition is necessarily inside the LISA
band. It can also reflect sensitivity to the broad transition shoulder or to
residual $\kappa_{\rm CS}$-dependent structure in the full signal. 
For still larger $\kappa_{\rm CS}$, the decay time can exceed the age of the
Universe, $t_s>t_0$. This corresponds to the theoretical stable-network limit:
the network survives until the present epoch and the leading spectrum becomes
effectively indistinguishable from that of stable cosmic strings. Reconstruction
of $\kappa_{\rm CS}$ in this region should therefore be interpreted with care.
Any apparent sensitivity near this boundary should be understood as sensitivity
to small deviations from the stable-string spectrum, rather than as a direct
measurement of a clearly resolved decay-induced transition.

The reconstruction-accuracy maps in Fig.~\ref{fig:rec_acc} provide a
complementary check by comparing the posterior median with the injected
parameters. For a parameter $X$, we define
\begin{equation}
    \Delta(\log_{10}X)
    \equiv
    \left|
    {\rm median}(\log_{10}X)
    -
    \log_{10}X_{\rm inj}
    \right| .
    \label{eq:median_bias}
\end{equation}
Values $\Delta(\log_{10}X)\lesssim 0.1$ correspond approximately to median
offsets below $25\%$ in the physical parameter.

Regions with both small marginalized uncertainties and small median offsets
represent genuine recovery of the injected parameters. By contrast, regions
with small apparent uncertainties but large median offsets indicate biased or
prior-influenced reconstruction, while regions with large uncertainties are
poorly localized even if the posterior median happens to lie near the injected
value. The accuracy maps therefore help distinguish true two-parameter
reconstruction from cases where the covariance diagnostics alone could be
misleading. 
\subsection{Best-constrained direction and posterior localization}

\begin{figure}
    \centering
    \includegraphics[width=.52\linewidth]{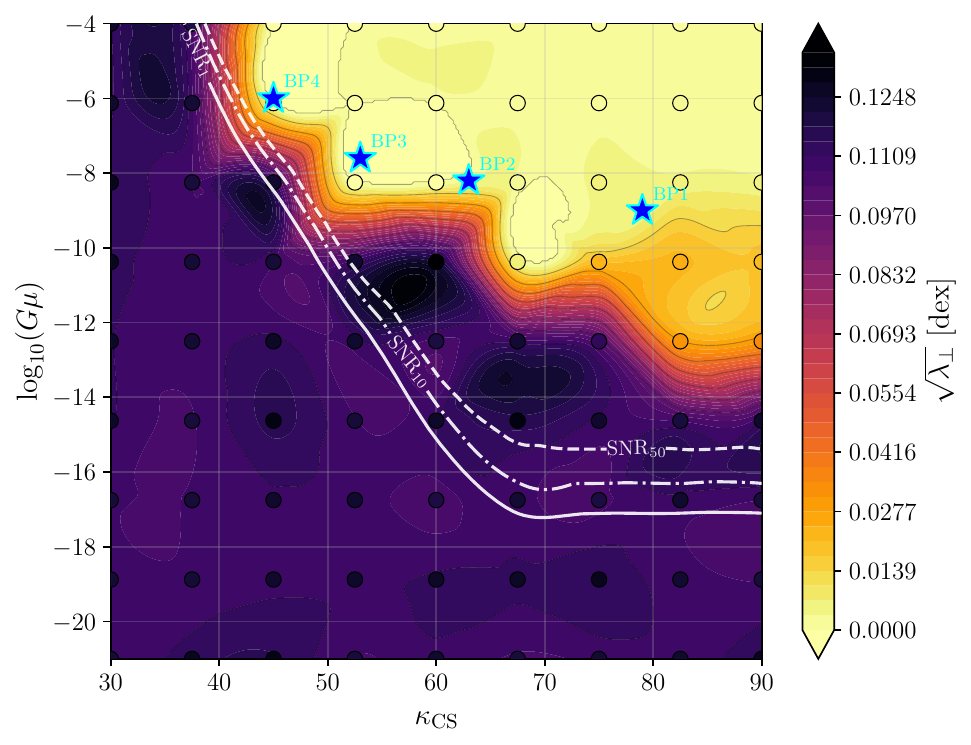}\includegraphics[width=.52\linewidth]{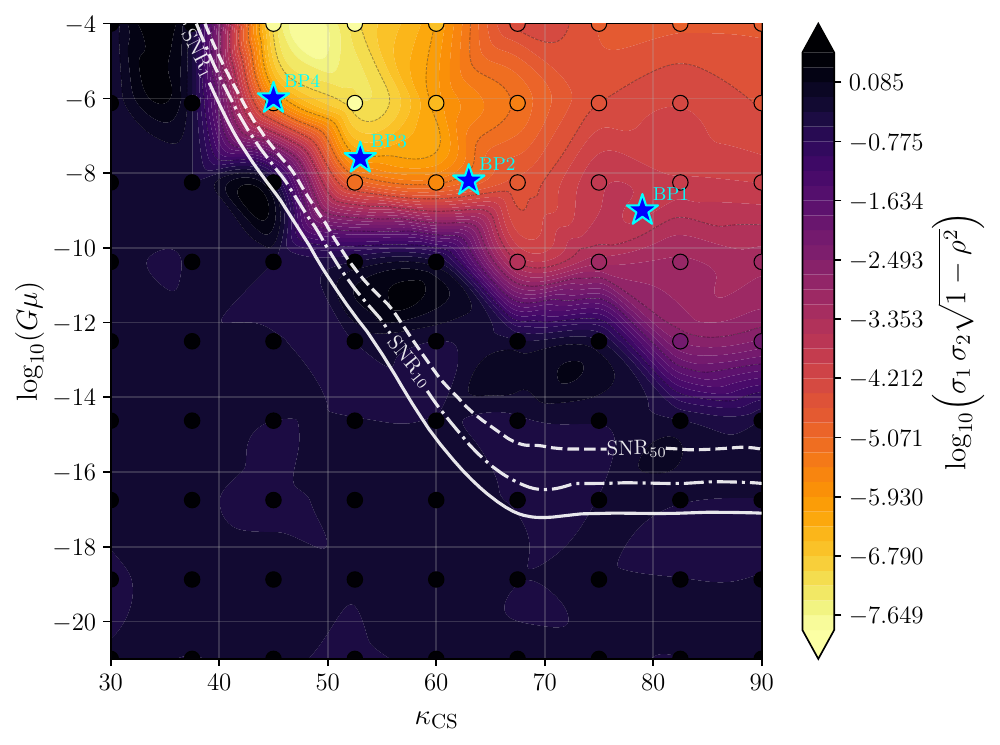}
   \caption{
Principal-space reconstruction diagnostics on the
$(G\mu,\kappa_{\rm CS})$ plane. Left: the smaller covariance eigenvalue
$\lambda_\perp$, which measures the posterior variance along the
best-constrained principal direction. Small values indicate that LISA measures
at least one local combination of $G\mu$ and $\kappa_{\rm CS}$ with high
precision. Right: the covariance-area diagnostic
$A_{\rm cov}\equiv\sqrt{\det C}
=\sigma_1\sigma_2\sqrt{1-\rho^2}
=\sqrt{\lambda_\parallel\lambda_\perp}$,
shown in the sampled $(\log_{10}G\mu,\log_{10}\kappa_{\rm CS})$ coordinates.
This quantity measures the overall posterior localization and is complementary
to $\kappa_{\rm deg}$, which measures posterior anisotropy. The comparison
distinguishes compact correlated posteriors from broad prior-dominated
posteriors: a region can have large $\kappa_{\rm deg}$ but small
$A_{\rm cov}$ if the constrained direction is sufficiently narrow, while an
approximately isotropic posterior can still have large $A_{\rm cov}$ if both
directions are poorly measured.
}
\label{fig:lambda_perp_area}
\end{figure}

Figure~\ref{fig:lambda_perp_area} summarizes two complementary diagnostics in
the principal-axis basis of the posterior covariance. The left panel shows the
smaller eigenvalue $\lambda_\perp$, which is the variance along the
best-constrained principal direction in the
$(\log_{10}G\mu,\log_{10}\kappa_{\rm CS})$ plane. Thus $\lambda_\perp$ is
measured in ${\rm dex}^2$, while $\sqrt{\lambda_\perp}$ gives the corresponding
posterior width in dex. This diagnostic identifies where LISA constrains at
least one local combination of $G\mu$ and $\kappa_{\rm CS}$ with high precision.

A small value of $\lambda_\perp$ does not, by itself, imply that both physical
parameters are independently reconstructed. In correlated trade-off regions,
$\lambda_\perp$ can be small while the full posterior remains elongated. This
means that LISA measures one approximately constant-spectrum combination of the
two parameters accurately, while the orthogonal combination is less well
constrained. Such regions represent informative but correlated reconstruction:
the data contain significant information about the metastable-string signal, but
the remaining uncertainty follows an amplitude--lifetime direction in parameter
space.

The right panel of Fig.~\ref{fig:lambda_perp_area} shows the covariance-area
diagnostic
\[
A_{\rm cov}
\equiv
\sqrt{\det C}
=
\sigma_1\sigma_2\sqrt{1-\rho^2}
=
\sqrt{\lambda_\parallel\lambda_\perp}.
\]
This quantity measures the total posterior localization in the sampled
log-parameter plane. It is complementary to $\kappa_{\rm deg}$, which measures
posterior anisotropy but not posterior size. A strongly correlated posterior can
still occupy a small area if the narrow direction is tightly constrained and the
broad direction remains bounded. Conversely, a nearly isotropic posterior can be
poorly localized if both eigenvalues are large.

The comparison between $\lambda_\perp$ and $A_{\rm cov}$ therefore separates
three qualitatively different cases. Regions with small $\lambda_\perp$ but
large $A_{\rm cov}$ indicate that only one parameter combination is accurately
measured, while the posterior remains broad along the orthogonal direction.
Regions with both small $\lambda_\perp$ and small $A_{\rm cov}$ correspond to
compact posteriors and therefore stronger localization in the
$(G\mu,\kappa_{\rm CS})$ plane. In the metastable-string case, these compact
regions can still have large $\kappa_{\rm deg}$ and large $|\rho|$, reflecting a
well-localized but correlated amplitude--lifetime trade-off. Finally, saturated
low-SNR regions generally have large covariance area or prior-controlled
structure, and should not be interpreted as physically informative
reconstruction.

The strongest reconstruction is therefore identified primarily by small
covariance area, small marginalized widths, and accurate recovery of the injected
parameters, rather than by posterior isotropy alone. A posterior with moderate
$\kappa_{\rm deg}$ and weak correlation is more balanced in shape, but it is not
necessarily better localized than a tilted posterior with smaller area. For
metastable strings, the most informative regions often correspond to compact
correlated posteriors: LISA constrains a narrow amplitude--lifetime combination
very precisely, while the remaining uncertainty is aligned with the physical
trade-off between $G\mu$ and $\kappa_{\rm CS}$. These regions mark where LISA
moves beyond detecting a metastable-string background and begins to reconstruct
the underlying network parameters.
\subsection{Benchmark posterior distributions}

\begin{figure}
    \centering
    \includegraphics[width=1\linewidth]{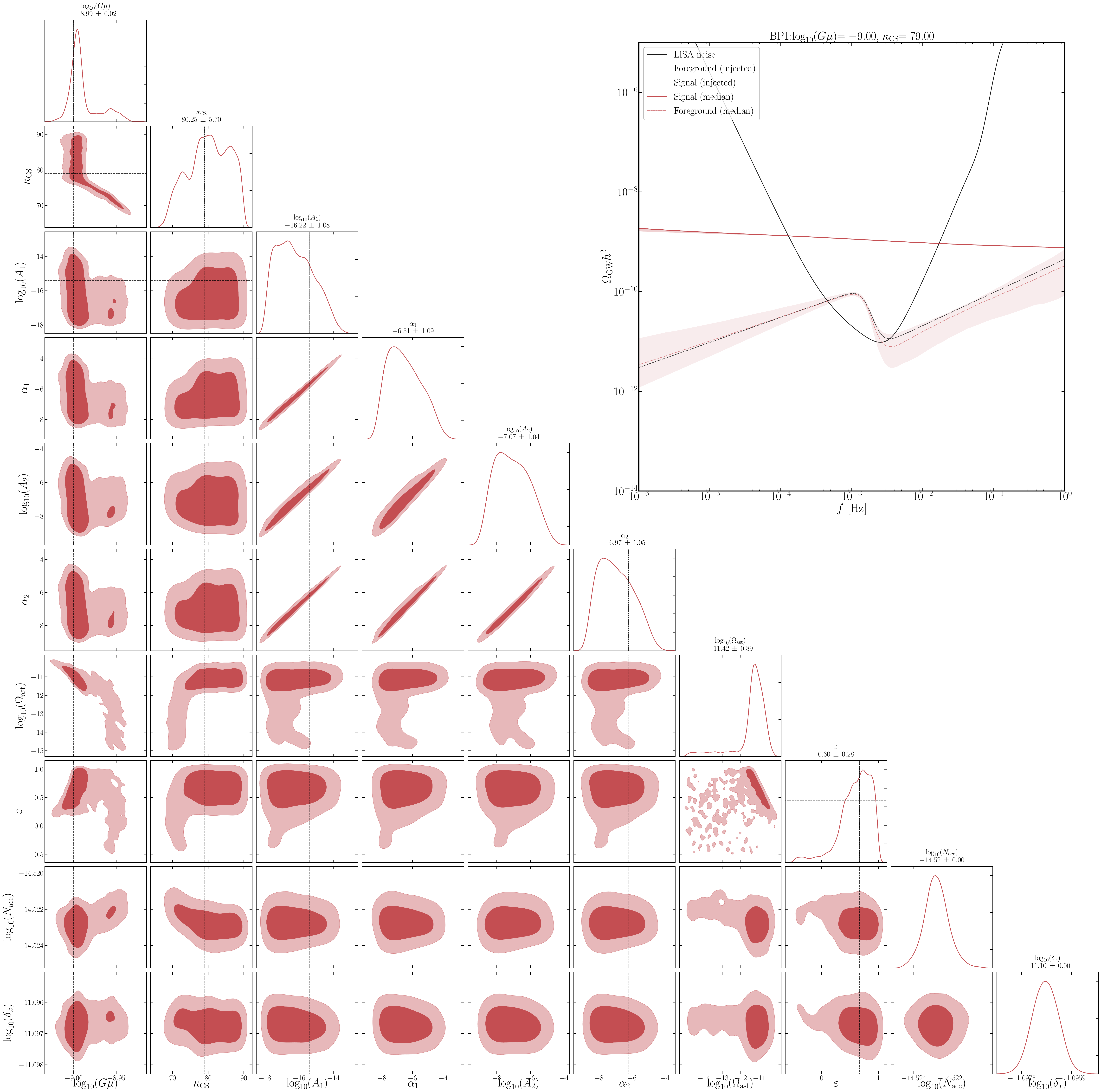}
   \caption{
Posterior distribution for benchmark point BP1. This point lies in a high-SNR
region where the dominant LISA-band spectrum is approximately plateau-like.
The posterior is extended mainly along the $\kappa_{\rm CS}$ direction,
indicating that $G\mu$ is constrained more directly through the overall
amplitude while $\kappa_{\rm CS}$ is less tightly identified. The appearance of
an anti-correlated ridge shows that LISA nevertheless retains residual
sensitivity to the metastable decay history, so changes in $G\mu$ and
$\kappa_{\rm CS}$ can partially compensate each other within part of the
posterior.
}
\label{fig:corner_bp1}
\end{figure}

\begin{figure}
    \centering
    \includegraphics[width=1\linewidth]{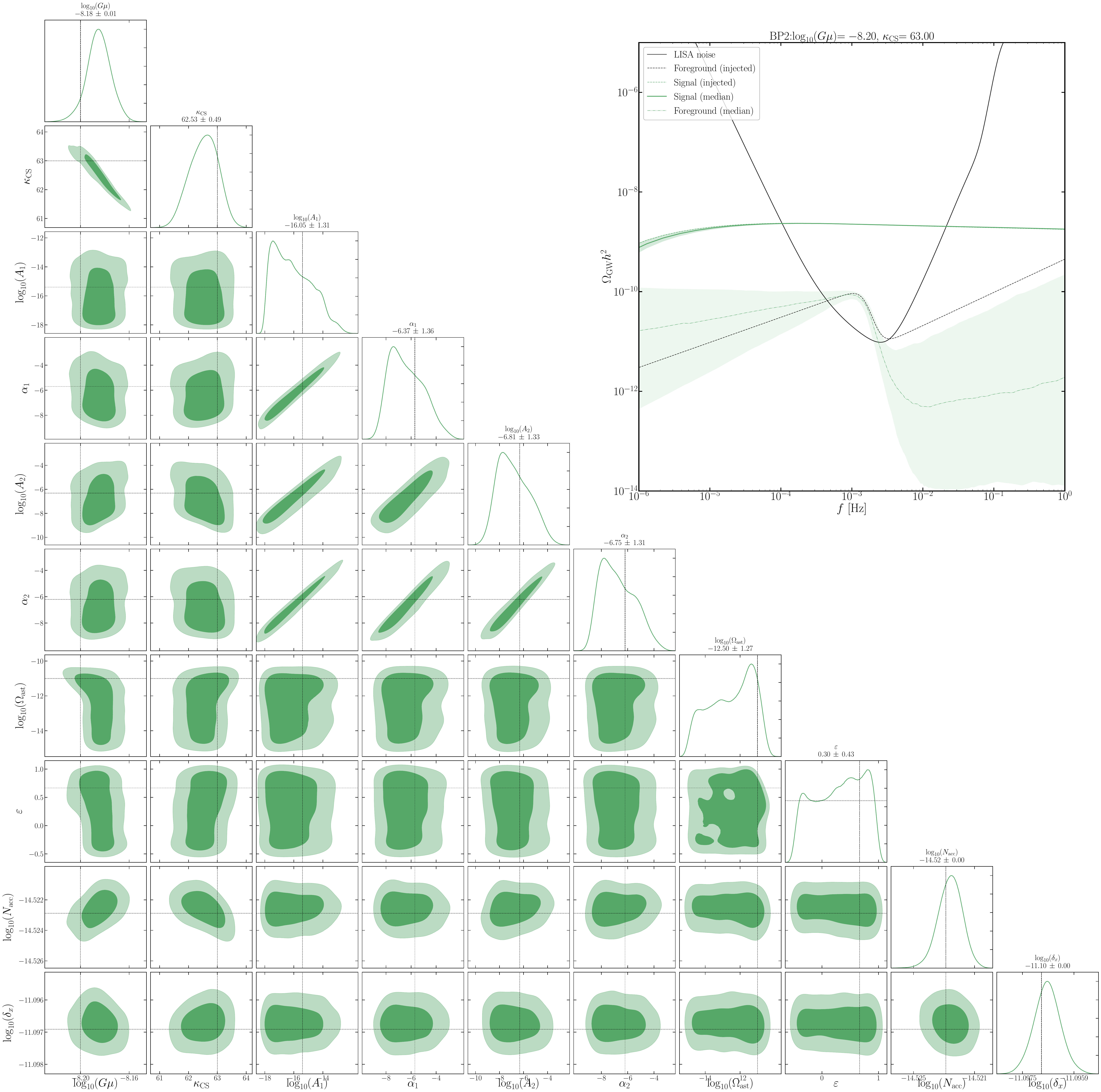}
   \caption{
Posterior distribution for benchmark point BP2. The posterior exhibits a clear
anti-correlated structure in the $(G\mu,\kappa_{\rm CS})$ plane, reflecting an
amplitude--lifetime trade-off between the two signal parameters. Despite this
correlation, the posterior occupies a relatively small area, showing that the
degeneracy is compact rather than prior dominated. LISA therefore constrains the
relevant parameter combination accurately, while the individual parameters
remain correlated.
}
\label{fig:corner_bp2}
\end{figure}

\begin{figure}
    \centering
    \includegraphics[width=1\linewidth]{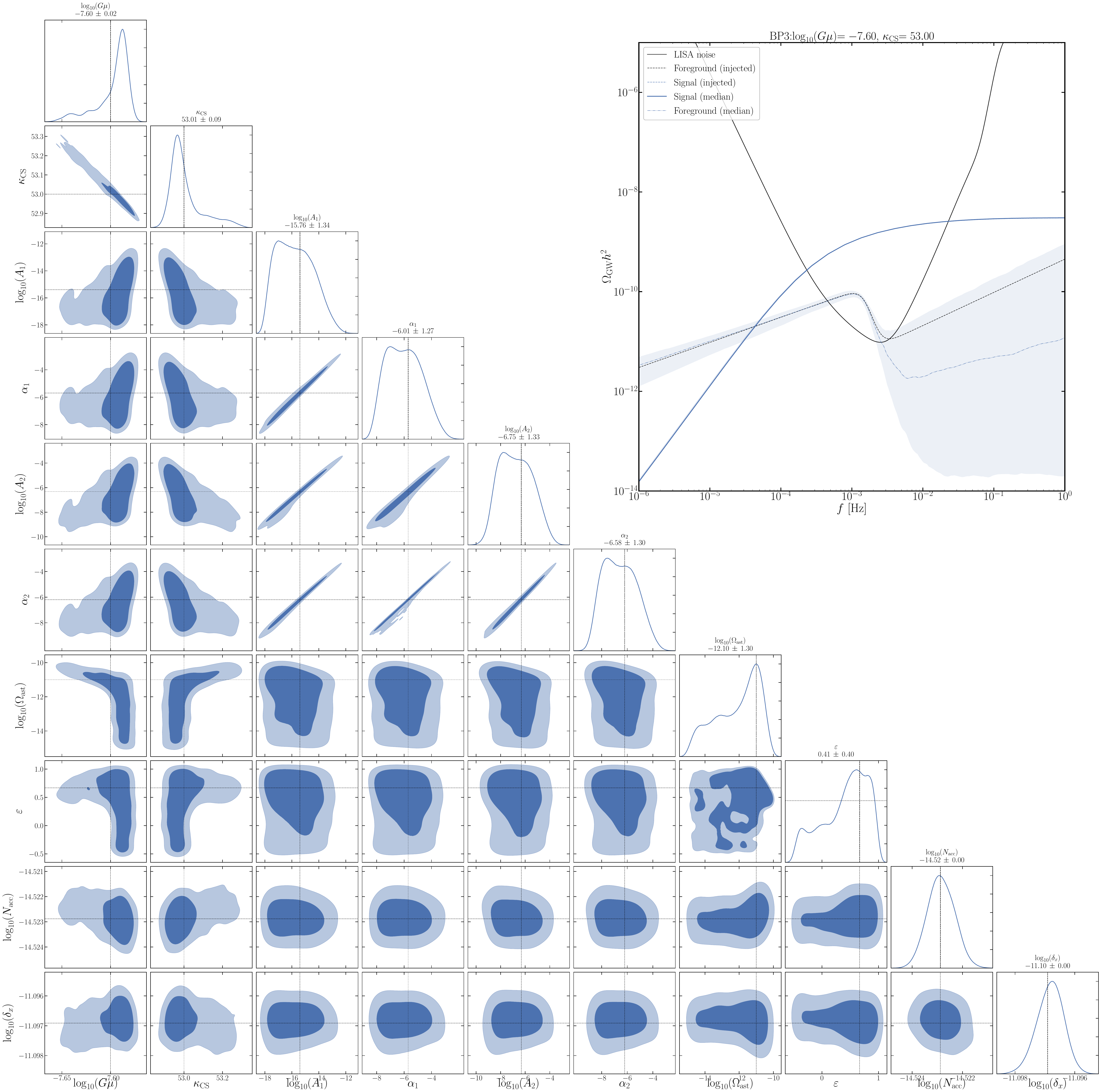}
  \caption{
Posterior distribution for benchmark point BP3. This benchmark illustrates a
compact but correlated reconstruction regime. The anti-correlated tilt reflects
the trade-off between the signal amplitude, controlled primarily by $G\mu$, and
the network lifetime, controlled by $\kappa_{\rm CS}$. The relatively small
posterior area shows that this is not an uninformative degeneracy: the LISA-band
spectrum provides significant constraining power, although the two physical
parameters are recovered with a correlated uncertainty.
}
\label{fig:corner_bp3}
\end{figure}

\begin{figure}
    \centering
    \includegraphics[width=1\linewidth]{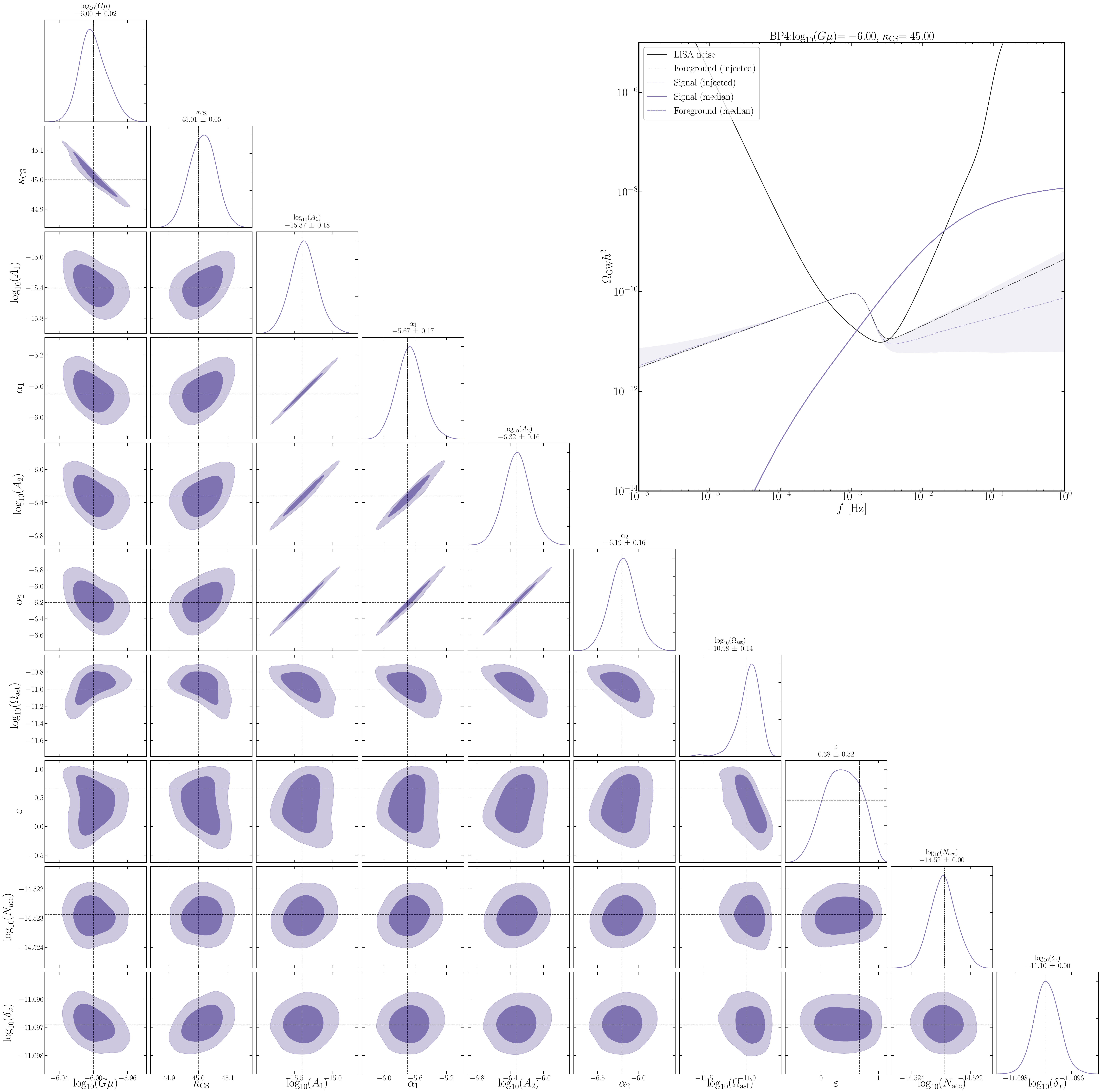}
  \caption{
Posterior distribution for benchmark point BP4. The posterior is strongly
anti-correlated, indicating that LISA constrains an approximately
constant-spectrum combination of $G\mu$ and $\kappa_{\rm CS}$. At the same time,
the posterior area remains small, so the reconstruction is localized rather than
prior dominated. This benchmark therefore represents a compact correlated
posterior: one principal direction is especially well measured, while the
orthogonal direction encodes the remaining amplitude--lifetime degeneracy.
}
\label{fig:corner_bp4}
\end{figure}

The four starred benchmark points in Fig.~\ref{fig:two_panel} illustrate the
main posterior morphologies encountered across the parameter space. Their corner
plots are shown in Figs.~\ref{fig:corner_bp1}--\ref{fig:corner_bp4}. These
examples are useful because the heat maps identify global trends, while the
corner plots show the actual posterior shapes and reveal whether a given region
corresponds to a tilted degeneracy, an axis-aligned weakly constrained
direction, or a compact two-dimensional posterior.

The strongly correlated benchmarks lie near regions where the metastable
spectral structure is only partially resolved by LISA. Their posteriors are
elongated and anti-correlated in the $(G\mu,\kappa_{\rm CS})$ plane, reflecting
an amplitude--lifetime trade-off. These points correspond to the diagonal
structures visible in the $1-\rho^2$ map. In such cases, the data constrain one
local combination of parameters more accurately than the orthogonal direction:
the posterior can therefore be strongly anisotropic while still remaining
informative.

A benchmark closer to the central transition region shows a less elongated
posterior. Here the likelihood has access to both amplitude and spectral-shape
information, allowing changes in $G\mu$ and $\kappa_{\rm CS}$ to be
distinguished more effectively. This case illustrates the transition from a
nearly one-dimensional trade-off to more direct two-parameter reconstruction.
However, as emphasized above, the quality of reconstruction should not be judged
from posterior isotropy alone. A correlated posterior with small covariance area
can be more localized than a nearly isotropic posterior with larger principal
widths.

A weakly correlated benchmark requires a more careful interpretation. If it lies
in a high-SNR region where the turnover or residual metastable structure is
visible, weak correlation can indicate successful reconstruction of both
parameters. If it lies in a plateau-dominated region, weak correlation can
instead reflect an axis-aligned posterior: $G\mu$ is constrained by the
amplitude, while $\kappa_{\rm CS}$ remains weakly identifiable. Thus the corner
plots are essential for interpreting regions where $\rho$, $\kappa_{\rm deg}$,
and the covariance area give complementary information.

A particularly interesting case is BP1 (Fig.\ref{fig:corner_bp1}), which lies close to the boundary between
the $f_{\rm cross}<f_{\rm LISA}^{\rm low}$ regime and the stable-network limit.
In this case the posterior can show both behaviors: an extended, nearly flat
direction associated with the approach to the stable-string limit, and an
anti-correlated branch associated with residual finite-lifetime sensitivity.
The flat direction appears because increasing $\kappa_{\rm CS}$ further makes
the spectrum increasingly insensitive to the precise decay time. The
anti-correlated part instead reflects the portion of the posterior where the
decay-sensitive residual is still resolved by the likelihood.

The sign of this anti-correlation can be understood locally from the
decay-sensitive contribution to the spectrum. In this regime,
the dominant stable-like part is nearly independent of $\kappa_{\rm CS}$, while
the residual finite-lifetime contribution still depends on both parameters.
Writing this contribution schematically as
\[
    \Omega_{\rm dec}
    =
    \Omega_{\rm dec}(G\mu,\kappa_{\rm CS}) ,
\]
a constant-residual direction satisfies
\[
    d\Omega_{\rm dec}
    \simeq
    \frac{\partial \Omega_{\rm dec}}{\partial \log_{10}G\mu}
    d\log_{10}G\mu
    +
    \frac{\partial \Omega_{\rm dec}}{\partial \kappa_{\rm CS}}
    d\kappa_{\rm CS}
    \simeq 0 .
\]
Therefore,
\[
    \frac{d\kappa_{\rm CS}}{d\log_{10}G\mu}
    \simeq
    -
    \frac{
    \partial \Omega_{\rm dec}/\partial \log_{10}G\mu
    }{
    \partial \Omega_{\rm dec}/\partial \kappa_{\rm CS}
    } .
\]
In the relevant part of the BP1 spectrum, decreasing $\kappa_{\rm CS}$ at fixed
$G\mu$ increases the decay-sensitive contribution, while increasing $G\mu$ at
fixed $\kappa_{\rm CS}$ decreases this contribution at the frequencies that
dominate the likelihood. The two derivatives therefore have the same sign in
the above expression, giving
\[
    \frac{d\kappa_{\rm CS}}{d\log_{10}G\mu}<0 .
\]
This produces an anti-correlated branch,
\[
    G\mu \uparrow
    \quad
    \Longleftrightarrow
    \quad
    \kappa_{\rm CS}\downarrow ,
\]
even though the dominant visible spectrum is approximately plateau-like. This
supports the interpretation that, in the hidden high-SNR regime, the remaining
$\kappa_{\rm CS}$ information is driven primarily by the decay-sensitive
finite-lifetime residual of the full metastable-string spectrum rather than by
a directly resolved turnover. 

\subsection{Summary of physical regimes}
\label{sec:sumpr}
The reconstruction behavior can be organized into a small number of physical
regimes. When the signal is weak, the posterior is prior dominated, and the
covariance diagnostics should not be interpreted as evidence for physical
reconstruction. When the LISA-band spectrum is effectively featureless, either
because it is dominated by the infrared tail or because it is approximately
plateau-like, the data mainly constrain limited combinations of the parameters.
In the plateau-dominated case, LISA primarily measures the overall normalization
and therefore constrains $G\mu$, while $\kappa_{\rm CS}$ can remain weakly
identifiable.

When LISA observes only part of the broad metastable transition, the posterior
can develop a correlated amplitude--lifetime degeneracy. In this regime the
signal may be detectable, and one principal parameter combination may be
measured accurately, but the remaining uncertainty is aligned with a trade-off
between $G\mu$ and $\kappa_{\rm CS}$. This produces large $|\rho|$ and often
large $\kappa_{\rm deg}$. Such regions should not automatically be classified as
poor reconstruction: if the covariance area and marginalized widths are small,
they correspond to compact but correlated reconstruction.

The most direct reconstruction occurs when the lifetime-dependent spectral
morphology is sampled by the sensitive part of the LISA band. This transition
window begins approximately when $f_{\rm cross}$ enters the band near
$f_{\rm LISA}^{\rm low}$ and extends until the infrared side of the transition
approaches the high-frequency edge, $f_{\rm low}\sim f_{\rm LISA}^{\rm high}$.
In this regime, the likelihood can use both amplitude and shape information to
separate the string tension from the decay scale. The resulting posterior may be
nearly balanced, or it may remain correlated; the key indicator of strong
reconstruction is not isotropy alone, but small covariance area, small
marginalized widths, and accurate recovery of the injected parameters.

The maps also show that sensitivity to $\kappa_{\rm CS}$ is not limited to
cases where the most visually apparent transition is directly observed. In
high-SNR plateau-like regions, the full metastable-string spectrum can retain
small coherent finite-lifetime distortions relative to the strict stable-string
limit. As discussed above, the decay-sensitive loop contribution can itself have
large weighted significance even when it is visually subdominant in the total
spectrum. When these residual shape differences are resolved, LISA can retain
partial sensitivity to the metastability scale even though the dominant spectrum
appears approximately plateau-like. This should be interpreted as lifetime
sensitivity of the full signal, rather than as evidence for a sharply resolved
turnover in the LISA band.

Overall, these results show that LISA can move beyond detection only when the
metastable-string signal is both sufficiently loud and sufficiently structured
within the detector band, or when high SNR makes residual finite-lifetime
distortions measurable. The combined interpretation of $\rho$,
$\kappa_{\rm deg}$, $\lambda_\perp$, the covariance area, the marginalized
widths, and the reconstruction accuracy separates prior domination,
one-parameter amplitude reconstruction, compact correlated reconstruction,
high-SNR residual lifetime sensitivity, and genuine two-parameter recovery
across the $(G\mu,\kappa_{\rm CS})$ plane.

\section{Reconstruction dependence on Galactic foreground templates}
\label{sec:foreground_template_comparison}

\begin{figure}
    \centering
    \includegraphics[width=.52\linewidth]{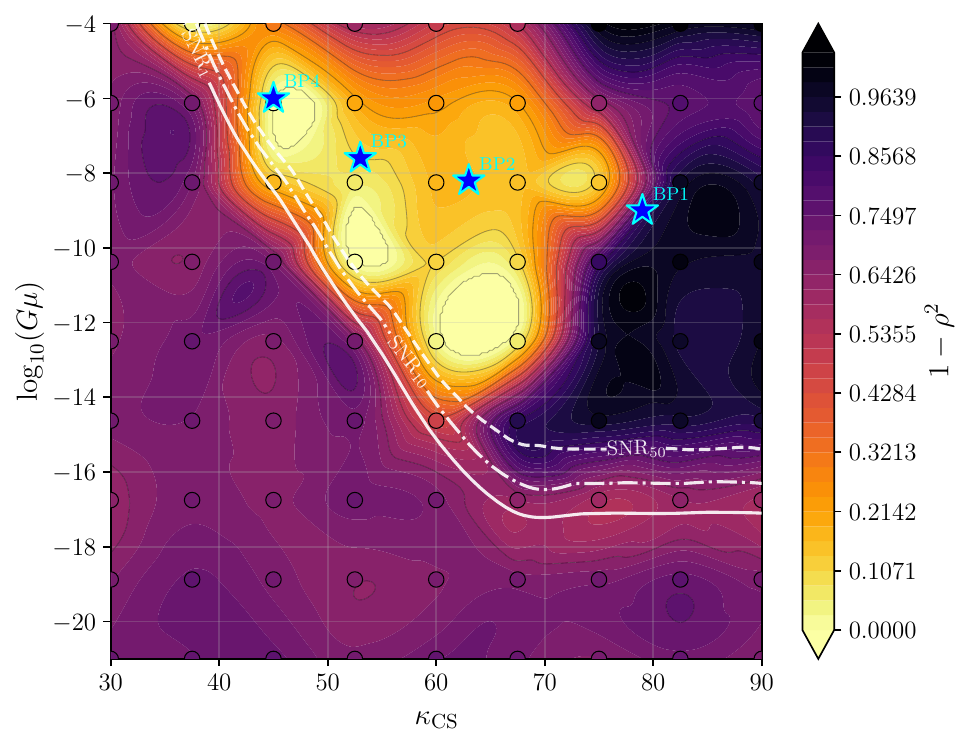}\includegraphics[width=.52\linewidth]{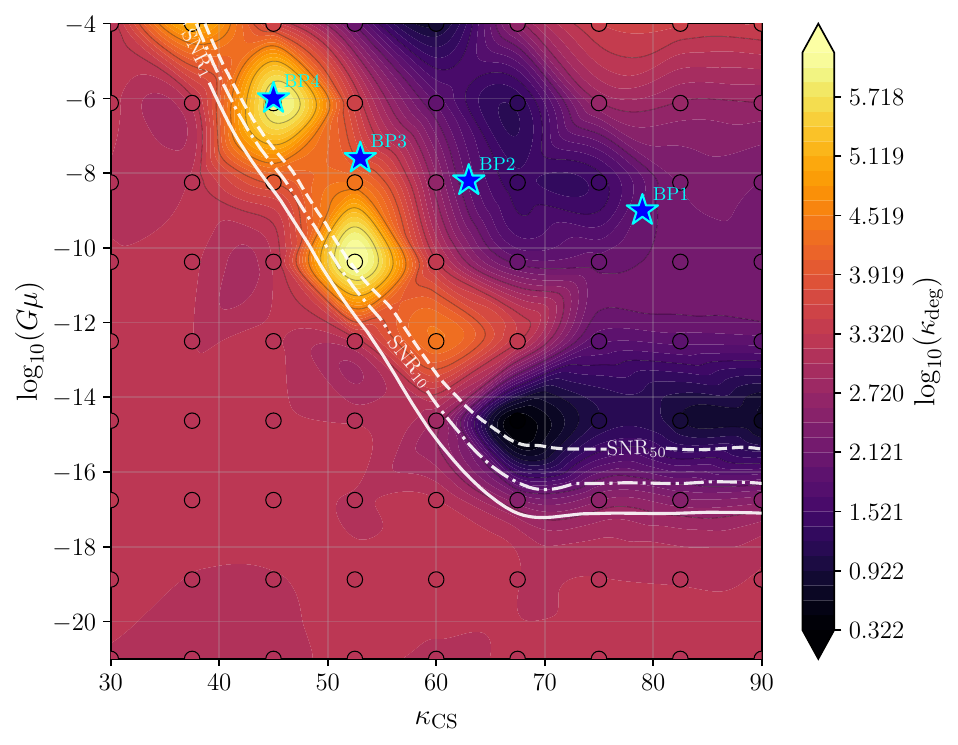}\\ \includegraphics[width=.52\linewidth]{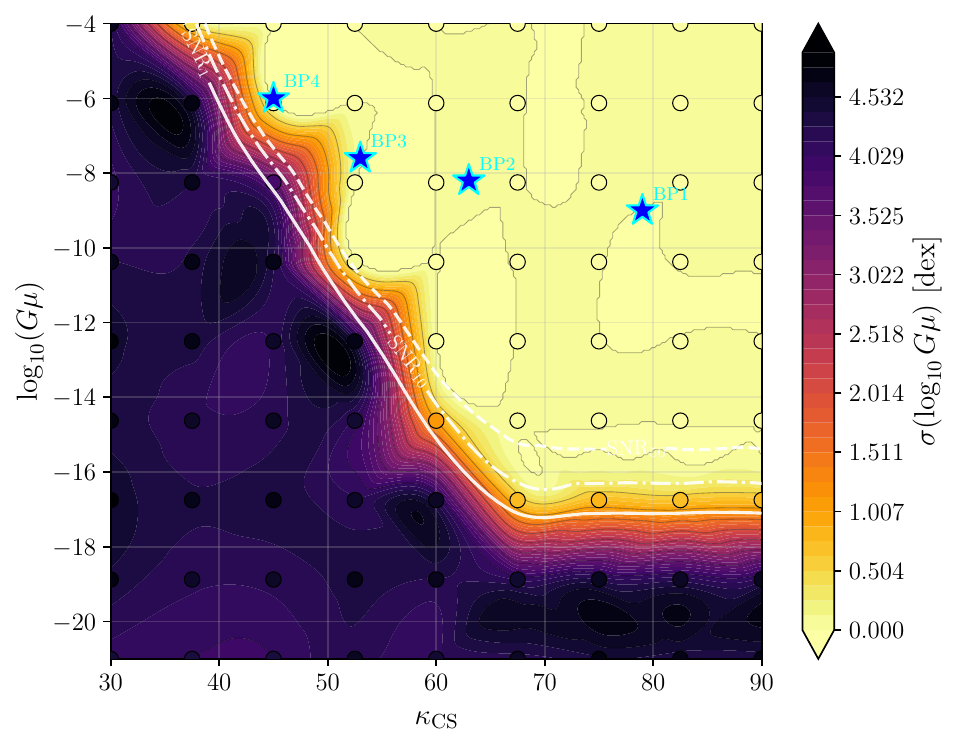}\includegraphics[width=.52\linewidth]{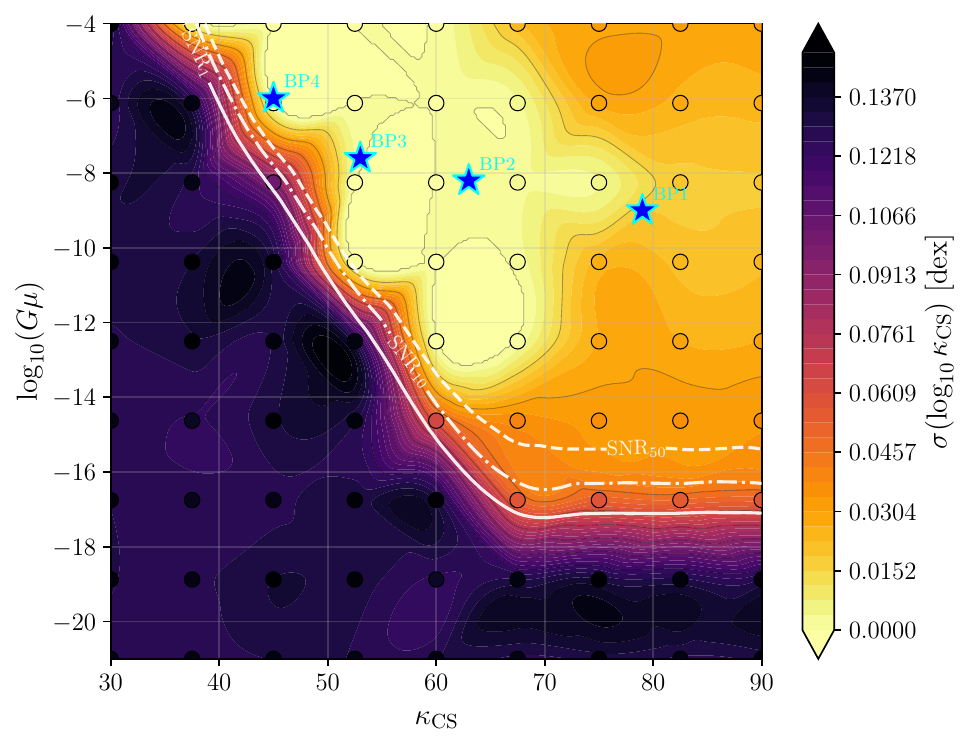}
  \caption{
Posterior-covariance diagnostics obtained using the reduced Galactic foreground
model of Eq.~\eqref{gal_tanh}. The panels show the same quantities as in
Fig.~\ref{fig:two_panel}: the correlation diagnostic $1-\rho^2$, the posterior
anisotropy $\kappa_{\rm deg}$, and the marginalized uncertainties in
$G\mu$ and $\kappa_{\rm CS}$. Compared with the baseline analysis using the more
flexible Galactic foreground model, the tanh template can substantially modify
the correlation and degeneracy structure. Since the foreground shape is fixed
and only its overall normalization is varied, the foreground has less freedom to
absorb metastable-string spectral curvature. As a result, the marginalized
parameter uncertainties improve in parts of the parameter space where the
cosmological signal overlaps with the Galactic foreground. In particular, the
region with $\mathcal{O}(10\%)$ reconstruction of $G\mu$ can extend to smaller
string tensions, improving from approximately $G\mu\sim 10^{-11}$--$10^{-12}$ in
the baseline foreground model to $G\mu\sim 10^{-14}$--$10^{-15}$ for favorable
values of $\kappa_{\rm CS}$.
}
\label{fig:foreground_comparison_tanh}
\end{figure}
\begin{figure}
    \centering
    \includegraphics[width=.52\linewidth]{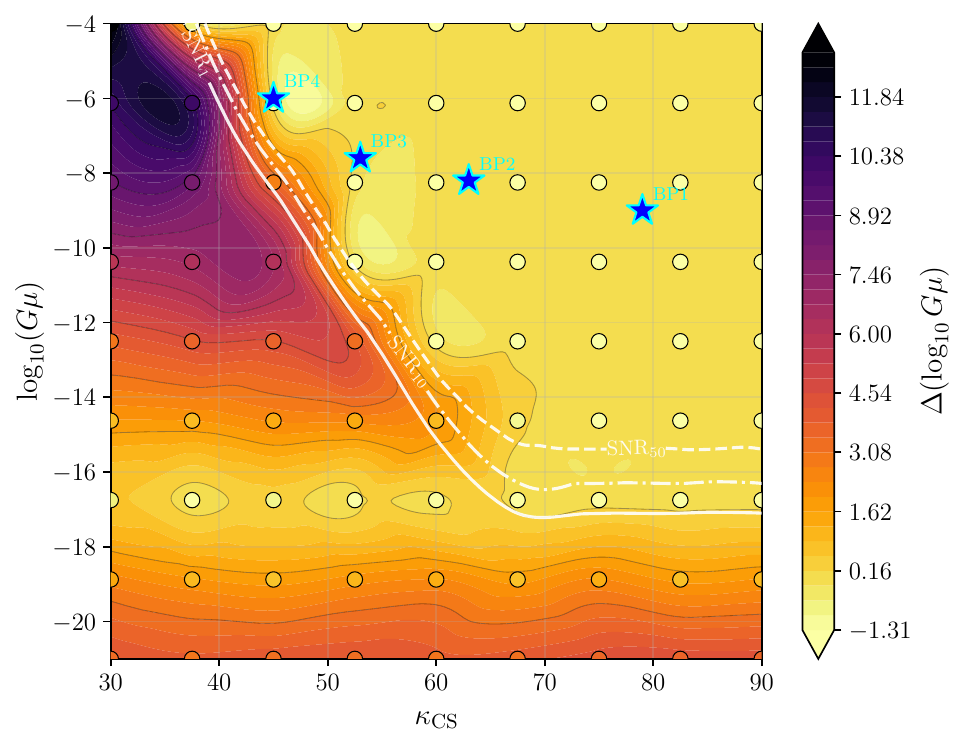}\includegraphics[width=.52\linewidth]{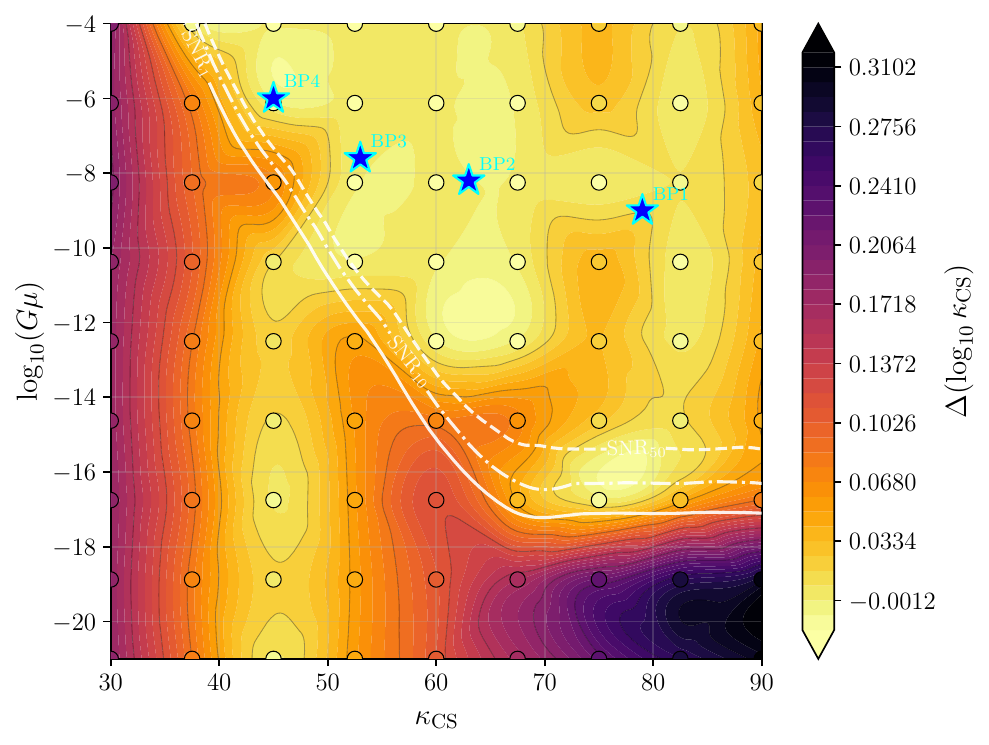}\\ \includegraphics[width=.52\linewidth]{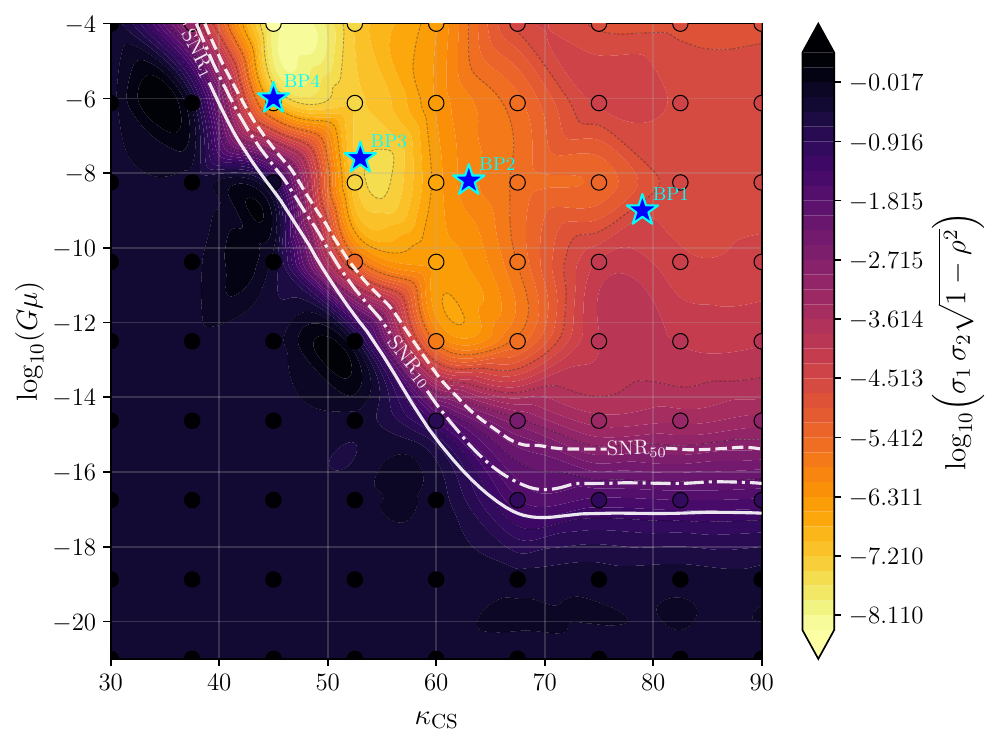}\includegraphics[width=.52\linewidth]{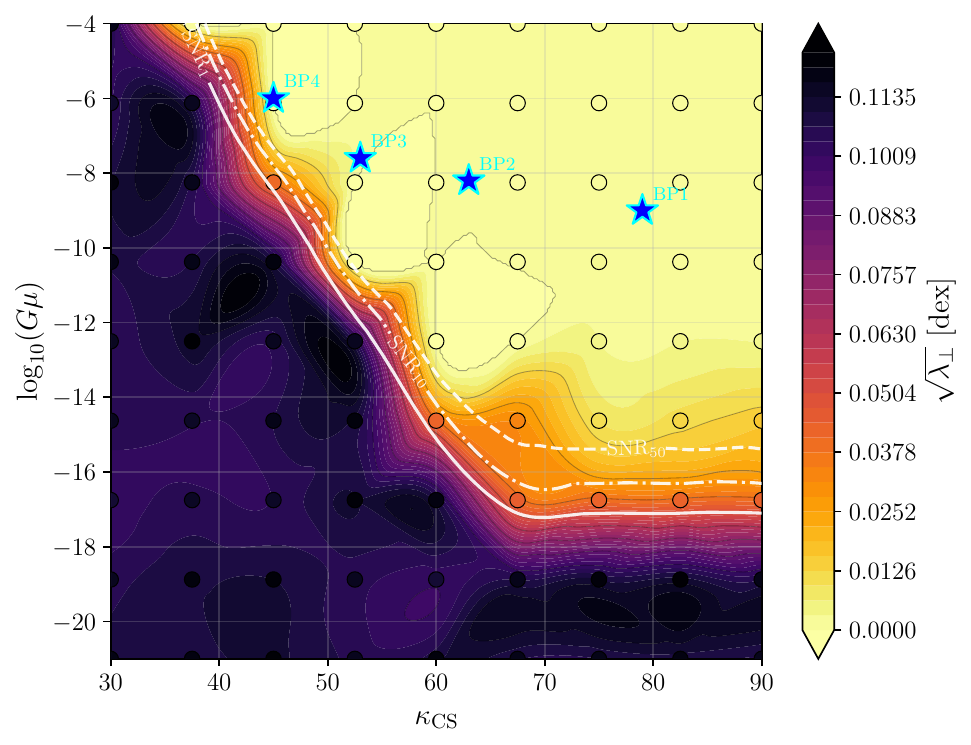}
\caption{
Reconstruction accuracy, posterior localization, and principal-width diagnostics
for the reduced Galactic foreground model of Eq.~\eqref{gal_tanh}. The top
panels show the median reconstruction offsets for $G\mu$ and
$\kappa_{\rm CS}$, defined as
$\Delta(\log_{10}X)\equiv
|\mathrm{median}(\log_{10}X)-\log_{10}X_{\rm inj}|$. These maps quantify
reconstruction accuracy and complement the marginalized posterior-width maps
shown in Fig.~\ref{fig:foreground_comparison_tanh}. The bottom-left panel shows
the covariance-area diagnostic
$A_{\rm cov}\equiv\sqrt{\det C}
=\sigma_1\sigma_2\sqrt{1-\rho^2}
=\sqrt{\lambda_\parallel\lambda_\perp}$,
which measures the overall localization of the posterior in the sampled
$(\log_{10}G\mu,\log_{10}\kappa_{\rm CS})$ plane. The bottom-right panel shows
$\lambda_\perp$, the smaller covariance eigenvalue, which measures the posterior
variance along the best-constrained principal direction. Together, these
diagnostics show that the improvement obtained with the tanh foreground template
is not only a reduction in posterior width, but also corresponds to accurate
recovery and stronger posterior localization in the informative regions of the
$(G\mu,\kappa_{\rm CS})$ plane.
}
\label{fig:foreground_accuracy_tanh}
\end{figure}

The reconstruction results presented in the main analysis were obtained using a
conservative phenomenological template for the unresolved Galactic
double-white-dwarf foreground. That template allows several foreground-shape
parameters to vary and is therefore flexible enough to absorb broad spectral
features in the LISA band. This choice is conservative: it reduces the risk of
overstating the reconstruction capability, but it can also weaken the inference
of cosmological signals whose spectral curvature overlaps with the Galactic
foreground.

To quantify the impact of this modeling assumption, we repeat the reconstruction
analysis using a more restrictive Galactic foreground template. Specifically, we
adopt the tanh model \cite{Caprini:2024hue,Blanco-Pillado:2024aca,Karnesis:2021tsh,Samanta:2025jec}
\begin{equation}
    h^2 \Omega_{\rm Gal}^{\rm GW}(f)
    =
    \frac{f^3}{2}
    \left(\frac{f}{1\,{\rm Hz}}\right)^{-7/3}
    \left[
    1+
    \tanh\left(
    \frac{f_{\rm knee}-f}{f_2}
    \right)
    \right]
    \exp\!\left[-\left(\frac{f}{f_1}\right)^\nu\right]
    h^2\Omega_{\rm Gal}.
    \label{gal_tanh}
\end{equation}
Here $h^2\Omega_{\rm Gal}$ is the only Galactic-foreground parameter varied in
the inference. The remaining shape parameters are fixed by the observation time,
with
\begin{align}
    \log_{10}\left(\frac{f_1}{{\rm Hz}}\right)
    &=
    a_1
    \log_{10}\left(\frac{T_{\rm obs}}{{\rm year}}\right)
    +
    b_1,
    \\
    \log_{10}\left(\frac{f_{\rm knee}}{{\rm Hz}}\right)
    &=
    a_k
    \log_{10}\left(\frac{T_{\rm obs}}{{\rm year}}\right)
    +
    b_k,
\end{align}
and
\begin{align}
    a_1 &= -0.15,
    &
    b_1 &= -2.72,
    \\
    a_k &= -0.37,
    &
    b_k &= -2.49,
    \\
    \nu &= 1.56,
    &
    f_2 &= 6.7\times 10^{-4}\,{\rm Hz}.
\end{align}
This type of tanh-suppressed foreground shape is commonly used in LISA
foreground studies and provides a useful limiting case in which the spectral
shape of the Galactic foreground is assumed to be well characterized. Compared
with the conservative foreground model, Eq.~\eqref{gal_tanh} removes most of
the foreground-shape freedom and allows only the overall foreground amplitude to
float.

This comparison is important because the metastable-string signal is
reconstructed through both its amplitude and its spectral curvature. In the
conservative foreground model, part of this curvature can be absorbed by the
foreground sector, especially when the metastable feature lies in the frequency
range where the Galactic foreground is significant. By contrast, the tanh
template has less freedom to mimic changes in the cosmological signal. We
therefore expect the reconstruction of $G\mu$ and $\kappa_{\rm CS}$ to improve
when Eq.~\eqref{gal_tanh} is used, particularly in the regions where foreground
and cosmological spectral structure overlap.

Figure~\ref{fig:foreground_comparison_tanh} shows the posterior-covariance
diagnostics obtained with the tanh Galactic foreground template. The panels
display the same quantities as in Fig.~\ref{fig:two_panel}: the correlation
diagnostic $1-\rho^2$, the posterior anisotropy $\kappa_{\rm deg}$, and the
marginalized uncertainties in $G\mu$ and $\kappa_{\rm CS}$. The overall
structure of the reconstruction landscape remains recognizable, but the
correlation and degeneracy patterns can change substantially once the foreground
shape is fixed. In particular, the tanh template reduces the ability of the
foreground model to absorb metastable-string curvature, allowing the likelihood
to attribute more of the observed spectral structure to the cosmological signal.

The improvement is especially visible in the marginalized uncertainties. With
the conservative Galactic foreground model, the reconstruction of $G\mu$ at the
$\mathcal{O}(10\%)$ level is limited to relatively high signal amplitudes. With
the tanh foreground template, this region can extend to significantly smaller
string tensions. Depending on the value of $\kappa_{\rm CS}$, the threshold for
$\mathcal{O}(10\%)$ reconstruction of $G\mu$ can move from approximately
$G\mu\sim 10^{-11}$--$10^{-12}$ in the baseline foreground model to
$G\mu\sim 10^{-14}$--$10^{-15}$ in the tanh-template analysis.

Figure~\ref{fig:foreground_accuracy_tanh} provides complementary diagnostics
for the same foreground model. The reconstruction-accuracy maps show the offset
between the posterior median and the injected values of $G\mu$ and
$\kappa_{\rm CS}$. These maps demonstrate that the improved posterior widths in
the informative regions correspond to accurate recovery of the injected
parameters, rather than simply to artificially narrow posteriors. The
$\lambda_\perp$ map further shows where LISA constrains the best-measured
principal parameter combination with high precision. Together, these diagnostics
confirm that the tanh template can improve both the precision and the accuracy
of the reconstruction in regions where the signal is sufficiently informative.

The comparison between the conservative foreground model and the tanh template
should be interpreted as a measure of the systematic impact of Galactic
foreground modeling. The conservative analysis provides a robust lower estimate
of the reconstruction capability, because it marginalizes over a flexible
foreground shape. The tanh analysis provides a more optimistic scenario, in
which the foreground spectral shape is assumed to be known more accurately and
only its normalization remains uncertain. The difference between the two cases
therefore quantifies how strongly metastable-string reconstruction depends on
foreground-shape uncertainty.

This result also emphasizes that the reconstruction landscape is not determined
only by the intrinsic LISA noise curve or by the cosmic-string signal amplitude.
It also depends on how much freedom is assigned to astrophysical foregrounds.
When the Galactic foreground is modeled flexibly, part of the metastable
spectral structure can be absorbed by foreground parameters, weakening the
constraints on $G\mu$ and $\kappa_{\rm CS}$. When the foreground shape is fixed
by Eq.~\eqref{gal_tanh}, the same spectral structure becomes more directly
available for cosmological reconstruction. The tanh-template analysis therefore
shows how improvements in foreground modeling can translate directly into a
larger region of the $(G\mu,\kappa_{\rm CS})$ plane where LISA can perform
two-parameter reconstruction.

\section{Conclusions}
\label{sec:conclusions}

Metastable cosmic strings provide a well-motivated example of a cosmological
stochastic gravitational-wave background whose spectrum carries information
beyond an overall amplitude. In the standard vacuum-tunneling scenario, the
network initially follows the scaling evolution of ordinary cosmic strings. At
later times, monopole--antimonopole pair nucleation on the string worldsheet
depletes the loop population and eventually suppresses the long-lived network.
This finite lifetime modifies the gravitational-wave spectrum, producing an
infrared turnover and a broad transition toward a stable-string-like
high-frequency contribution. The string tension $G\mu$ primarily controls the
overall amplitude, while the metastability parameter $\kappa_{\rm CS}$ controls
the decay time of the network.

In this work, we studied whether LISA can reconstruct the parameters of such a
metastable cosmic-string background in the presence of instrumental noise and
unresolved astrophysical foregrounds. We performed a Bayesian analysis in the
two-dimensional signal-parameter space $(G\mu,\kappa_{\rm CS})$, including
Galactic double-white-dwarf and extragalactic compact-binary foregrounds in the
data model. 

Our results show that reconstruction is controlled by the amount of
lifetime-dependent spectral structure accessible in the LISA band. When the
observed spectrum is effectively featureless, either because LISA observes only
the infrared tail or because the signal is approximately plateau-like, the data
mainly constrain limited combinations of the parameters. In plateau-dominated
regions, LISA primarily measures the overall normalization of the stochastic
background and therefore constrains $G\mu$ more readily than
$\kappa_{\rm CS}$. In this regime, the background can be detectable while the
metastability parameter remains weakly identifiable. The most direct
reconstruction occurs when the broad metastable transition is sampled by the
sensitive part of the LISA band. This transition window begins approximately
when $f_{\rm cross}$ enters the band near $f_{\rm LISA}^{\rm low}$ and extends
until the infrared side of the transition approaches the high-frequency edge,
$f_{\rm low}\sim f_{\rm LISA}^{\rm high}$. In this regime, the data contain both
amplitude and shape information, allowing LISA to separate the string tension
from the network lifetime.

The covariance diagnostics in Figs.~\ref{fig:two_panel} and
\ref{fig:lambda_perp_area} make this reconstruction landscape explicit. The
correlation map identifies diagonal regions where changes in $G\mu$ and
$\kappa_{\rm CS}$ compensate each other, while the eigenvalue-ratio map
quantifies the anisotropy of the posterior. Together with $\lambda_\perp$, the
covariance area, and the marginalized widths, these diagnostics show that
posterior anisotropy has more than one physical origin. It can arise from a
correlated amplitude--lifetime trade-off, from an axis-aligned loss of
identifiability of $\kappa_{\rm CS}$ in plateau-dominated regions, or from the
geometry of the prior in low-SNR regions. Importantly, a large
$\kappa_{\rm deg}$ does not by itself imply poor reconstruction. In the
metastable-string case, some of the smallest posterior areas occur in regions
with both large $\kappa_{\rm deg}$ and large $|\rho|$, corresponding to compact
but correlated reconstruction. In these regions, LISA measures a narrow
amplitude--lifetime combination with high precision, while the remaining
uncertainty is oriented along the correlated trade-off direction. A joint
interpretation of all diagnostics is therefore essential for separating
detection, weak identifiability, compact correlated reconstruction, and genuine
two-parameter recovery.

A notable feature of the metastable-string case is that sensitivity to
$\kappa_{\rm CS}$ is not limited to the direct observation of the most visually
apparent spectral transition. In high-SNR regions, the total spectrum can appear
approximately plateau-like while still retaining small coherent
finite-lifetime distortions relative to the strict stable-string limit. As
discussed above, the decay-sensitive loop contribution can itself have large
weighted significance even when it is visually subdominant in the total
spectrum. When these residual shape differences are resolved, they can
contribute to the reconstruction of $\kappa_{\rm CS}$. Thus, the loss of a
visually obvious turnover or crossover inside the LISA band does not by itself
imply the complete loss of lifetime information, although the interpretation
becomes more subtle near the stable-network boundary.

The benchmark posterior distributions shown in
Figs.~\ref{fig:corner_bp1}--\ref{fig:corner_bp4} confirm the interpretation of
the heat maps. Strongly correlated benchmarks exhibit elongated
anti-correlated posteriors, corresponding to spectra that are nearly
indistinguishable under compensating changes of $G\mu$ and $\kappa_{\rm CS}$.
However, these correlated posteriors can still be compact if the covariance
area and marginalized widths are small. Benchmarks in more informative regions
show improved localization of the signal parameters, while plateau-dominated
cases demonstrate that a precise measurement of $G\mu$ does not automatically
imply a precise measurement of $\kappa_{\rm CS}$. The best reconstruction
regions are therefore not simply the highest-SNR regions or the most isotropic
posterior regions, but those where the signal is sufficiently loud,
spectrally informative, and localized in the full
$(G\mu,\kappa_{\rm CS})$ plane.

We also examined the dependence of the reconstruction on the Galactic foreground
model. The conservative foreground template used in the main analysis allows
several shape parameters to vary and therefore gives a robust, but conservative,
estimate of the reconstruction capability. Repeating the analysis with the
reduced tanh Galactic template of Eq.~\eqref{gal_tanh}, shown in
Figs.~\ref{fig:foreground_comparison_tanh} and
\ref{fig:foreground_accuracy_tanh}, demonstrates that improved foreground
knowledge can substantially enlarge the region of parameter space where
$G\mu$ and $\kappa_{\rm CS}$ are accurately reconstructed. This comparison
quantifies the extent to which Galactic foreground uncertainty can degrade
metastable-string inference.

The analysis presented here focuses on the standard one-timescale metastable
benchmark, in which the onset of loop breaking and the collapse of the
loop-producing network are identified as
$t_s\equiv t_{\rm LB}\equiv t_{\rm NC}$. More general metastable scenarios may
involve distinct timescales, finite-temperature enhancement of monopole-pair
nucleation, modified loop-production histories, or additional emission from
finite string segments. Such effects can change the infrared tail, shift the
spectral transition, or introduce features that are partially degenerate with
the standard vacuum-tunneling template. Extending the present Bayesian
reconstruction framework to these generalized scenarios is an important
direction for future work.

Additional extensions include more flexible astrophysical foreground models,
alternative loop distributions, nonstandard cosmological expansion histories,
and joint analyses with other gravitational-wave frequency bands. Each of these
effects can alter either the amplitude or the shape of the stochastic
background, and hence can affect the inferred reconstruction regions. The
diagnostic framework developed here provides a systematic way to determine
whether such effects produce genuine parameter degeneracies, prior-dominated
posteriors, compact correlated reconstruction, or improved spectral
reconstruction. Our results therefore identify where LISA can move beyond the
detection of a stochastic background and begin to probe the lifetime of an
underlying metastable cosmic-string network.
 \section*{Acknowledgments}
The work of RS is supported by the research project TAsP (Theoretical Astroparticle Physics) funded by the Istituto Nazionale di Fisica Nucleare (INFN).

\bibliography{bibliography}

@article{Buchmuller:2021mbb,
    author = "Buchmuller, Wilfried and Domcke, Valerie and Schmitz, Kai",
    title = "{Stochastic gravitational-wave background from metastable cosmic strings}",
    eprint = "2107.04578",
    archivePrefix = "arXiv",
    primaryClass = "hep-ph",
    reportNumber = "CERN-TH-2021-107, DESY 21-101",
    doi = "10.1088/1475-7516/2021/12/006",
    journal = "JCAP",
    volume = "12",
    number = "12",
    pages = "006",
    year = "2021"
}

@article{Buchmuller:2023aus,
    author = "Buchmuller, Wilfried and Domcke, Valerie and Schmitz, Kai",
    title = "{Metastable cosmic strings}",
    eprint = "2307.04691",
    archivePrefix = "arXiv",
    primaryClass = "hep-ph",
    reportNumber = "CERN-TH-2023-118, MS-TP-23-37, DESY-23-117",
    doi = "10.1088/1475-7516/2023/11/020",
    journal = "JCAP",
    volume = "11",
    pages = "020",
    year = "2023"
}

@article{Preskill:1992ck,
    author = "Preskill, John and Vilenkin, Alexander",
    title = "{Decay of metastable topological defects}",
    eprint = "hep-ph/9209210",
    archivePrefix = "arXiv",
    reportNumber = "HUTP-92-A018, CALT-68-1786",
    doi = "10.1103/PhysRevD.47.2324",
    journal = "Phys. Rev. D",
    volume = "47",
    pages = "2324--2342",
    year = "1993"
}

@article{Monin:2008mp,
    author = "Monin, A. and Voloshin, M. B.",
    title = "{The Spontaneous breaking of a metastable string}",
    eprint = "0808.1693",
    archivePrefix = "arXiv",
    primaryClass = "hep-th",
    reportNumber = "FTPI-MINN-08-31, UMN-TH-2713-08",
    doi = "10.1103/PhysRevD.78.065048",
    journal = "Phys. Rev. D",
    volume = "78",
    pages = "065048",
    year = "2008"
}

@article{Tranchedone:2026lav,
    author = "Tranchedone, Lorenzo and Carragher, Ethan and Hardy, Edward and Koscelansk{\'a} van IJcken, Nat{\'a}lie",
    title = "{Metastable cosmic strings are broken at the start}",
    eprint = "2601.04320",
    archivePrefix = "arXiv",
    primaryClass = "hep-ph",
     doi = "",
    journal = "",
    volume = "",
    month = "1",
    year = "2026"
}

@article{Asl:2026zpj,
    author = "Asl, Doa Hashemi and Schmitz, Kai",
    title = "{New gravitational-wave templates for metastable cosmic strings: Loop breaking versus network collapse}",
    eprint = "2604.28097",
    archivePrefix = "arXiv",
    primaryClass = "hep-ph",
    reportNumber = "MS-TP-26-15",
     doi = "",
    journal = "",
    volume = "",
    month = "4",
    year = "2026"
}

@article{deGiorgi:2026fyx,
    author = "de Giorgi, Arturo and Ingoldby, James and Khoze, Valentin V. and Turner, Jessica",
    title = "{Thermal Metastable Strings in One-Scale Models and Gravitational Waves}",
    eprint = "2606.02689",
    archivePrefix = "arXiv",
    primaryClass = "hep-ph",
    reportNumber = "IPPP/26/41",
     doi = "",
    journal = "",
    volume = "",
    month = "6",
    year = "2026"
}

@article{Monin:2008uj,
    author = "Monin, A. and Voloshin, M. B.",
    title = "{Breaking of a metastable string at finite temperature}",
    eprint = "0809.5286",
    archivePrefix = "arXiv",
    primaryClass = "hep-th",
    reportNumber = "FTPI-MINN-08-37, UMN-TH-2719-08",
    doi = "10.1103/PhysRevD.78.125029",
    journal = "Phys. Rev. D",
    volume = "78",
    pages = "125029",
    year = "2008"
}

@article{Monin:2009ch,
    author = "Monin, A. and Voloshin, M. B.",
    title = "{Destruction of a metastable string by particle collisions}",
    eprint = "0902.0407",
    archivePrefix = "arXiv",
    primaryClass = "hep-th",
    reportNumber = "FTPI-MINN-09-06, UMN-TH-2736-09",
    doi = "10.1134/S1063778810040162",
    journal = "Phys. Atom. Nucl.",
    volume = "73",
    pages = "703--710",
    year = "2010"
}

@article{Monin:2009gi,
    author = "Monin, A. and Voloshin, M. B.",
    title = "{Spontaneous and Induced Decay of Metastable Strings and Domain Walls}",
    eprint = "0904.1728",
    archivePrefix = "arXiv",
    primaryClass = "hep-th",
    reportNumber = "FTPI-MINN-09-15, UMN-TH-2743-09",
    doi = "10.1016/j.aop.2009.07.007",
    journal = "Annals Phys.",
    volume = "325",
    pages = "16--48",
    year = "2010"
}

@article{Datta:2026fav,
    author = "Datta, Satyabrata and Samanta, Rome",
    title = "{PTA-Compatible Domain Walls at LISA and Taiji: Bayesian Reconstruction and Multiband Inference}",
    eprint = "2606.09713",
    archivePrefix = "arXiv",
    primaryClass = "hep-ph",
    doi = "",
    journal = "",
    volume = "",
    pages = "",
    month = "6",
    year = "2026"
}

@article{Vilenkin:1982hm,
    author = "Vilenkin, A.",
    title = "{COSMOLOGICAL EVOLUTION OF MONOPOLES CONNECTED BY STRINGS}",
    doi = "10.1016/0550-3213(82)90037-2",
    journal = "Nucl. Phys. B",
    volume = "196",
    pages = "240--258",
    year = "1982"
}

@article{Vachaspati:1984gt,
    author = "Vachaspati, Tanmay and Vilenkin, Alexander",
    title = "{Gravitational Radiation from Cosmic Strings}",
    reportNumber = "HUTP-84/A065",
    doi = "10.1103/PhysRevD.31.3052",
    journal = "Phys. Rev. D",
    volume = "31",
    pages = "3052",
    year = "1985"
}

@article{Antusch:2023zjk,
    author = "Antusch, Stefan and Hinze, Kevin and Saad, Shaikh and Steiner, Jonathan",
    title = "{Singling out SO(10) GUT models using recent PTA results}",
    eprint = "2307.04595",
    archivePrefix = "arXiv",
    primaryClass = "hep-ph",
    doi = "10.1103/PhysRevD.108.095053",
    journal = "Phys. Rev. D",
    volume = "108",
    number = "9",
    pages = "095053",
    year = "2023"
}

@article{Lazarides:2023rqf,
    author = "Lazarides, George and Maji, Rinku and Moursy, Ahmad and Shafi, Qaisar",
    title = "{Inflation, superheavy metastable strings and gravitational waves in non-supersymmetric flipped SU(5)}",
    eprint = "2308.07094",
    archivePrefix = "arXiv",
    primaryClass = "hep-ph",
    doi = "10.1088/1475-7516/2024/03/006",
    journal = "JCAP",
    volume = "03",
    pages = "006",
    year = "2024"
}

@article{Afzal:2023cyp,
    author = "Afzal, Adeela and Mehmood, Maria and Rehman, Mansoor Ur and Shafi, Qaisar",
    title = "{Supersymmetric hybrid inflation and current-carrying metastable cosmic strings in SU(4)c{\texttimes}SU(2)L{\texttimes}U(1)R}",
    eprint = "2308.11410",
    archivePrefix = "arXiv",
    primaryClass = "hep-ph",
    doi = "10.1103/d7zx-c7f7",
    journal = "Phys. Rev. D",
    volume = "112",
    number = "8",
    pages = "083545--23",
    year = "2025"
}

@article{Ahmed:2023pjl,
    author = "Ahmed, Waqas and Chowdhury, Talal Ahmed and Nasri, Salah and Saad, Shaikh",
    title = "{Gravitational waves from metastable cosmic strings in the Pati-Salam model in light of new pulsar timing array data}",
    eprint = "2308.13248",
    archivePrefix = "arXiv",
    primaryClass = "hep-ph",
    doi = "10.1103/PhysRevD.109.015008",
    journal = "Phys. Rev. D",
    volume = "109",
    number = "1",
    pages = "015008",
    year = "2024"
}

@article{Afzal:2023kqs,
    author = "Afzal, Adeela and Shafi, Qaisar and Tiwari, Amit",
    title = "{Gravitational wave emission from metastable current-carrying strings in E6}",
    eprint = "2311.05564",
    archivePrefix = "arXiv",
    primaryClass = "hep-ph",
    doi = "10.1016/j.physletb.2024.138516",
    journal = "Phys. Lett. B",
    volume = "850",
    pages = "138516",
    year = "2024"
}

@article{Lazarides:2024niy,
    author = "Lazarides, George and Maji, Rinku and Shafi, Qaisar",
    title = "{Quantum tunneling in the early universe: stable magnetic monopoles from metastable cosmic strings}",
    eprint = "2402.03128",
    archivePrefix = "arXiv",
    primaryClass = "hep-ph",
    doi = "10.1088/1475-7516/2024/05/128",
    journal = "JCAP",
    volume = "05",
    pages = "128",
    year = "2024"
}

@article{Ahmed:2024iyd,
    author = "Ahmed, Waqas and Mehmood, Maria and Rehman, Mansoor Ur and Zubair, Umer",
    title = "{Inflation, proton decay and gravitational waves from metastable strings in SU(4)$_{C}$ {\texttimes} SU(2)$_{L}$ {\texttimes} U(1)$_{R}$ model}",
    eprint = "2404.06008",
    archivePrefix = "arXiv",
    primaryClass = "hep-ph",
    doi = "10.1088/1475-7516/2025/09/035",
    journal = "JCAP",
    volume = "09",
    pages = "035",
    year = "2025"
}

@article{Antusch:2024nqg,
    author = "Antusch, Stefan and Hinze, Kevin and Saad, Shaikh",
    title = "{Explaining PTA results by metastable cosmic strings from SO(10) GUT}",
    eprint = "2406.17014",
    archivePrefix = "arXiv",
    primaryClass = "hep-ph",
    doi = "10.1088/1475-7516/2024/10/007",
    journal = "JCAP",
    volume = "10",
    pages = "007",
    year = "2024"
}

@article{Datta:2025owx,
    author = "Datta, Satyabrata and Samanta, Rome",
    title = "{Multifaceted supercooling: from PTA to LIGO}",
    eprint = "2506.21397",
    archivePrefix = "arXiv",
    primaryClass = "hep-ph",
    doi = "10.1007/JHEP12(2025)161",
    journal = "JHEP",
    volume = "12",
    pages = "161",
    year = "2025"
}

@article{Maji:2024cwv,
    author = "Maji, Rinku and Moursy, Ahmad and Shafi, Qaisar",
    title = "{Induced gravitational waves, metastable cosmic strings and primordial black holes in GUTs}",
    eprint = "2409.13584",
    archivePrefix = "arXiv",
    primaryClass = "hep-ph",
    doi = "10.1088/1475-7516/2025/01/106",
    journal = "JCAP",
    volume = "01",
    pages = "106",
    year = "2025"
}

@article{Pallis:2024joc,
    author = "Pallis, C.",
    title = "{T-model Higgs inflation and metastable cosmic strings}",
    eprint = "2409.14338",
    archivePrefix = "arXiv",
    primaryClass = "hep-ph",
    doi = "10.1007/JHEP01(2025)178",
    journal = "JHEP",
    volume = "01",
    pages = "178",
    year = "2025"
}

@article{Hill:1987qx,
    author = "Hill, Christopher T. and Hodges, Hardy M. and Turner, Michael S.",
    title = "{Bosonic Superconducting Cosmic Strings}",
    reportNumber = "FERMILAB-PUB-87-063-A",
    doi = "10.1103/PhysRevD.37.263",
    journal = "Phys. Rev. D",
    volume = "37",
    pages = "263",
    year = "1988"
}

@article{Emond:2021vts,
    author = "Emond, William T. and Ramazanov, Sabir and Samanta, Rome",
    title = "{Gravitational waves from melting cosmic strings}",
    eprint = "2108.05377",
    archivePrefix = "arXiv",
    primaryClass = "hep-ph",
    doi = "10.1088/1475-7516/2022/01/057",
    journal = "JCAP",
    volume = "01",
    number = "01",
    pages = "057",
    year = "2022"
}

@article{Blanco-Pillado:2013qja,
    author = "Blanco-Pillado, Jose J. and Olum, Ken D. and Shlaer, Benjamin",
    title = "{The number of cosmic string loops}",
    eprint = "1309.6637",
    archivePrefix = "arXiv",
    primaryClass = "astro-ph.CO",
    doi = "10.1103/PhysRevD.89.023512",
    journal = "Phys. Rev. D",
    volume = "89",
    number = "2",
    pages = "023512",
    year = "2014"
}

@article{Blanco-Pillado:2017oxo,
    author = "Blanco-Pillado, Jose J. and Olum, Ken D.",
    title = "{Stochastic gravitational wave background from smoothed cosmic string loops}",
    eprint = "1709.02693",
    archivePrefix = "arXiv",
    primaryClass = "astro-ph.CO",
    doi = "10.1103/PhysRevD.96.104046",
    journal = "Phys. Rev. D",
    volume = "96",
    number = "10",
    pages = "104046",
    year = "2017"
}

@article{Martins:1996jp,
    author = "Martins, C. J. A. P. and Shellard, E. P. S.",
    title = "{Quantitative string evolution}",
    eprint = "hep-ph/9602271",
    archivePrefix = "arXiv",
    reportNumber = "DAMTP-R-96-5",
    doi = "10.1103/PhysRevD.54.2535",
    journal = "Phys. Rev. D",
    volume = "54",
    pages = "2535--2556",
    year = "1996"
}

@article{Martins:2000cs,
    author = "Martins, C. J. A. P. and Shellard, E. P. S.",
    title = "{Extending the velocity dependent one scale string evolution model}",
    eprint = "hep-ph/0003298",
    archivePrefix = "arXiv",
    doi = "10.1103/PhysRevD.65.043514",
    journal = "Phys. Rev. D",
    volume = "65",
    pages = "043514",
    year = "2002"
}

@article{Sousa:2013aaa,
    author = "Sousa, L. and Avelino, P. P.",
    title = "{Stochastic Gravitational Wave Background generated by Cosmic String Networks: Velocity-Dependent One-Scale model versus Scale-Invariant Evolution}",
    eprint = "1304.2445",
    archivePrefix = "arXiv",
    primaryClass = "astro-ph.CO",
    doi = "10.1103/PhysRevD.88.023516",
    journal = "Phys. Rev. D",
    volume = "88",
    number = "2",
    pages = "023516",
    year = "2013"
}

@article{Auclair:2019wcv,
    author = "Auclair, Pierre and others",
    title = "{Probing the gravitational wave background from cosmic strings with LISA}",
    eprint = "1909.00819",
    archivePrefix = "arXiv",
    primaryClass = "astro-ph.CO",
    doi = "10.1088/1475-7516/2020/04/034",
    journal = "JCAP",
    volume = "04",
    pages = "034",
    year = "2020"
}

@article{Damour:2001bk,
    author = "Damour, Thibault and Vilenkin, Alexander",
    title = "{Gravitational wave bursts from cusps and kinks on cosmic strings}",
    eprint = "gr-qc/0104026",
    archivePrefix = "arXiv",
    reportNumber = "IHES-P-01-15",
    doi = "10.1103/PhysRevD.64.064008",
    journal = "Phys. Rev. D",
    volume = "64",
    pages = "064008",
    year = "2001"
}

@article{Matsunami:2019fss,
    author = "Matsunami, Daiju and Pogosian, Levon and Saurabh, Ayush and Vachaspati, Tanmay",
    title = "{Decay of Cosmic String Loops Due to Particle Radiation}",
    eprint = "1903.05102",
    archivePrefix = "arXiv",
    primaryClass = "hep-ph",
    doi = "10.1103/PhysRevLett.122.201301",
    journal = "Phys. Rev. Lett.",
    volume = "122",
    number = "20",
    pages = "201301",
    year = "2019"
}

@article{Auclair:2019jip,
    author = "Auclair, Pierre and Steer, Dani{\`e}le A. and Vachaspati, Tanmay",
    title = "{Particle emission and gravitational radiation from cosmic strings: observational constraints}",
    eprint = "1911.12066",
    archivePrefix = "arXiv",
    primaryClass = "hep-ph",
    doi = "10.1103/PhysRevD.101.083511",
    journal = "Phys. Rev. D",
    volume = "101",
    number = "8",
    pages = "083511",
    year = "2020"
}

@article{Balaji:2023ehk,
    author = "Balaji, Shyam and Dom{\`e}nech, Guillem and Franciolini, Gabriele",
    title = "{Scalar-induced gravitational wave interpretation of PTA data: the role of scalar fluctuation propagation speed}",
    eprint = "2307.08552",
    archivePrefix = "arXiv",
    primaryClass = "gr-qc",
    doi = "10.1088/1475-7516/2023/10/041",
    journal = "JCAP",
    volume = "10",
    pages = "041",
    year = "2023"
}

@article{Leblond:2009fq,
    author = "Leblond, Louis and Shlaer, Benjamin and Siemens, Xavier",
    title = "{Gravitational Waves from Broken Cosmic Strings: The Bursts and the Beads}",
    eprint = "0903.4686",
    archivePrefix = "arXiv",
    primaryClass = "astro-ph.CO",
    reportNumber = "NSF-KITP-09-36, MIFP-09-16",
    doi = "10.1103/PhysRevD.79.123519",
    journal = "Phys. Rev. D",
    volume = "79",
    pages = "123519",
    year = "2009"
}

@article{Chitose:2025qyt,
    author = "Chitose, Akifumi and Ibe, Masahiro and Neda, Shunsuke and Shirai, Satoshi",
    title = "{Do Cosmic String Segments Emit Gravitational Waves?}",
    eprint = "2507.12386",
    archivePrefix = "arXiv",
    primaryClass = "hep-ph",
    reportNumber = "IPMU25-0039",
    doi = "",
    journal = "",
    volume = "",
    number = "",
    month = "7",
    year = "2025"
}

@article{Borah:2022byb,
    author = "Borah, Debasish and Jyoti Das, Suruj and Saha, Abhijit Kumar and Samanta, Rome",
    title = "{Probing WIMP dark matter via gravitational waves{\textquoteright} spectral shapes}",
    eprint = "2202.10474",
    archivePrefix = "arXiv",
    primaryClass = "hep-ph",
    doi = "10.1103/PhysRevD.106.L011701",
    journal = "Phys. Rev. D",
    volume = "106",
    number = "1",
    pages = "L011701",
    year = "2022"
}

@article{Chianese:2024gee,
    author = "Chianese, Marco and Datta, Satyabrata and Miele, Gennaro and Samanta, Rome and Saviano, Ninetta",
    title = "{Probing flavored regimes of leptogenesis with gravitational waves from cosmic strings}",
    eprint = "2406.01231",
    archivePrefix = "arXiv",
    primaryClass = "hep-ph",
    doi = "10.1103/PhysRevD.111.L041305",
    journal = "Phys. Rev. D",
    volume = "111",
    number = "4",
    pages = "L041305",
    year = "2025"
}

@article{Cui:2018rwi,
    author = "Cui, Yanou and Lewicki, Marek and Morrissey, David E. and Wells, James D.",
    title = "{Probing the pre-BBN universe with gravitational waves from cosmic strings}",
    eprint = "1808.08968",
    archivePrefix = "arXiv",
    primaryClass = "hep-ph",
    reportNumber = "KCL-PH-TH/2018-47",
    doi = "10.1007/JHEP01(2019)081",
    journal = "JHEP",
    volume = "01",
    pages = "081",
    year = "2019"
}

@article{Guedes:2018afo,
    author = "Guedes, G. S. F. and Avelino, P. P. and Sousa, L.",
    title = "{Signature of inflation in the stochastic gravitational wave background generated by cosmic string networks}",
    eprint = "1809.10802",
    archivePrefix = "arXiv",
    primaryClass = "astro-ph.CO",
    doi = "10.1103/PhysRevD.98.123505",
    journal = "Phys. Rev. D",
    volume = "98",
    number = "12",
    pages = "123505",
    year = "2018"
}

@article{Gouttenoire:2019kij,
    author = "Gouttenoire, Yann and Servant, G{\'e}raldine and Simakachorn, Peera",
    title = "{Beyond the Standard Models with Cosmic Strings}",
    eprint = "1912.02569",
    archivePrefix = "arXiv",
    primaryClass = "hep-ph",
    reportNumber = "DESY-19-204",
    doi = "10.1088/1475-7516/2020/07/032",
    journal = "JCAP",
    volume = "07",
    pages = "032",
    year = "2020"
}

@article{Antusch:2024ypp,
    author = "Antusch, Stefan and Hinze, Kevin and Saad, Shaikh and Steiner, Jonathan",
    title = "{Probing SUSY at gravitational wave observatories}",
    eprint = "2405.03746",
    archivePrefix = "arXiv",
    primaryClass = "hep-ph",
    doi = "10.1016/j.physletb.2024.138924",
    journal = "Phys. Lett. B",
    volume = "856",
    pages = "138924",
    year = "2024"
}

@article{Chitose:2023dam,
    author = "Chitose, Akifumi and Ibe, Masahiro and Nakayama, Yuhei and Shirai, Satoshi and Watanabe, Keiichi",
    title = "{Revisiting metastable cosmic string breaking}",
    eprint = "2312.15662",
    archivePrefix = "arXiv",
    primaryClass = "hep-ph",
    reportNumber = "IPMU23-0054",
    doi = "10.1007/JHEP04(2024)068",
    journal = "JHEP",
    volume = "04",
    pages = "068",
    year = "2024"
}

@article{Chitose:2024pmz,
    author = "Chitose, Akifumi and Ibe, Masahiro and Neda, Shunsuke and Shirai, Satoshi",
    title = "{Gravitational waves from metastable cosmic strings in supersymmetric new inflation model}",
    eprint = "2411.13299",
    archivePrefix = "arXiv",
    primaryClass = "hep-ph",
    reportNumber = "IPMU24-0043",
    doi = "10.1088/1475-7516/2025/04/010",
    journal = "JCAP",
    volume = "04",
    pages = "010",
    year = "2025"
}

@article{Caprini:2019pxz,
    author = "Caprini, Chiara and Figueroa, Daniel G. and Flauger, Raphael and Nardini, Germano and Peloso, Marco and Pieroni, Mauro and Ricciardone, Angelo and Tasinato, Gianmassimo",
    title = "{Reconstructing the spectral shape of a stochastic gravitational wave background with LISA}",
    eprint = "1906.09244",
    archivePrefix = "arXiv",
    primaryClass = "astro-ph.CO",
    reportNumber = "LISA-CosWG-19-02",
    doi = "10.1088/1475-7516/2019/11/017",
    journal = "JCAP",
    volume = "11",
    pages = "017",
    year = "2019"
}

@article{Flauger:2020qyi,
    author = "Flauger, Raphael and Karnesis, Nikolaos and Nardini, Germano and Pieroni, Mauro and Ricciardone, Angelo and Torrado, Jes{\'u}s",
    title = "{Improved reconstruction of a stochastic gravitational wave background with LISA}",
    eprint = "2009.11845",
    archivePrefix = "arXiv",
    primaryClass = "astro-ph.CO",
    doi = "10.1088/1475-7516/2021/01/059",
    journal = "JCAP",
    volume = "01",
    pages = "059",
    year = "2021"
}

@article{Caprini:2024hue,
    author = "Caprini, Chiara and Jinno, Ryusuke and Lewicki, Marek and Madge, Eric and Merchand, Marco and Nardini, Germano and Pieroni, Mauro and Roper Pol, Alberto and Vaskonen, Ville",
    collaboration = "LISA Cosmology Working Group",
    title = "{Gravitational waves from first-order phase transitions in LISA: reconstruction pipeline and physics interpretation}",
    eprint = "2403.03723",
    archivePrefix = "arXiv",
    primaryClass = "astro-ph.CO",
    reportNumber = "LISA-COSWG-24-01, CERN-TH-2024-029",
    doi = "10.1088/1475-7516/2024/10/020",
    journal = "JCAP",
    volume = "10",
    pages = "020",
    year = "2024"
}

@article{Blanco-Pillado:2024aca,
    author = "Blanco-Pillado, Jose J. and Cui, Yanou and Kuroyanagi, Sachiko and Lewicki, Marek and Nardini, Germano and Pieroni, Mauro and Rybak, Ivan Yu. and Sousa, Lara and Wachter, Jeremy M.",
    collaboration = "LISA Cosmology Working Group",
    title = "{Gravitational waves from cosmic strings in LISA: reconstruction pipeline and physics interpretation}",
    eprint = "2405.03740",
    archivePrefix = "arXiv",
    primaryClass = "astro-ph.CO",
    reportNumber = "LISA-COSWG-24-02, CERN-TH-2024-085",
    doi = "10.1088/1475-7516/2025/05/006",
    journal = "JCAP",
    volume = "05",
    pages = "006",
    year = "2025"
}

@article{Guan:2025idx,
    author = "Guan, Shuo and Guo, Huai-Ke and Jiao, Dian and Liang, Qingyuan and Wu, Lei and Zhang, Yang",
    title = "{Measuring gravitational wave spectrum from electroweak phase transition and Higgs self-couplings}",
    eprint = "2511.00996",
    archivePrefix = "arXiv",
    primaryClass = "hep-ph",
    doi = "10.1103/2xy8-9y1r",
    journal = "Phys. Rev. D",
    volume = "113",
    number = "7",
    pages = "075014",
    year = "2026"
}

@article{LISACosmologyWorkingGroup:2025vdz,
    author = "Gammal, Jonas El and others",
    collaboration = "LISA Cosmology Working Group",
    title = "{Reconstructing primordial curvature perturbations via scalar-induced gravitational waves with LISA}",
    eprint = "2501.11320",
    archivePrefix = "arXiv",
    primaryClass = "astro-ph.CO",
    reportNumber = "CERN-TH-2024-217",
    doi = "10.1088/1475-7516/2025/05/062",
    journal = "JCAP",
    volume = "05",
    pages = "062",
    year = "2025"
}

@article{Gowling:2021gcy,
    author = "Gowling, Chloe and Hindmarsh, Mark",
    title = "{Observational prospects for phase transitions at LISA: Fisher matrix analysis}",
    eprint = "2106.05984",
    archivePrefix = "arXiv",
    primaryClass = "astro-ph.CO",
    doi = "10.1088/1475-7516/2021/10/039",
    journal = "JCAP",
    volume = "10",
    pages = "039",
    year = "2021"
}

@article{Gowling:2022pzb,
    author = "Gowling, Chloe and Hindmarsh, Mark and Hooper, Deanna C. and Torrado, Jes{\'u}s",
    title = "{Reconstructing physical parameters from template gravitational wave spectra at LISA: first order phase transitions}",
    eprint = "2209.13551",
    archivePrefix = "arXiv",
    primaryClass = "astro-ph.CO",
    doi = "10.1088/1475-7516/2023/04/061",
    journal = "JCAP",
    volume = "04",
    pages = "061",
    year = "2023"
}

@article{Giese:2021dnw,
    author = "Giese, Felix and Konstandin, Thomas and van de Vis, Jorinde",
    title = "{Finding sound shells in LISA mock data using likelihood sampling}",
    eprint = "2107.06275",
    archivePrefix = "arXiv",
    primaryClass = "astro-ph.CO",
    reportNumber = "DESY-21-109, DESY 2021-02966",
    doi = "10.1088/1475-7516/2021/11/002",
    journal = "JCAP",
    volume = "11",
    pages = "002",
    year = "2021"
}

@article{Dvali:2022vwh,
    author = "Dvali, Gia and Valbuena-Berm{\'u}dez, Juan Sebasti{\'a}n and Zantedeschi, Michael",
    title = "{Dynamics of confined monopoles and similarities with confined quarks}",
    eprint = "2210.14947",
    archivePrefix = "arXiv",
    primaryClass = "hep-th",
    doi = "10.1103/PhysRevD.107.076003",
    journal = "Phys. Rev. D",
    volume = "107",
    number = "7",
    pages = "076003",
    year = "2023"
}

@article{Vilenkin:1981bx,
    author = "Vilenkin, A.",
    title = "{Gravitational radiation from cosmic strings}",
    doi = "10.1016/0370-2693(81)91144-8",
    journal = "Phys. Lett. B",
    volume = "107",
    pages = "47--50",
    year = "1981"
}

@article{Adams:2013qma,
    author = "Adams, Matthew R. and Cornish, Neil J.",
    title = "{Detecting a Stochastic Gravitational Wave Background in the presence of a Galactic Foreground and Instrument Noise}",
    eprint = "1307.4116",
    archivePrefix = "arXiv",
    primaryClass = "gr-qc",
    doi = "10.1103/PhysRevD.89.022001",
    journal = "Phys. Rev. D",
    volume = "89",
    number = "2",
    pages = "022001",
    year = "2014"
}

@article{Boileau:2021sni,
    author = "Boileau, Guillaume and Lamberts, Astrid and Christensen, Nelson and Cornish, Neil J. and Meyer, Renate",
    title = "{Spectral separation of the stochastic gravitational-wave background for LISA in the context of a modulated Galactic foreground}",
    eprint = "2105.04283",
    archivePrefix = "arXiv",
    primaryClass = "gr-qc",
    doi = "10.1093/mnras/stab2575",
    journal = "Mon. Not. Roy. Astron. Soc.",
    volume = "508",
    number = "1",
    pages = "803--826",
    year = "2021",
    note = "[Erratum: Mon.Not.Roy.Astron.Soc. 508, 5554--5555 (2021)]"
}

@article{Korol:2021pun,
    author = "Korol, Valeriya and Hallakoun, Na'ama and Toonen, Silvia and Karnesis, Nikolaos",
    title = "{Observationally driven Galactic double white dwarf population for LISA}",
    eprint = "2109.10972",
    archivePrefix = "arXiv",
    primaryClass = "astro-ph.HE",
    doi = "10.1093/mnras/stac415",
    journal = "Mon. Not. Roy. Astron. Soc.",
    volume = "511",
    number = "4",
    pages = "5936--5947",
    year = "2022"
}

@article{Korol:2020lpq,
    author = "Korol, V. and others",
    title = "{Populations of double white dwarfs in Milky Way satellites and their detectability with LISA}",
    eprint = "2002.10462",
    archivePrefix = "arXiv",
    primaryClass = "astro-ph.GA",
    doi = "10.1051/0004-6361/202037764",
    journal = "Astron. Astrophys.",
    volume = "638",
    pages = "A153",
    year = "2020"
}

@article{Liu:2023qap,
    author = "Liu, Chang and Ruan, Wen-Hong and Guo, Zong-Kuan",
    title = "{Confusion noise from Galactic binaries for Taiji}",
    eprint = "2301.02821",
    archivePrefix = "arXiv",
    primaryClass = "astro-ph.IM",
    doi = "10.1103/PhysRevD.107.064021",
    journal = "Phys. Rev. D",
    volume = "107",
    number = "6",
    pages = "064021",
    year = "2023"
}

@article{Regimbau:2011rp,
    author = "Regimbau, Tania",
    title = "{The astrophysical gravitational wave stochastic background}",
    eprint = "1101.2762",
    archivePrefix = "arXiv",
    primaryClass = "astro-ph.CO",
    doi = "10.1088/1674-4527/11/4/001",
    journal = "Res. Astron. Astrophys.",
    volume = "11",
    pages = "369--390",
    year = "2011"
}

@article{Babak:2023lro,
    author = "Babak, Stanislav and Caprini, Chiara and Figueroa, Daniel G. and Karnesis, Nikolaos and Marcoccia, Paolo and Nardini, Germano and Pieroni, Mauro and Ricciardone, Angelo and Sesana, Alberto and Torrado, Jes{\'u}s",
    title = "{Stochastic gravitational wave background from stellar origin binary black holes in LISA}",
    eprint = "2304.06368",
    archivePrefix = "arXiv",
    primaryClass = "astro-ph.CO",
    doi = "10.1088/1475-7516/2023/08/034",
    journal = "JCAP",
    volume = "08",
    pages = "034",
    year = "2023"
}

@article{Lehoucq:2023zlt,
    author = "Lehoucq, Leonard and Dvorkin, Irina and Srinivasan, Rahul and Pellouin, Clement and Lamberts, Astrid",
    title = "{Astrophysical uncertainties in the gravitational-wave background from stellar-mass compact binary mergers}",
    eprint = "2306.09861",
    archivePrefix = "arXiv",
    primaryClass = "astro-ph.HE",
    doi = "10.1093/mnras/stad2917",
    journal = "Mon. Not. Roy. Astron. Soc.",
    volume = "526",
    number = "3",
    pages = "4378--4387",
    year = "2023"
}

@article{Phinney:2001di,
    author = "Phinney, E. S.",
    title = "{A Practical theorem on gravitational wave backgrounds}",
    eprint = "astro-ph/0108028",
    archivePrefix = "arXiv",
    journal = "",
    month = "7",
    year = "2001"
}

@article{Tinto:2004wu,
    author = "Tinto, Massimo and Dhurandhar, Sanjeev V.",
    title = "{TIME DELAY}",
    eprint = "gr-qc/0409034",
    archivePrefix = "arXiv",
    doi = "10.12942/lrr-2005-4",
    journal = "Living Rev. Rel.",
    volume = "8",
    pages = "4",
    year = "2005"
}

@article{McNamara:2008zz,
    author = "McNamara, P. and Vitale, S. and Danzmann, K.",
    editor = "Scott, Susan M. and McClelland, David E.",
    collaboration = "LISA",
    title = "{LISA Pathfinder}",
    doi = "10.1088/0264-9381/25/11/114034",
    journal = "Class. Quant. Grav.",
    volume = "25",
    pages = "114034",
    year = "2008"
}

@article{Armano:2018kix,
    author = "Armano, M. and others",
    title = "{Beyond the Required LISA Free-Fall Performance: New LISA Pathfinder Results down to 20  $\mu$Hz}",
    doi = "10.1103/PhysRevLett.120.061101",
    journal = "Phys. Rev. Lett.",
    volume = "120",
    number = "6",
    pages = "061101",
    year = "2018"
}

@article{Cornish:2001bb,
    author = "Cornish, Neil J.",
    title = "{Detecting a stochastic gravitational wave background with the Laser Interferometer Space Antenna}",
    eprint = "gr-qc/0106058",
    archivePrefix = "arXiv",
    doi = "10.1103/PhysRevD.65.022004",
    journal = "Phys. Rev. D",
    volume = "65",
    pages = "022004",
    year = "2002"
}

@article{Karnesis:2021tsh,
    author = "Karnesis, Nikolaos and Babak, Stanislav and Pieroni, Mauro and Cornish, Neil and Littenberg, Tyson",
    title = "{Characterization of the stochastic signal originating from compact binary populations as measured by LISA}",
    eprint = "2103.14598",
    archivePrefix = "arXiv",
    primaryClass = "astro-ph.IM",
    doi = "10.1103/PhysRevD.104.043019",
    journal = "Phys. Rev. D",
    volume = "104",
    number = "4",
    pages = "043019",
    year = "2021"
}

@article{Chen:2023zkb,
    author = "Chen, Zu-Cheng and Huang, Qing-Guo and Liu, Chang and Liu, Lang and Liu, Xiao-Jin and Wu, You and Wu, Yu-Mei and Yi, Zhu and You, Zhi-Qiang",
    title = "{Prospects for Taiji to detect a gravitational-wave background from cosmic strings}",
    eprint = "2310.00411",
    archivePrefix = "arXiv",
    primaryClass = "astro-ph.IM",
    doi = "10.1088/1475-7516/2024/03/022",
    journal = "JCAP",
    volume = "03",
    pages = "022",
    year = "2024"
}

@article{speagle2020dynesty,
  title={dynesty: a dynamic nested sampling package for estimating Bayesian posteriors and evidences},
  author={Speagle, Joshua S},
  journal={Monthly Notices of the Royal Astronomical Society},
  volume={493},
  number={3},
  pages={3132--3158},
  year={2020},
  publisher={Oxford University Press}
}

@article{Ashton:2018jfp,
    author = "Ashton, Gregory and others",
    title = "{BILBY: A user-friendly Bayesian inference library for gravitational-wave astronomy}",
    eprint = "1811.02042",
    archivePrefix = "arXiv",
    primaryClass = "astro-ph.IM",
    doi = "10.3847/1538-4365/ab06fc",
    journal = "Astrophys. J. Suppl.",
    volume = "241",
    number = "2",
    pages = "27",
    year = "2019"
}

@article{Samanta:2025jec,
    author = "Samanta, Rome",
    title = "{Probing leptogenesis at LISA: a Fisher analysis}",
    eprint = "2503.09884",
    archivePrefix = "arXiv",
    primaryClass = "hep-ph",
    doi = "10.1088/1475-7516/2025/08/095",
    journal = "JCAP",
    volume = "08",
    pages = "095",
    year = "2025"
}

@article{Kume:2024xvh,
    author = "Kume, Jun'ya and Peloso, Marco and Pieroni, Mauro and Ricciardone, Angelo",
    title = "{Assessing the impact of unequal noises and foreground modeling on SGWB reconstruction with LISA}",
    eprint = "2410.10342",
    archivePrefix = "arXiv",
    primaryClass = "gr-qc",
    reportNumber = "RESCEU-14/24, CERN-TH-2024-170",
    doi = "10.1088/1475-7516/2025/06/030",
    journal = "JCAP",
    volume = "06",
    pages = "030",
    year = "2025"
}

@article{Dimitriou:2025bvq,
    author = "Dimitriou, Androniki and Figueroa, Daniel G. and Simakachorn, Peera and Zaldivar, Bryan",
    title = "{Cosmic string gravitational wave backgrounds at LISA: I. Signal survey, template reconstruction, and model comparison}",
    eprint = "2508.05395",
    archivePrefix = "arXiv",
    primaryClass = "astro-ph.CO",
    doi = "10.1088/1475-7516/2026/05/037",
    journal = "JCAP",
    volume = "05",
    pages = "037",
    year = "2026"
}

@article{Dimitriou:2026agw,
    author = "Dimitriou, Androniki and Figueroa, Daniel G. and Simakachorn, Peera and Stomberg, Isak and Zaldivar, Bryan",
    title = "{Cosmic string gravitational wave backgrounds at LISA: II. Reconstruction of conventional signals over astrophysical foregrounds}",
    eprint = "2607.15219",
    archivePrefix = "arXiv",
    primaryClass = "astro-ph.CO",
    doi = "",
    journal = "",
    volume = "",
    pages = "",
    month = "7",
    year = "2026"
}

@article{Boileau:2020rpg,
    author = "Boileau, Guillaume and Christensen, Nelson and Meyer, Renate and Cornish, Neil J.",
    title = "{Spectral separation of the stochastic gravitational-wave background for LISA: Observing both cosmological and astrophysical backgrounds}",
    eprint = "2011.05055",
    archivePrefix = "arXiv",
    primaryClass = "gr-qc",
    doi = "10.1103/PhysRevD.103.103529",
    journal = "Phys. Rev. D",
    volume = "103",
    number = "10",
    pages = "103529",
    year = "2021"
}

@article{Romano:2016dpx,
    author = "Romano, Joseph D. and Cornish, Neil J.",
    title = "{Detection methods for stochastic gravitational-wave backgrounds: a unified treatment}",
    eprint = "1608.06889",
    archivePrefix = "arXiv",
    primaryClass = "gr-qc",
    doi = "10.1007/s41114-017-0004-1",
    journal = "Living Rev. Rel.",
    volume = "20",
    number = "1",
    pages = "2",
    year = "2017"
}

@article{Smith:2019wny,
    author = "Smith, Tristan L. and Caldwell, Robert",
    title = "{LISA for Cosmologists: Calculating the Signal-to-Noise Ratio for Stochastic and Deterministic Sources}",
    eprint = "1908.00546",
    archivePrefix = "arXiv",
    primaryClass = "astro-ph.CO",
    doi = "10.1103/PhysRevD.100.104055",
    journal = "Phys. Rev. D",
    volume = "100",
    number = "10",
    pages = "104055",
    year = "2019",
    note = "[Erratum: Phys.Rev.D 105, 029902 (2022)]"
}

@article{Babichev:2023pbf,
    author = "Babichev, E. and Gorbunov, D. and Ramazanov, S. and Samanta, R. and Vikman, A.",
    title = "{NANOGrav spectral index {\ensuremath{\gamma}}=3 from melting domain walls}",
    eprint = "2307.04582",
    archivePrefix = "arXiv",
    primaryClass = "hep-ph",
    doi = "10.1103/PhysRevD.108.123529",
    journal = "Phys. Rev. D",
    volume = "108",
    number = "12",
    pages = "123529",
    year = "2023"
}

@article{ligoo5,
    author = "Abbott, R. and others",
    collaboration = "KAGRA, Virgo, LIGO Scientific",
    title = "{Upper limits on the isotropic gravitational-wave background from Advanced LIGO and Advanced Virgo\textquoteright{}s third observing run}",
    eprint = "2101.12130",
    archivePrefix = "arXiv",
    primaryClass = "gr-qc",
    reportNumber = "LIGO-DCC-P2000314",
    doi = "10.1103/PhysRevD.104.022004",
    journal = "Phys. Rev. D",
    volume = "104",
    number = "2",
    pages = "022004",
    year = "2021"
}

@article{LIGOScientific:2016aoc,
    author = "Abbott, B. P. and others",
    collaboration = "LIGO Scientific, Virgo",
    title = "{Observation of Gravitational Waves from a Binary Black Hole Merger}",
    eprint = "1602.03837",
    archivePrefix = "arXiv",
    primaryClass = "gr-qc",
    reportNumber = "LIGO-P150914",
    doi = "10.1103/PhysRevLett.116.061102",
    journal = "Phys. Rev. Lett.",
    volume = "116",
    number = "6",
    pages = "061102",
    year = "2016"
}

@article{Datta:2026ffs,
    author = "Datta, Satyabrata and Samanta, Rome",
    title = "{Mapping Domain-Wall Bayesian Reconstruction with LISA}",
    eprint = "2606.15387",
    archivePrefix = "arXiv",
    primaryClass = "astro-ph.HE",
     doi = "1",
    journal = "",
    volume = "",
    number = "",
    pages = "",
    month = "6",
    year = "2026"
}

@article{lisa,
    author = "Amaro-Seoane, Pau and others",
    collaboration = "LISA",
    title = "{Laser Interferometer Space Antenna}",
    eprint = "1702.00786",
    archivePrefix = "arXiv",
    primaryClass = "astro-ph.IM",
    journal = "",
    month = "2",
    year = "2017"
}

@article{ng1,
    author = "Agazie, Gabriella and others",
    collaboration = "NANOGrav",
    title = "{The NANOGrav 15 yr Data Set: Evidence for a Gravitational-wave Background}",
    eprint = "2306.16213",
    archivePrefix = "arXiv",
    primaryClass = "astro-ph.HE",
    doi = "10.3847/2041-8213/acdac6",
    journal = "Astrophys. J. Lett.",
    volume = "951",
    number = "1",
    pages = "L8",
    year = "2023"
}

@article{ng2,
    author = "Antoniadis, J. and others",
    collaboration = "EPTA, InPTA:",
    title = "{The second data release from the European Pulsar Timing Array - III. Search for gravitational wave signals}",
    eprint = "2306.16214",
    archivePrefix = "arXiv",
    primaryClass = "astro-ph.HE",
    doi = "10.1051/0004-6361/202346844",
    journal = "Astron. Astrophys.",
    volume = "678",
    pages = "A50",
    year = "2023"
}

@article{ng3,
    author = "Reardon, Daniel J. and others",
    title = "{Search for an Isotropic Gravitational-wave Background with the Parkes Pulsar Timing Array}",
    eprint = "2306.16215",
    archivePrefix = "arXiv",
    primaryClass = "astro-ph.HE",
    doi = "10.3847/2041-8213/acdd02",
    journal = "Astrophys. J. Lett.",
    volume = "951",
    number = "1",
    pages = "L6",
    year = "2023"
}

@article{ng4,
    author = "Xu, Heng and others",
    title = "{Searching for the Nano-Hertz Stochastic Gravitational Wave Background with the Chinese Pulsar Timing Array Data Release I}",
    eprint = "2306.16216",
    archivePrefix = "arXiv",
    primaryClass = "astro-ph.HE",
    doi = "10.1088/1674-4527/acdfa5",
    journal = "Res. Astron. Astrophys.",
    volume = "23",
    number = "7",
    pages = "075024",
    year = "2023"
}

@article{ng5,
    author = "Afzal, Adeela and others",
    collaboration = "NANOGrav",
    title = "{The NANOGrav 15 yr Data Set: Search for Signals from New Physics}",
    eprint = "2306.16219",
    archivePrefix = "arXiv",
    primaryClass = "astro-ph.HE",
    reportNumber = "FERMILAB-PUB-23-589-T",
    doi = "10.3847/2041-8213/acdc91",
    journal = "Astrophys. J. Lett.",
    volume = "951",
    number = "1",
    pages = "L11",
    year = "2023"
}
\end{document}